%% file: Draft_2026-11-22_Arxiv.tex
\documentclass[12pt,a4paper]{article}

\input{preamble.tex}

\usepackage[final]{microtype}
\usepackage{libertine}

\usepackage{placeins}

\usepackage[
  backend=biber,
  style=authoryear,
  natbib=true,      
  giveninits=true,  
  doi=false,
  isbn=false,
  url=false,
  eprint=false,
  uniquelist=false,
  backref=false
]{biblatex}
\renewbibmacro{in:}{}
\newcommand\posscite[1]{\citeauthor{#1}'s (\citeyear{#1})}

\usepackage{tikz}
\usetikzlibrary{positioning,calc,fit,arrows.meta}

\title{Ancestral origins of environmental (in)attention}
\author{C\'esar Barilla \hspace{.1cm} \textcircled{r} \hspace{.1cm}Palaash Bhargava}

\date{ \today }

\makeatletter
\let\thetitle\@title
\let\theauthor\@author
\let\thedate\@date
\makeatother

\titleformat{\section}{\centering\scshape\large}{\thesection}{0.7em}{}
\titleformat{\subsection}{\centering\normalfont\normalsize\scshape}{\thesubsection}{0.7em}{}
\titleformat*{\paragraph}{\scshape\bfseries}

\titlespacing*{\section}{0pt}{1.5ex plus .5ex}{1ex}
\titlespacing*{\subsection}{0pt}{1ex plus .5ex}{0.5ex}
\titlespacing*{\paragraph}{0pt}{.75ex plus 1ex minus .2ex}{.75em}           

\begin{document}

    ~\vspace*{-2em}
    \begin{center}
        {\noindent
        \Large
        \bfseries 
        \scshape
        \thetitle
        }
    
        \vspace*{1.5em}
        
        \makebox[\textwidth][c]{
        \begin{minipage}{1.2\linewidth}
        \centering
        \begin{minipage}{5cm}
            \centering
            {\large \scshape César Barilla}
        \end{minipage}
        \hspace{1em} \textcircled{\scriptsize r} \hspace{.6em}
        \setcounter{footnote}{0}
        \begin{minipage}{5cm}
            \centering
            {\large \scshape Palaash Bhargava}
        \end{minipage}
        \end{minipage}
        }

    \end{center}

    


    \vspace*{2em}    

    \noindent\makebox[\textwidth][c]{
        \begin{minipage}{.9\textwidth}
            \noindent
            {\scshape \bfseries Abstract} \hspace{.5em} 
            How do ancestral climate experiences shape today’s attitudes towards environmental issues? Combining multiple contemporary surveys and ethnic group-level cultural records with a model of cultural selection on costly attention, we show that ancestral climate anomalies have a lasting U-shaped effect on environmental concern. Descendants of groups exposed to the most regular and the most anomalous climates attach greater importance to the environment, with a dip in between; a pattern mirrored in cultural narratives. But why should contemporary environmental concern remain anchored in ancestral climates when lessons about the value of environmental attention can be learned beyond the conditions in which they were initially acquired? Our model shows that because attention is chosen before conditions are realized, cultural selection rewards perceptions that perform well on average without fully correcting their accuracy across conditions. Ancestral climates can thus leave lasting imprints, even as their influence weakens when environments change. 
            The combination of motives to exploit typical conditions and to protect against extremes gives rise to the U-shape, reconciling anthropological accounts that tie environmental concern to regular and to anomalous climates.
            \\[.5em]
        \end{minipage}
    }

    \blfootnote{
        This version: \textcolor{red!50!black}{\today}.
        \textcircled{\scriptsize r} stands for random order author naming convention.
    }

    \blfootnote{
        César Barilla: University of Oxford \& Nuffield College, \hyperlink{cesar.barilla@economics.ox.ac.uk}{cesar.barilla@economics.ox.ac.uk}; Palaash Bhargava: Max Planck Institute for Behavioral Economics, \hyperlink{bhargava@econ.mpg.de}{bhargava@econ.mpg.de}.
        We thank 
        Michael Best,
        Shivangi Bishnoi,
        Sandra Black, 
        Laura Boudreau,
        Yeon-Koo Che, 
        Mark Dean,
        Edgard Dewitte,
        Léa Dousset,
        Ruben Durante,
        Alex Eble,
        Dalila Figueiredo,
        Duarte Gon\c{c}alves,
        Florian Grosset,
        Navin Kartik,
        Ajinkya Keskar,
        Tushar Kundu,
        Elliot Lipnowski,
        Nicolas Longuet-Marx,
        Margaret Meyer,
        Nathan Nunn, 
        Cristian Pop-Eleches, 
        Nishith Prakash,
        Michael Price,
        Bernard Salanié,
        Matthias Sch\"{u}ndeln,
        Matthias Sutter,
        Miguel Urquiola, 
        Akanksha Vardani, 
        and Jack Willis
        for helpful comments and discussions. 
        We are grateful to seminar participants at Columbia, Oxford, University of Alabama, SUNY Binghamton, University of Rochester, Historical Danish Political Economy, MPWZ Workshop, Max Planck Alumni Workshop, NHH FAIR and IPWSD for feedback. 
        Uday Bangabash, Pari Bishen, Kevin Li Grant, Gina Nemade and Smriti Raj Sharma provided excellent research assistance. 
        We acknowledge the support of Columbia's Program for Economic Research. 
    }

\vspace{-1em}
            
\section{Introduction}\label{sec:introduction}

    Climate change raises the stakes of understanding the heterogeneity of environmental attitudes: the effectiveness of climate policies depends on the importance individuals and communities attach to environmental issues. While existing explanations for such variation usually focus on contemporary factors, select anthropological studies trace communities' environmental concern to their ancestors' experiences. 
    
    Yet these accounts associate such concern with contrasting ancestral climates.
    The development of environmental stewardship is sometimes attributed to the ritualization of ecological knowledge  (for ethnic groups such as the Toraja, Manobo and Balinese) which is more valuable when climate realizations are closer to typical conditions \citep{lansing1993emergent,berkes2017sacred,atran2002folkecology}.
    By contrast, other accounts point to ancestral climates marked by frequent or intense departures from typical conditions (for, e.g., the Sámi, Quechua and Māori), increasing the need for protection against extreme events \citep{berkes1998linking, gomez2013reinterpreting, bwambale2018traditional}.
    Both observations, however, single out the same feature of ancestral environments: the intensity of deviations from typical conditions.

    Combining global-scale evidence with a theory of cultural selection on costly attention, we ask whether these patterns generalize: \emph{do ancestral climatic conditions systematically shape the importance descendants attach to environmental issues, and if so, why and in which direction?} 
    Anthropological evidence suggests that the value of internalizing environmental considerations is tied to returns from adaptation. 
    This points to a ``dual motive'': regular climates reward careful adaptation to typical conditions, while highly variable climates reward preparedness for extreme events. Both may therefore foster environmental concern, albeit for different reasons. If these attitudes were to persist across generations, we would expect a non-monotonic relationship, with greater concern among descendants of populations exposed to either very regular or highly anomalous ancestral climates.

    While it is intuitive that ancestral climatic experiences shaped behavior in the past, why should the importance individuals attach to environmental issues today remain anchored to the climates their ancestors experienced generations ago? Although persistence is often the default expectation in studies of cultural transmission, there are good reasons to expect such linkages to fade in our setting. 
    First, what is transmitted here is a perception of when and how much environmental attention matters, not merely specific imitated practices. Different climatic conditions may call for different levels of attention, but individuals can, in principle, learn which conditions warrant greater attention and which do not. Given the stakes involved, such knowledge should be valuable and applicable across environments: individuals recognize the value of environmental attention in different conditions, regardless of where their ancestors lived. Second, descendants today live in environments fundamentally different from those of their pre-industrial ancestors, where technology, markets, and institutions have transformed how people interact with nature. As \citet{nunn_giuliano} suggest, inherited cultural traits lose relevance as the conditions that shaped them change. Why, then, should ancestral climates continue to influence contemporary environmental concern, even when individuals have incentives to learn and adapt? Our theory offers a distinct explanation: because attention must be chosen before environmental conditions are realized, cultural selection rewards perceptions that perform well on average under ancestral conditions, without necessarily correcting their assessment across the range. As a result, ancestral climates can leave a lasting imprint on environmental concern. This influence may weaken as descendants' environments diverge from those of their ancestors, but need not disappear.

    We construct a dataset with novel components combining eight sources of contemporary environmental attitudes and cultural narratives globally, ancestral population characteristics, and historical climate records (see \Cref{fig:synthetic-overview} for an overview). Our analysis operates at two complementary levels: individual-level survey measures of environmental concern and ethnic (linguistic) group-level measures of environmental themes in cultural narratives. At the individual level, our primary outcome comes from the fifth and sixth waves of the World Values Survey (WVS), which directly measure how much individuals value caring for the environment.\footnote{Conducted between 2005 and 2014, covering over 150,000 individuals across 78 countries and 202 ethnic groups.} We corroborate these results using composite indices of environmental attitudes, opinions, and actions constructed from the Integrated Values Survey and the Environmental Module of the European Social Survey.\footnote{These cover approximately 300,000 individuals across 93 countries and 30,000 individuals across 23 countries.} At the ethnic group level, we measure the prevalence of environmental themes in folklore using dictionary-based methods applied to \citet{folklore}, complemented by natural language processing measures constructed from ethnographic descriptions in Wikipedia and the Joshua Project.

    We link outcomes at both levels to ancestral environmental conditions by matching individuals and cultural narratives to ethnic groups through spoken language, using the Ancestral Characteristics of Modern Populations database of \citet{nunn_giu_database}.\footnote{Our analysis incorporates ethnic groups represented in multiple ethnographic databases, including \citet{Murdock}.} We then combine the historical locations of these groups with reconstructed temperature records from \citet{mann_2009}, allowing us to characterize the climatic conditions experienced by their ancestors between 1600 and 1920 C.E. Our primary measure of ancestral climate anomalies captures the average magnitude of deviations from typical climatic conditions within each generation, aggregated across generations. 

    \begin{figure}[t!]
        \centering
        \begin{tikzpicture}[
          every node/.style={align=center},
          xscale=4, 
          yscale=2.5,
          >=Stealth,
          tips=proper,
          ] 
          
            \node[draw,TolMutedIndigo!80!black] (value) at (-1.5,0) {
                \footnotesize \textit{Individual-level outcomes}
                \\[.25em]
                \small \textbf{Perceived value of attention} 
                \\[.15em] 
                \small to environmental considerations
                \\ \small today \& across situations
                };

            \node[draw,TolMutedRose] (ancestors) at (0,0) {
                \small Ancestral Ethnic Groups
                };
            \draw [thick,dashed,gray, <->] (value) edge[] node [midway, label={[label distance=-.1cm, rotate=0] above:{\tiny language}}] {} (ancestors) {} ;
    
            \node[draw,TolMutedGreen] (temperature) at (1.5,0) {
              \footnotesize \textit{Explanatory variable}
              \\[.25em]
              \small \textbf{Ancestral Climate Conditions}
              \\[.15em] 
              \small intensity of "anomalies"
              };
            \draw [thick,dashed,gray, <->] (ancestors) edge[] node [midway,  label={[label distance=-.1cm, rotate=0] below:{\tiny location}}] {} (temperature) {} ;

            \draw [thick,TolMutedTeal, ->] (ancestors.north) edge[bend right=45] node [pos = .6,  label={[label distance=-.05cm, rotate=0] above:{\footnotesize cultural transmission}}] {} (value.north) {} ;

            \draw [thick,TolMutedTeal, ->] (temperature.north) edge[bend right=45] node [pos = .4,  label={[label distance=-.05cm, rotate=0] above:{\footnotesize experiences}}] {} (ancestors.north) {} ;

            \node[draw,TolMutedPurple!80!black] (folklore) at (-1.2,-1.2) {
              \footnotesize \textit{Group-level outcomes}
              \\[.25em]
              \small \textbf{Environmental themes} 
              \\[.15em] 
              \small in cultural material 
              \\
              \small (folklore, ethnographic descriptions)
              }; 
            \draw [thick,dashed,gray, <->] (folklore) edge[bend left =-30] node [midway, label={[label distance=-.1cm, rotate=45] above:{\tiny language}}] {} (ancestors) {} ;
          
            \draw [thick,TolMutedTeal, ->] (ancestors.south) edge[bend right=-45] node [pos = .5,  label={[label distance=-.05cm, rotate=45] below:{\footnotesize cultural records}}] {} (folklore.east) {} ;
        \end{tikzpicture}
        \caption{Overview of variables, theoretical channels, and matching strategy.}
        \label{fig:synthetic-overview}
        {\footnotesize Solid arrows represent the channels that we formalize in our theory and test in the data; dashed arrows represent the matching channels that enable us to link outcomes to ancestral ethnic/linguistic groups to experienced conditions.}
    \end{figure}

    We develop a model combining cultural transmission with costly attention to explain why ancestral climates continue to shape contemporaneous environmental concern. Individuals, within a generation, decide how much attention to devote to environmental considerations, in the various situations they face. They do so knowing the \emph{distribution} of the climate they face but before knowing the specific \emph{conditions} that will realize. Their decisions depend on their beliefs about when and how much such attention matters, which we refer to as their cultural type. Each generation learns from multiple cultural models; cultural types that yield better outcomes are more likely to be taken as models and become more prevalent, subject to the additional influence of random mutations and a tendency to revert toward a default type. Yet this process does not necessarily produce accurate beliefs about every possible condition. Because attention is chosen in advance, selection rewards beliefs that perform well on average, without correcting how individuals perceive its value in other circumstances. \Cref{thm:convergence} characterizes long-run outcomes: when reversion is weak, the prevailing cultural type is the one closest to the default among those that correctly capture the \emph{average} value of attention under ancestral conditions. Skewed beliefs can therefore persist across generations without being equally accurate in other environments. 
    
    We then introduce the two motives suggested by anthropological accounts: adapting to typical environmental conditions (exploitation) and preparing for extreme events (protection). The former makes attention most valuable when conditions are close to normal, while the latter makes it most valuable when conditions deviate substantially from normal. On their own, these motives generate no lasting effect of ancestral climate: with perfect learning, descendants would correctly perceive the value of environmental attention under the conditions they face today, regardless of their ancestors' experiences. Only when these motives interact with coarse cultural selection do ancestral climates leave a lasting imprint. \Cref{thm:U-shape} shows that this produces an approximately U-shaped relationship between descendants' perceived value of environmental attention and the average intensity of climate anomalies in the \emph{distribution} of ancestral climates. The shape of this relationship depends on the underlying motives: adaptation to typical conditions alone predicts declining concern as ancestral anomalies increase, while protection against extremes alone predicts increasing concern. Their combination produces the U-shape. The model takes these \emph{condition-level} motives from the anthropological evidence and explains how they can generate persistent, empirically distinguishable patterns in environmental attitudes across populations today based on \emph{distribution-level} ancestral heterogeneity.
    
    Turning to the empirical evidence, we examine whether the patterns predicted by our theoretical framework are reflected in contemporary environmental attitudes. Our individual-level results document the U-shaped relationship between ancestral climate anomalies and contemporary environmental concern that our theory formalizes. Descendants of populations that experienced either very regular or highly anomalous climates attach greater importance to environmental issues today, while those whose ancestors experienced intermediate levels of climatic variability attach the least importance. Through the lens of our theory, this pattern reflects the different incentives for environmental attention across ancestral climates. In particular, intermediate variability represents conditions where neither motive is strongly rewarded: typical conditions are not sufficiently regular for fine-tuned adaptation to consistently pay off, while extreme events are not sufficiently frequent or intense to strongly reward protective behavior. Importantly, this U-shaped relationship is remarkably consistent across independent datasets, alternative measures of environmental concern, different constructions of ancestral climate anomalies, and a wide range of empirical specifications. The relationship remains evident after accounting for individual characteristics, historical differences across ethnic groups, geographic conditions, and changes in temperature across ancestral generations \citep{nunn_giuliano}. It is also reproduced using alternative measures of environmental attitudes and actions from the Integrated Values Survey and the European Social Survey.
    
    The same U-shape appears at the ethnic group level, in the prevalence of environmental themes in folklore. This is consistent with a higher perceived value of environmental attention being reflected in cultural discourse and retained in the collective memory of successive generations. The relationship holds across alternative measures constructed using natural language processing of rich ethnographic descriptions from Wikipedia and the Joshua Project, expanding both the depth of environmental themes captured and the coverage of ethnic groups across the world. Together, the two levels support the model's account: the pattern appears not only in individual attitudes but also in the cultural material through which they may be transmitted.

    The cultural selection framework generates further predictions beyond the U-shape, which we examine empirically. First, while ancestral environments can leave lasting imprints on descendants' perceptions, these effects should be weaker when contemporary climatic conditions differ substantially from those experienced by their ancestors. Consistent with this prediction, we find a more pronounced U-shaped relationship among individuals whose current and ancestral climatic conditions are relatively similar, although the pattern remains evident even when they differ. Second, if environmental attention is shaped by the returns to adaptation, ancestral populations that were more economically developed and less reliant on the natural environment should have faced weaker incentives to develop such attitudes. A comparative statics result formalizes this prediction, and our empirical heterogeneity analysis corroborates it across several measures of ancestral ethnic group development. We also find suggestive evidence that contemporaneous exposure to damages from climate shocks accentuates the relationship. Finally, the U-shaped pattern is nearly identical among children of immigrants and non-immigrants, consistent with the role of language-based cultural channels rather than current geographic location.

    \paragraph{Related Literature} 
    Our paper bridges ecological anthropology and the economics of cultural transmission. Several works in ecological anthropology study how environmental norms and practices emerge from ancestral ecological experiences \citep{berkes2017sacred, lansing1993emergent, atran2002folkecology, gomez2013reinterpreting}. 
    We generalize these insights at a global scale by bringing them to the literature on the historical origins of culture \citep{nunn2009importance, fernandez2025understanding}, which has documented the enduring influence of ancestral experiences on religious practices \citep{mccleary2006religion}, behavioral traits \citep{lowes2017evolution}, human capital decisions \citep{becker2020forced}, and norms related to prosociality, cooperation, gender, and economic preferences \citep{alesina2013origins, lowes_2018, becker2020ancient}. 
    Close to our work are studies linking ancestral conditions to contemporary culture. \citet{nunn_giuliano} show how differences in environmental conditions across generations influence the persistence of traditional practices. We shift attention from whether inherited traditions remain useful to what individuals learn about the value of environmental considerations itself and why these perceptions persist even when environments change. Similarly, \citet{buggle2021climate} document how historical climatic variability fostered cooperation and trust in European regions. 
    The driving force between conditions and attention is different; we document a non-monotonic relationship and provide a distinct explanation for its persistence. Finally, we relate to work on narratives as vehicles of collective memory and values \citep{shiller2017narrative, akerlof2016bread, folklore}, including ecological knowledge \citep{osemeobo1994role, ghana_folklore, sanganyado2018impact}, by showing that environmental themes exhibit the same historical patterns as individual attitudes.

    Our model contributes to theories of cultural transmission by examining why historically shaped perceptions persist, rather than taking persistence as given, and drawing a new connection between (in)attention and evolutionary dynamics. 
    We build on the use of tools from evolutionary biology to model culture \citep{cavalli1981cultural, boyd1988culture}.
    In economics, a large literature following \citet{bisinverdier2001} studies endogenous socialization incentives and institutions \citep[see][for a review]{BisinVerdier2023}; another related strand uses evolutionary game theory to examine the dynamics of social norms, see e.g. \citet{jindani2026dynamics,gulesci2025stepping} for recent contributions \citep[see also][for a review]{young2015evolution}. 
    We connect the canonical rational inattention framework \citep[][for a review]{Sims2003,MackowiakMatejkaWiederholt2023} with cultural transmission shaping the \emph{prior perception of value} in the attention problem. Although previous work emphasizes the role of information in cultural transmission \citep[e.g.][]{Adriani2009parents}, to the best of our knowledge, the formal combination of rational inattention and cultural transmission is novel.

    Finally, we contribute to research on the determinants of environmental and climate attitudes, which emphasizes contemporary factors such as emotion-based learning \citep{leiserowitz}, personal experience and morality \citep{weber1997perception, hansen2004role}, misperceived norms \citep{falk_climate}, political ideology, and economic preferences \citep{luo_zhao, shi2016knowledge}. Particularly relevant is \citet{Dewitte2024}, who demonstrates how historical exposure to fossil fuel extraction shaped economic identities and contemporary climate change denial in the United States. 
    Closer to the mechanisms of intergenerational transmission, \citet{Prakash2025} provide experimental evidence that environmental education can shape attitudes and behaviors within families.
    
    The paper is structured as follows. 
    \Cref{sec:data} describes the data and provides a non-technical overview of the theoretical framework.
    \Cref{sec:empirical-results} presents the empirical strategy and the results. 
    \Cref{sec:theory} introduces the model and the theoretical results.
    \Cref{sec:robustness-checks} provides robustness checks, and \Cref{sec:heterogeneity} the heterogeneity analysis.
    \Cref{sec:conclusion} concludes.

\section{Data and conceptual framework}\label{sec:data}


    We first define the empirical objects: contemporary perceptions of the value of environmental concerns, environmental themes in cultural material, links between individuals and ancestral ethnic groups, and historical climate anomalies. 
    We introduce the key variables we study and the eight different data sources used to construct them (\Cref{subsec:construction-of-variables}, see \Cref{datasources} for an overview). 
    Having defined our objects of interest, we then build them into a conceptual framework (\Cref{subsec:theory-overview}, see \Cref{fig:mechanism-conceptual-map} for an overview) to analyze the impact of ancestral experiences on descendants’ perceptions via cultural transmission. This foreshadows and gives an intuitive overview of the model (\Cref{sec:theory}).

    \subsection{Construction of variables of interest}\label{subsec:construction-of-variables}
    
    Our main outcome of interest is the \emph{subjective value of attention to environmental issues} at the individual level -- for which we provide multiple proxies relying on the Integrated Value Surveys \citep{ivs} and European Social Survey \citep{ess8}.
    Our second key outcome is \emph{the prevalence of environmental themes in cultural material} (folklore data from \citet{folklore}, written records from The Joshua Project and Wikipedia), capturing group level beliefs, practices and traditions.
    For each outcome, we highlight one main variable that will serve as our baseline to illustrate the results, which are then corroborated across multiple alternative measures.
    Our central explanatory variable is \emph{the average intensity of ancestral climate anomalies}, and linking is done through language-based matching.
    This and additional variables capturing ancestral climatic conditions are constructed using historical temperature data from \citet{mann_2009}.

    \begin{figure}[t!]
		\centering
        \resizebox{\textwidth}{!}{
          \begin{tikzpicture}[
                x=1cm, y=1cm,
                font=\small,
                >=Stealth,
                flow/.style={dashed, draw=gray!75, line width=0.6pt, -{Stealth[length=2.2mm,width=2.2mm]}},
                box/.style={draw, line width=1pt, fill=white, align=center, minimum height=9mm},
                big/.style={box, fill=gray!10, minimum height=12mm},
              ]
              
              \begin{scope}[local bounding box=diagram]
              
              \def\yM{0.0}     
              \def\yN{-1.6}    
              \def\yS{-3.2}    
              \def\yT{-4.6}    
              \def\yB{-6.4}    
              
              \def\xM{1.5}
              \def\xN{-4.0}
              
              \def\xIVS{-6.1}
              \def\xESS{-3.0}
              \def\xWIKI{-0.2}
              \def\xFOLK{4.0}
              \def\xJOSH{9.0}
              
              \def\xWVS{-6.9}
              \def\xEVS{-5.3}
              
              \def\xIND{-4.5}
              \def\xGRP{5.2}
              
              \node[box, draw=TolMutedGreen, text=TolMutedGreen, minimum width=16.5cm]
                (mann) at (\xM,\yM)
                {Historical Temperature Series (Mann et al., 2009) $\rightsquigarrow$ \textbf{avg. anomalies index}};
              
              \node[box, draw=TolMutedRose, text=TolMutedRose, minimum width=7.9cm]
                (nunn) at (\xN,\yN)
                {Ancestral Characteristics of Modern Populations\\(Giuliano and Nunn, 2018)};
              
              \node[box, draw=TolMutedIndigo, text=TolMutedIndigo, minimum width=3.8cm]
                (ivs) at (\xIVS,\yS) {IVS};
              
              \node[box, draw=TolMutedIndigo, text=TolMutedIndigo, minimum width=2.1cm]
                (ess) at (\xESS,\yS) {ESS};
              
              \node[box, draw=TolMutedPurple, text=TolMutedPurple, minimum width=2.6cm]
                (wiki) at (\xWIKI,\yS) {Wikipedia};
              
              \node[box, draw=TolMutedPurple, text=TolMutedPurple, minimum width=5.1cm]
                (folk) at (\xFOLK,\yS) {Folklore \\ (Michalopoulos and Xue, 2021)};
              
              \node[box, draw=TolMutedPurple, text=TolMutedPurple, minimum width=4.3cm]
                (josh) at (\xJOSH,\yS) {The Joshua Project};
              
              \node[box, draw=TolMutedIndigo, text=TolMutedIndigo, minimum width=1.5cm]
                (wvs) at (\xWVS,\yT) {WVS};
              
              \node[box, draw=TolMutedIndigo, text=TolMutedIndigo, minimum width=1.5cm]
                (evs) at (\xEVS,\yT) {EVS};
              
              \node[big, draw=TolMutedIndigo!80!black, text=TolMutedIndigo!80!black, minimum width=6.9cm]
                (indiv) at (\xIND,\yB)
                {\textbf{Individual-level outcomes} \\ value of environmental concerns\\ \& pro-environmental action indices};
              
              \node[big, draw=TolMutedPurple!80!black, text=TolMutedPurple!80!black, minimum width=11.3cm]
                (group) at (\xGRP,\yB)
                {\textbf{Group-level outcomes} \\ prevalence of environmental themes\\captured via textual classification \& narrative analysis};
              
              
              \draw[flow] ($(mann.south west)!(nunn.north)!(mann.south east)$) -- (nunn.north);
              \draw[flow] ($(mann.south west)!(folk.north)!(mann.south east)$) -- (folk.north);
              \draw[flow] ($(mann.south west)!(josh.north)!(mann.south east)$) -- (josh.north);
              
              \draw[flow] ($(nunn.south west)!(ivs.north)!(nunn.south east)$) -- (ivs.north);
              \draw[flow] ($(nunn.south west)!(ess.north)!(nunn.south east)$) -- (ess.north);
              \draw[flow] ($(nunn.south west)!(wiki.north)!(nunn.south east)$) -- (wiki.north);
              
              \draw[flow] ($(ivs.south west)!(wvs.north)!(ivs.south east)$) -- (wvs.north);
              \draw[flow] ($(ivs.south west)!(evs.north)!(ivs.south east)$) -- (evs.north);
              
              \draw[flow] (wvs.south) -- ($(indiv.north west)!(wvs.south)!(indiv.north east)$);
              \draw[flow] (evs.south) -- ($(indiv.north west)!(evs.south)!(indiv.north east)$);
              \draw[flow] (ess.south) -- ($(indiv.north west)!(ess.south)!(indiv.north east)$);
              
              \draw[flow] (wiki.south) -- ($(group.north west)!(wiki.south)!(group.north east)$);
              \draw[flow] (folk.south) -- ($(group.north west)!(folk.south)!(group.north east)$);
              \draw[flow] (josh.south) -- ($(group.north west)!(josh.south)!(group.north east)$);
              
              \end{scope}
              
              
          \end{tikzpicture}
          }
		\caption{Datasets and merging strategy} \label{datasources}
	\end{figure} 

    \paragraph{Individual-level data -- environmental concern}\label{env_pref}
    Our individual level measures are proxies for the subjective belief about \emph{the overall value of environmental concerns}, which guides costly effort decisions in factoring in environmental concern across situations.
    Our main dependent variable is individual responses on whether \textit{``Looking after the environment is important to this person; to care for nature and save life resources.''}, from $1$ (not important at all) to $6$ (extremely important to me),
    in waves 5 and 6 of \textbf{World Values Surveys} section of the Integrated Value Surveys.\footnote{
        The Integrated Values Surveys (IVS) is a harmonized global dataset that integrates and standardizes responses from the World Values Survey (WVS) and the European Values Study (EVS), and is designed to facilitate cross-country and within-country analyses over time of human values, beliefs, and preferences using consistent measures.
        WVS covers nationally representative samples from over 100 countries between 1981 and 2022. 
        Individuals are asked questions along many thematic categories (perceptions of life, environment, work, family, politics, society, religion and national identity). 
        For our lead question of interest, we have a total of 157,142 respondents from 78 countries in waves 5 and 6 (2005-2014).
    }
	The broad and context-free framing of the question makes it a suitable proxy for an individual's assessment of the value of making efforts to accommodate environmental concerns (and figure out how to optimally do so) across situations.
    Additionally, the Integrated Value Survey provides an array of demographic information on the respondents, including language spoken at home. This is later used to link individuals to their ethnic ancestors, via the language or dialect associated to the ethnic groups available in the ancestral characteristics database.
	
	We construct alternative proxies for the value of environmental attention.
    For our second set of proxies, we use other questions from the World Values Survey (WVS) and the European Values Study (EVS) that capture self-reported engagement in pro-environmental actions and support for environmental causes.
    Since responses to these questions may reflect both demand- and supply-side considerations and often require individuals to make explicit trade-offs, we recover a latent measure of intrinsic environmental concern by conducting a principal component analysis (PCA), and use the first principal component as our variable.\footnote{
        See \Cref{subsec:IVS-details} for further details about the questions we use for our PCA, data construction, sample size and additional demographic characteristics we use as control variables in our regressions on variables constructed from IVS.
    }
	Lastly, we perform a similar PCA exercise on the climate module of the 8\textsuperscript{th} round of the European Social Survey (ESS) to construct a third set of proxy variables.\footnote{
        Similar to IVS, the ESS is a cross-national repeated social survey that measures public attitudes, beliefs, values, and behaviors, primarily across European countries. 
        The 8\textsuperscript{th} round contains a dedicated climate module that elicits respondents' views and behaviors related to climate change, energy use, and environmental policy. See \Cref{subsec:ESS-details} for details on the climate module, additional demographic variables used as controls and our associated data construction strategy.
        }
    We construct four alternative measures of the value of environmental attention: the first three gradually expand the set of included questions (simultaneously reducing the subsample), where questions concern various environmental attitudes and tradeoffs; the fourth PCA is based solely on climate-change related beliefs.\footnote{
        Information on constructed proxies available for approximately 9,000 to 33,000 respondents across 14 to 23 countries.
        }
	As with the IVS, the ESS provides rich demographic information on respondents, which we similarly use to link individuals to their ethnic ancestry (via language) and provide controls.
    For all constructions across different datasets, individual survey responses and the first principal component are normalized to lie between 0 and 1, with higher numbers indicating greater subjective value.

    \paragraph{Group level data -- environmental themes}
    Our lead group-level variable of interest is a measure of the \emph{relative proportion of environmentally-related motifs in folklore}, constructed using the data from \citet{folklore}. 
	This folklore database builds on \posscite{berezkin} catalog of motifs and links the linguistic groups therein to the ethnic groups in \posscite{Murdock} Ethnographic Atlas, providing a distribution of oral traditions and folktales associated to each ethnic group (see \Cref{subsec:folklore-details} for details). 
    Following the approach from \citet{folklore}, we use \emph{ConceptNet}\footnote{
        ConceptNet is a knowledge representation project designed to reflect human knowledge and its expression in language.
        } 
    \citep{conceptnet} to classify motifs associated to each ethnic group as an environment-related or non-environment-related motif. 
    We use the seed words: weather, climate, temperature, environment and natural disaster. 
    We then retrieve all associated words and phrases for each seed that have high relevance score and manually verify the relevance of each item. 
    This yields a list of 228 words which we employ in a dictionary-based approach to classify motifs as environment vs non-environment-related.
    Using this information, our main measure of environment focus in folklore is: 
	\[
	\mathfrak{F}_e := \log\left(1 + \dfrac{\text{No. of environment-related motifs}}{\text{Total no. of motifs}}\right)
	\]
	for each ethnic group $e$ in the Ethnographic Atlas sample. 
	
	For robustness, we construct alternative versions of our main outcome using (i) the raw share of environmentally related motifs among all motifs associated with an ethnic group, (ii) an inverse hyperbolic sine transformation of this share, and (iii) a log-link transformation of the share (for log-Poisson regressions). 
    We also construct an alternative classification based on subject–verb–object triplets, identifying environmental motifs through manually classified subjects and objects. This provides a syntactic alternative to the ConceptNet-based dictionary approach.
    \Cref{subsec:folklore-details} provides full details on parsing, term classification, and construction of the index.
    
    To complement the folklore-based measure, we construct group-level indices of environmental narratives using ethnographic descriptions from the Joshua Project, as well as textual descriptions from English Wikipedia pages on ethnic groups. In both cases, we collect standardized narrative texts for all ethnic groups with available content,\footnote{For the Joshua Project, we exclude sections explicitly framed for missionary purposes. For Wikipedia, we manually compile the relevant set of pages corresponding to groups in the ancestral characteristics database and scrape their full narrative content, relying on closely related entries (e.g., language or associated political entities) when a dedicated page is unavailable.} and use a manual review of a random subset to identify conceptually grounded markers of environmental salience. Guided by this exercise, we implement a BART-based zero-shot classification procedure \citep[fine-tuned on MultiNLI,][]{williams2018broad} to score each group’s text along predefined environmental dimensions.\footnote{Including: ecological references, climate exposure, subsistence–environment linkages, resistance to environmental degradation, and nature-related spirituality (Joshua Project); and: the centrality of nature to identity and livelihood, nature-linked spiritual systems, traditional ecological knowledge, and narratives of resistance to environmentally threatening external actors (Wikipedia). We additionally collect information on the number of Wikipedia language editions covering each group as a proxy for documentation breadth.} These scores are aggregated into a normalized composite index. Geographic identifiers in the database allow us to link groups to historical temperature anomaly data and additional controls. \Cref{subsec:Joshua-details} and \Cref{subsec:Wikipedia-details} provide full details on data sources, text processing, classification design, and sample construction.

    \paragraph{Ancestral ethnic group data}
	We rely on \posscite{nunn_giu_database} \emph{Ancestral Characteristics of Modern Populations} database to obtain information on the pre-industrial traits of individuals’ ancestors. The dataset links ethnographic information on ethnic groups—drawn from Murdock’s \emph{Ethnographic Atlas}, \posscite{Bondarenko} dataset on the Peoples of Easternmost Europe, and \posscite{Murdock} \emph{World Ethnographic Sample}—to over 7,000 contemporary languages and dialects listed in \posscite{gordon} \emph{Ethnologue: Languages of the World}, which we use to map individuals in the IVS and ESS to their corresponding ethnic ancestors.\footnote{For our analysis, we also use key pre-industrial characteristics of ethnic groups as controls. Specifically, we consider measures capturing primary subsistence patterns; indicators of settlement and economic complexity; agricultural intensity; community size; and local and supra-local jurisdictional hierarchy.} The geographic coordinates provided for ethnic groups in the ethnographic sources allow us to match ancestral locations to grid-level climate data from \citet{mann_2009}, enabling the construction of measures of ancestral climatic exposure. Additional details about the dataset and corresponding variables are in \Cref{subsec:ancestors-details}.

    \paragraph{Historical temperature data and anomaly measure}\label{temp-construction}
	\citet{mann_2009} provide information on global patterns of detrended surface temperature for a long historical period using a climate field reconstruction approach (see \Cref{sec:mann} for details).
    The dataset reports the average annual temperature anomalies (deviations from the 1961-1995 reference-period average measured in degree celsius) at the 5degree-by-5degree ($\approx 555\mathrm{km} \times 555\mathrm{km}$) grid cell level since 500 CE. 
	We restrict attention to the temperature data between 1600 CE and 1920 CE, to improve reliability (by restricting to years for which at least two proxies are available) and ensure that our measure is independent from contemporaneous confounds (by retaining a gap of at least 85 years with survey responses).
	
	The main explanatory variable we construct is the \emph{average intensity of deviations from typical conditions} faced by ethnic ancestors. 
    This captures the extent to which weather conditions concentrate on their (locally) ``regular'' range, or stray into ``anomalous'' extreme events.
	Our baseline measure assumes that each generation lives for 20 years, without overlaps.\footnote{
        Our measure of ancestral climate anomalies should not be confused with \citet{nunn_giuliano}'s measure of climate instability. They use across generation changes in average temperatures to capture the difference in life experiences of different generations, whereas we are interested in within-generation experiences of relative anomalies, i.e. short-term weather changes and the intensity of those changes from ``normal'' range of expected weather conditions.}
	We also assume that all the historical generations associated to a particular ethnic group occupy the same grid cell.\footnote{
        This is reasonable even with minor historical movements as cells cover an area of approximately 308,025km$^2$. 
        } 
	Therefore, for our primary specification, each ethnic group has a total of 16 generations associated to them. Fixing a grid cell / ethnic group, use $g$ to denote its $g^{\text{th}}$ generation; the associated avg. anomaly measure is constructed as follows:
	\begin{enumerate}
		\item Let $\tau_{g,t}$ the temperature faced by generation $g$ in year $t$ of their lifetime $\mathbb{T}_g$; denote $\hat{F}_g$ the empirical CDF of $\{\tau_{g,t}\}_{t \in \mathbb{T}_g}$.
        Recast temperature realizations in terms of \emph{deviations from the ``typical'' range}, defined as the range between the $\alpha^\text{th}$ and the $(1-\alpha)^\text{th}$ percentile of the empirical distribution of temperature values within a generation's lifetime ($\alpha=.2$ in our main specification):
        \[
        z_{t,g} :=  
		 \mathds{1}_{\tau_{g,t}<\hat{F}_g^{-1}(\alpha)}
		\bigl| \tau_{g,t}-\hat{F}_g^{-1}(\alpha) \bigr|
		+ 
		\mathds{1}_{\tau_{g,t}>\hat{F}_g^{-1}(1-\alpha)}
		\bigl| \tau_{g,t}-\hat{F}_g^{-1}(1-\alpha) \bigr|
        \]
		\item For generation $g$, let $\hat{\theta}_g$ the average intensity of deviations from the typical range of conditions:
		\(
		\hat{\theta}_g = \frac{1}{K_g}
		\sum_{t \in \mathbb{T}_g} 
		z_{t,g},
		\)
		where $K_g := \#\bigl\{ \tau_{g,t} \bigm| \tau_{g,t}<\hat{F}_g^{-1}(\alpha) \text{ or } \tau_{g,t}>\hat{F}_g^{-1}(1-\alpha)\bigr\} $ is the normalizing constant that counts the number of excursions outside the $(\alpha,1-\alpha)$ percentile range. 
		\item Define the average ancestral climatic anomalies for an ethnic group as $\text{Avg. anomalies} = \frac{1}{G}\sum_{g=1}^{G} \hat{\theta}_g$, where $G$ is the number of generations ($16$ is our baseline specification).
	\end{enumerate}
	
	$\hat{\theta}_g$ captures the intensity of deviations (anomalies) from regular climate conditions in generation $g$'s lifetime.\footnote{
        This approach is similar to the exercise carried out by climatologists in extreme weather attribution and anomaly analysis \citep{smith2008improvements,nationalcommittee}. 
        } 
	Averaging over generations captures the overall \emph{scale} of anomalies experienced by one's ancestors.\footnote{
        Variation in our variable of interest at the grid level is reported in \Cref{temp_map}.
    }
	In \Cref{sec:robustness-checks}, we show our results are robust to the choice of the time span of generations, thresholds for extreme events, and alternative measures of anomalies.

    \subsection{Conceptual framework}\label{subsec:theory-overview}

        We embed these objects in a model of cultural transmission of the perceived value of caring about environmental issues.
        The present subsection provides an intuitive overview of the theory to guide the presentation of empirical results; the complete analysis is postponed to \Cref{sec:theory}. 
        
        Individuals repeatedly face new situations in which incorporating environmental considerations into their decisions requires costly effort (attention). 
        How much they “care” in a given situation depends on their perceived stakes: what they believe they stand to gain from taking environmental consequences into account, or to lose by ignoring them.
        In any given decision, the true stakes are a function of realized climate conditions: in some realizations, environmental attention matters little; in others, it is decisive. 
        Crucially, however, attention must be chosen before climate conditions are observed. 
        What matters ex ante is therefore the \emph{expected} value of attention, taken under the specific climate distribution individuals face.
        Our main object of interest is the underlying perceived stakes function, which specifies the value of attention as a function of climate conditions one might face. 
        The resulting aggregate perceived importance of environmental care (across situations and under the climate conditions individuals anticipate facing) is the object we measure in the data.
        
        To explain how such perceptions form, we combine two ingredients. 
        First, the individual level effort choice is modeled as a canonical rational inattention problem; this provides a flexible approach that can capture ``care'' across a variety of independent situations while remaining abstract about details of specific problems.
        Expected stakes translate into attention effort: individuals optimally choose how much costly information to acquire given their perceived stakes function.
        This gives a precise foundation to the ``value of caring''.
        Second, we embed this attention problem into population-level evolutionary dynamics that capture cultural transmission. 
        Within each generation, individuals differ in their perceived stakes functions: this defines cultural types. 
        Individuals behave optimally given their perceptions, realize gains and losses under the actual climate conditions they face. 
        More successful types are more likely to be taken as models hence transmit more effectively to the next generation, subject to noise. 
        Conversely, cultural types that perform worse die out.
        
        If attention were chosen after climate realizations are observed, evolutionary pressure would select the correct stakes function at every point.
        In that case, ancestral climate conditions would have no effect on descendants' perceptions. 
        But, because attention is chosen ex ante, evolutionary selection is necessarily coarse. 
        Fitness depends only on average realized losses, and therefore only on the expected value of the stakes function under experienced conditions. Evolution attempts to discipline a function but can identify only a moment of that function. 
        The remaining degrees of freedom are shaped by historical contingency.
        This generates our central theoretical prediction: descendants’ perceived value of environmental attention reflects the true average value of attention under the climate distribution experienced by their ancestors.
        \Cref{fig:mechanism-conceptual-map} illustrates the mechanism.

        \begin{figure}[t!]
          \centering
          \resizebox{.7\textwidth}{!}{
          \begin{tikzpicture}[
            box/.style={draw, rounded corners, align=center, minimum width=3cm, minimum height=0.9cm},
            arrow/.style={->, thick}
            ]
        
              \node[box] (climate) {Climate conditions distribution \\ $q \in \Delta(Z)$};
              \node[box, below=1.2cm of climate] (realized) {Realized climate \\ $z \sim q$};
              \node[box, below=1.2cm of realized] (true) {True stakes \\ $v^\star(z)$};
        
              \node[box, right=4cm of climate] (belief) {Belief function \\ $f:Z \rightarrow \reals_+$};
              \node[box, below=1.2cm of belief] (mean) {Expected value \\ $\mathbb E_q[f(z)]$};
              \node[box, below=1.2cm of mean] (attention) {Attention choice};
        
              \node[box, below=1.2cm of attention] (fitness) {Realized gains \\ and losses};
              \node[box, right=4cm of attention, red!50!black] (evolution) {Cultural transmission \\ (selection + mutation)};
        
              \draw[arrow] (climate) -- (realized);
              \draw[arrow] (realized) -- (true);
              \draw[arrow] (belief) -- (mean);
              \draw[arrow] (mean) -- (attention);
              \draw[arrow] (attention) -- (fitness);
              \draw[arrow] (true) -- (fitness);

              \draw[->,dotted,bend right=40,red!50!black,thick] (fitness) to (evolution);
              \draw[->,dotted,bend right=40,red!50!black,thick] (evolution) to (belief);
        
              \draw[arrow, dashed] (climate) -- (mean);
        
              \node[align=left, font=\small] at ($(mean)+(2.6,0)$) {};
        
            \end{tikzpicture}
            }
          \caption{The cultural transmission mechanism -- selection depends only on the average}
          \label{fig:mechanism-conceptual-map}
    \end{figure}    
        
        A simple micro-foundation for the primitive shape of the true value of attention, as a function of \emph{conditions}, gives testable implications of \emph{distributional} heterogeneity.
        Guided by evidence from anthropological accounts, we assume that environmental considerations combine two broad motives within any generation.
        The first is exploitation: attention helps individuals make better use of typical climate conditions, so its value declines as conditions move away from the normal range. 
        The second is protection: attention reduces losses from extreme events, so its value rises with the severity of deviations. 
        Aggregating across heterogeneous situations yields a U-shaped relationship between realized climate intensity and the value of attention.
        Because evolutionary dynamics select only on averages, this primitive U-shape approximately transfers to contemporaneous aggregate perceived value: ancestral populations (and subsequently their descendants) exposed either to very stable or to highly variable climates converge to higher perceived value of environmental attention, while intermediate variability induces lower perceived value. 

\section{Empirical results}\label{sec:empirical-results}

    \subsection{Individual level analysis -- value of environmental attention}
	
    	We begin by examining the impact of average ancestral climatic anomalies on an individual's perceived value of attention to environmental issues.
    	We run a simple OLS specification, controlling for individual demographic characteristics, historical ethnic groups' social and political characteristics, historical ethnic groups' geographical characteristics and respondent's country-by-year fixed effects.
    	Our empirical strategy is equivalent to a continuous treatment framework where the treatment is the average intensity of anomalies in ancestral climate conditions and the assignment mechanism is driven by lottery of birth, i.e. an individual doesn't choose which ethnicity they are born in. 
    	Our regression equation reads as:
    	\begin{align}\label{main_eq}
    		y_{iect} = \beta_1 \text{Avg. Anomalies}_e + \beta_2 (\text{Avg. Anomalies}_e)^2 + \mathbf{X_{ict}\Gamma} + \mathbf{X_{e}\Omega} + \alpha_{ct} + \epsilon_{iect},
    	\end{align}
    	where $i$ indexes the individual, $e$ indexes the historical ethnic group an individual belongs to, $c$ indexes the country of the individual and $t$ indexes year. Specifically:
    	\begin{itemize}
    		\item $y_{iect}$ refers to the dependent variable of interest, i.e. an individual's perception of the value of attention, captured by the multiple proxies created from WVS, IVS, and ESS (see \Cref{subsec:construction-of-variables});
    		\item $\text{Avg. Anomalies}_e$ is the index of anomaly intensity in ancestral generations as in \Cref{subsec:construction-of-variables};
    		\item $\mathbf{X_{ict}}$ is the vector of individual level demographic characteristics which include Age, Gender, Income, Educational level and dummies for Occupation categories;
    		\item $\mathbf{X_e}$ corresponds to the vector of historical ethnic group controls (see \Cref{subsec:ancestors-details} for details about coding) which include primary mode of subsistence, economic development indicators (complexity of settlement, size of local community and intensity of agriculture and level of local and global jurisdictional development), as well as ethnicity level geographical variables such as distance to closest coast and equator and K\"{o}ppen climate classification of the spatial grid the ethnic group was historically located in;
    		\item $\alpha_{ct}$ corresponds to country-year fixed effects which capture any contemporaneous variables such as level of economic development of the country (of residence), general population's level of schooling, awareness and perception of climate issues\footnote{This can capture all sorts of general equilibrium factors and spillovers within the current generation that can emanate from the country of residence} and contemporary climate shocks faced by the country's population.
    	\end{itemize}
    	We cluster the standard errors at the ethnic group level. Our coefficients of interest are $\beta_1$ and $\beta_2$.
    	Our theory suggests that the level of attention should initially decrease and then increase with the intensity of anomalies faced by ancestral generations. 
    	As a result, the expected signs on $\beta_1$ and $\beta_2$ should be negative and positive respectively, capturing that the perceived value is (i) decreasing over the lower range of values of ancestral climatic variability because the diminution of benefits from exploitation dominates evolutionary dynamics and (ii) increasing over higher values because the increase of the loss-protection motive for attention starts to dominate.

    \begin{table}[t!]\centering
    \footnotesize
    \begin{threeparttable}
    \caption{Coefficients for the impact of average anomalies in ancestral climatic conditions on individual's self-reported attention to the environment \label{table:main-regression}}
    \begin{tabular}{lcccc}
    \hline
    \multicolumn{5}{c}{Dependent variable: Self reported measure of attention to environment}\\
                        &         (1)   &         (2)   &         (3)   &         (4)   \\
    \hline
    \hline
    Avg. anomalies    &      -6.291***&      -7.095***&      -9.194***&      -9.877***\\
                        &     (1.587)   &     (1.902)   &     (2.087)   &     (2.003)   \\
    Avg. anomalies sq.&      71.038***&      78.984***&     100.939***&     111.767***\\
                        &    (18.330)   &    (21.756)   &    (22.608)   &    (21.346)   \\
    Demographic Controls & N & Y & Y & Y \\
    Historical ethnic group characteristics & N & N & Y & Y \\
    Historical topographic characteristics & N & N & N & Y \\
    Country-year Fixed effects & Y & Y & Y & Y \\
    \hline
    Mean of dep var     &       0.702   &       0.704   &       0.704   &       0.704   \\
    St. Dev. of dep var &       0.252   &       0.250   &       0.250   &       0.250   \\
    Min value Avg. anomalies&       0.015   &       0.015   &       0.015   &       0.015   \\
    Max value Avg. anomalies&       0.093   &       0.093   &       0.093   &       0.093   \\
    R-sq                &       0.097   &       0.113   &       0.113   &       0.113   \\
    Adj. R-sq           &       0.097   &       0.112   &       0.112   &       0.113   \\
    N                   &      157142   &      138067   &      138067   &      138067   \\
    \hline
    \multicolumn{5}{c}{ {*}{*}{*} p$<$0.01, {*}{*}p$<$0.05, {*} p$<$0.1, {+} p$<$0.15}\\
    \hline
    \hline
    \end{tabular}
    \begin{tablenotes}
    \footnotesize
    \item \textbf{Note: } Unit of observation is an individual. Dependent variable is the individual's value of environmental attention, ranging between 0 and 1 and increasing with reported importance. The variable is constructed by rescaling the answer to the prompt: On a scale of 1 to 6, how important is it for this individual to take care of the environment. Avg. anomalies refers to the average intensity of deviations from the typical climate conditions (\textit{specific to each ancestral generation}) across generations. Avg. anomalies sq. refers to the square of the average anomalies term. Average anomalies range between 0.015 and 0.093 within the sample. Demographic controls include dummies for income deciles, occupation categories, gender, education level and age. Historical ethnic group characteristics include measure of development such as agricultural intensity, complexity of settlement, level of political hierarchies, size of the local community and main source of subsistence. Historical topographic characteristics include controls for distance to the equator, distance to the closest coast and geographical K\"{o}ppen climate classification for the location of the ethnic group obtained from Ethnographic Atlas. Standard errors are clustered at ethnicity level.
    \end{tablenotes}
    \end{threeparttable}
    \end{table}
            
            \begin{figure}[t!]
    		\centering
    		\includegraphics[scale=0.3]{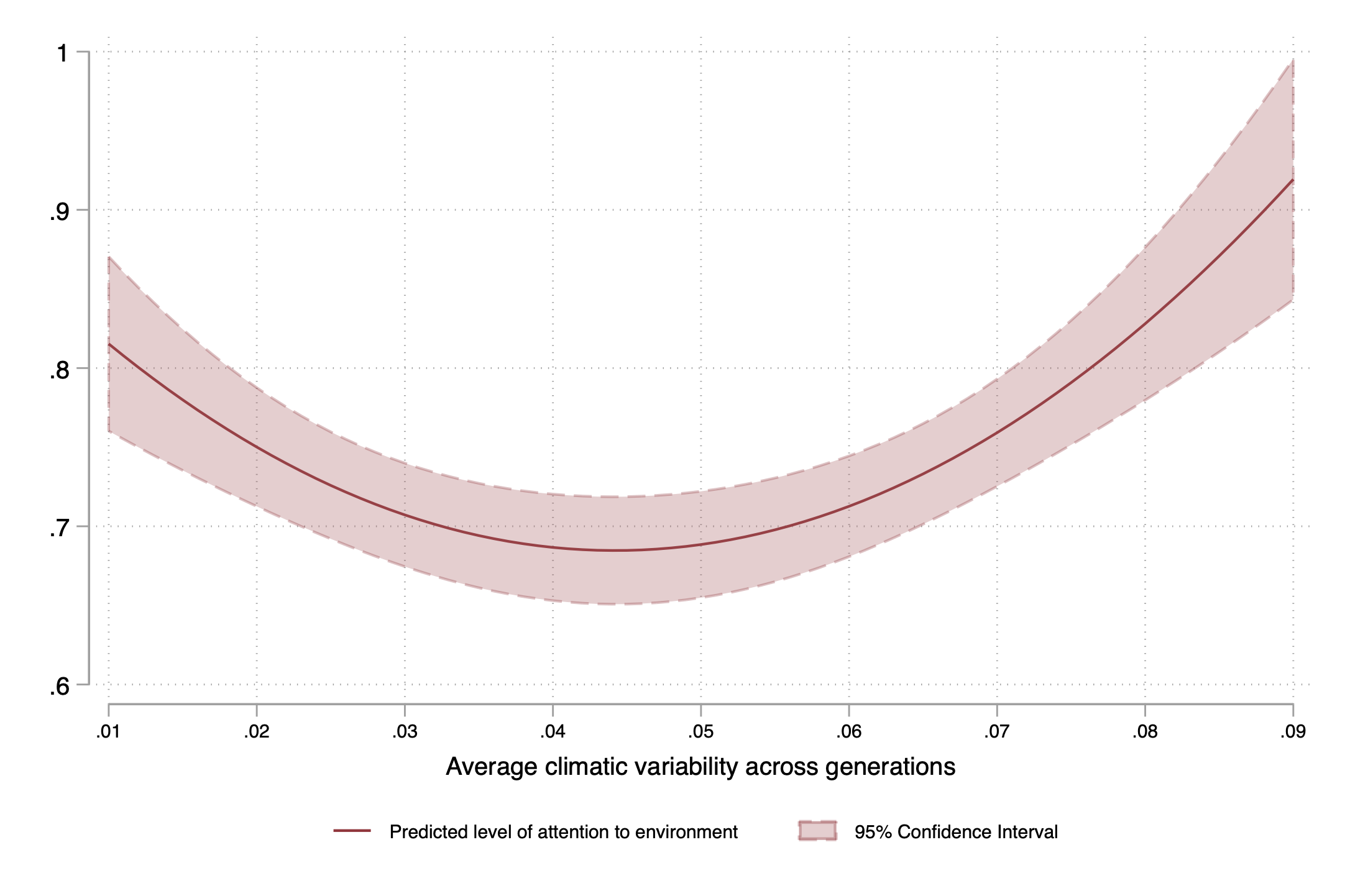}
    		\caption{Margins plot - Individual's level of attention as a function of average ancestral climatic anomalies} \label{fig1}
    		\fnote{\textbf{Note:} N = 138067. Margins plot corresponding to \Cref{main_eq}. Average value of climatic anomalies range from 0.01 to 0.09. 
            For the margins plot, continuous controls are fixed at their mean values. Discrete / categorical controls are fixed at Male = 1, Historically agrarian society = 1 and K\"{o}ppen climate classification = D.}
    	\end{figure} 
    
        \Cref{table:main-regression} reports the results from \Cref{main_eq} for our lead variable of interest, i.e. normalized responses to the question: \textit{``How important is it to you to take care of the environment?''}.
        Column 1 reports the results we obtain from regressing the variable on historical measures of ancestral climatic volatility controlling for country-year fixed effects. Subsequent columns add individual level and historical controls. Across all specifications, our results corroborate the hypothesis. 
	The results we find are all statistically significant at 1\% level and consistent across different choice of sample or method of construction of our ancestral climate anomalies index.
	
	For the purpose of a stylized illustration, assume that, on our index of climate anomalies, the most stable ancestral climates scored 0.01 and the most unstable ancestral climates scored 0.09.\footnote{For the sample in consideration, the average ancestral climate anomalies captured by our constructed index ranges from 0.015 to 0.093. We restrict the sample to remove a small mass of outliers lying outside the $99^\text{th}$ percentile of the distribution of our values of average anomalies.} In that case, the coefficients $\beta_1$ and $\beta_2$ suggest that an increase of 0.01 ($\sim 10\%$) in the average deviations from the typical climate conditions in the most stable ancestral environments would have led to a 10.9\% \textit{decrease} in attention to environment today relative to the sample mean and an increase of 0.01 in the average deviations in the most unstable ancestral environments would have led to a 12.9\% percentage point \textit{increase} in attention to environment today. 
	
	We provide additional evidence on this relationship, using our second proxy, constructed from data on self-reported engagement in pro-environmental actions and support for national environmental causes from the Integrated Values Surveys. 
	We present results from regressing the environmental attention index on historical measures of ancestral climatic volatility, controlling for individual-level characteristics, historical ethnic-group-level controls, and country–year fixed effects in \Cref{table:ivs_reg}. Across Columns 1 to 4, we vary the choice of sample and variables used to construct the PCA whose normalized first component we use as our index (see \Cref{subsec:IVS-details} for more details). 
	Across all specifications, we observe a statistically significant U-shaped relationship between ancestral climatic anomalies and environmental attention, with the estimated coefficients satisfying $\beta_1 < 0$ and $\beta_2 > 0$.

    We finally replicate this analysis with our third proxy constructed from the climate and energy module included in the 8\textsuperscript{th} round of the European Social Survey (ESS). We estimate \Cref{main_eq} with the pro-environment indices as the dependent variables, controlling for the same set of individual-level characteristics and historical ethnic-group-level controls as in the previous specifications. Because all ESS observations are drawn from a single survey year (2016), we include country fixed effects rather than country–year fixed effects.
	\Cref{table:ess-reg} reports the results. Across Columns 1 to 4, we vary our choice of variables to construct the PCA whose normalized first component we use as our index (see \Cref{subsec:ESS-details} for more details). Across all specifications, we again observe a statistically significant U-shaped relationship between ancestral climatic anomalies and pro-environmental attention, with estimated coefficients satisfying $\beta_1 < 0$ and $\beta_2 > 0$.

    \subsection{Group level analysis -- environmental themes in culture}

        Beyond individual level analysis, the impact of ancestors' experiences on descendants' perceptions can be captured at the group level through the stock of knowledge and collective memory related to environmental themes. 
	    The stories, legends, songs, etc. which constitute folklore and narratives are transmitted within language groups across times and locations. 
        Hence this measure serves two possible purposes.
    	First, if successive generations of a particular ethnic group face climate conditions which increase each generation's level of attention paid to the environment, then increased attention should result in a higher occurrence of environmental themes in folklore and associated narratives.
        Second, this corroborates the idea that transmission across generation passes through cultural material.
    	
    	We test this theory by first regressing the level of environmental focus in ethnic folklore on our variable of average ancestral climatic volatility, using the data described in \Cref{subsec:construction-of-variables}.
    	We run the following specification at the ethnicity level:
    	\begin{equation}\label{folklore_eq}
    		\begin{aligned}
    			\mathfrak{F}_{ec}  = \beta_0 & +  \beta_1 \text{Avg. anomalies}_{ec} + \beta_2 (\text{Avg. anomalies})^2_{ec} + \mathbf{X_{ec}\Omega} + \alpha_c + \epsilon_{ec}
    		\end{aligned}
    	\end{equation}
    	where $e$ indexes an ethnic group, $c$ indexes the country whose current geography would have historically included the ethnic group and $\mathfrak{F}_{ec}$ denotes the previously constructed outcome variable for prevalence of environmental motifs in folklore, i.e:
    	\[
    	\mathfrak{F}_{ec} := \log\left(1+\dfrac{\text{No. of related motifs}_{ec}}{\text{Total no. of motifs}_{ec}}\right) 
    	\]
    	$\mathbf{X_{ec}}$ captures historical ethnic group controls as before such as primary mode of subsistence, economic development indicators (complexity of settlement, size of local community and intensity of agriculture) and level of jurisdictional development (both locally and globally). 
    	It also includes ethnicity level geographical variables such as distance to closest coast and equator and K\"{o}ppen climate classification of the spatial grid the ethnic group was historically located in. 
    	In line with \citet{folklore}, we additionally control for the first year of publication and the total number of publications, authors, publishers and languages of the text associated to each ethnic group. 
    	Country fixed effects are captured by $\alpha_c$. 
    	We cluster the standard errors at the language group level as specified by \citet{berezkin}. 
    	Results from this specification are reported in \Cref{folklore_reg}.

    \begin{table}[t!]\centering
    \footnotesize
    \begin{threeparttable}
    \caption{Coefficients for the impact of ancestral climatic anomalies on occurrences of environment-related ethnic folklore \label{folklore_reg}}
    \begin{tabular}{lccccc}
    \hline
    \multicolumn{5}{c}{Dependent variable: $\ln(1+\text{No. of environment-related motifs}/\text{Total no. of motifs})$}\\
                        &         (1)   &         (2)   &         (3)   &         (4)   \\
    \hline
    \hline
    Avg. anomalies    &      -3.233***&      -3.172***&      -2.493***&      -1.548** \\
                        &     (0.944)   &     (0.735)   &     (0.784)   &     (0.764)   \\
    Avg. anomalies sq.&      33.930***&      31.539***&      25.341***&      17.237** \\
                        &     (9.526)   &     (7.684)   &     (8.047)   &     (7.112)   \\
    Folklore document characteristics & N & Y & Y & Y \\
    Historical ethnic group characteristics & N & N & Y & Y \\
    Historical topographic characteristics & N & N & Y & Y \\
    Country Fixed effects & N & N & N & Y \\
    \hline
    Mean of dep var     &       0.276   &       0.276   &       0.276   &       0.277   \\
    St. Dev. of dep var &       0.097   &       0.097   &       0.097   &       0.096   \\
    Min value Avg. anomalies&       0.010   &       0.010   &       0.010   &       0.010   \\
    Max value Avg. anomalies&       0.097   &       0.097   &       0.097   &       0.097   \\
    R-sq                &       0.020   &       0.116   &       0.198   &       0.439   \\
    Adj. R-sq           &       0.018   &       0.110   &       0.180   &       0.366   \\
    N                   &        1037   &        1037   &        1037   &         982   \\
    \hline
    \multicolumn{5}{c}{ {*}{*}{*} p$<$0.01, {*}{*}p$<$0.05, {*} p$<$0.1, {+} p$<$0.15}\\
    \hline
    \hline
    \end{tabular}
    \begin{tablenotes}
    \footnotesize
    \item \textbf{Note:} Unit of observation is an ethnic group. Environment-related motifs are classified using ConceptNet. Avg. anomalies refers to the average intensity of deviations from the typical climate conditions (\textit{specific to each ancestral generation}) across generations. Avg. anomalies sq. refers to the square of the average variability term. Average anomalies range between 0.01 and 0.097 within the sample. Folklore document characteristics include the first year of publication, number of authors that have covered the motifs in their works, number of languages, publishers and publications that contain the motif records. Historical ethnic group characteristics include measure of development such as agricultural intensity, complexity of settlement, level of political hierarchies and main source of subsistence. Historical topographic characteristics include controls for distance to the equator, distance to the closest coast and geographical K\"{o}ppen climate classification for the location of the ethnic group obtained from Ethnographic Atlas. Standard errors are clustered at language group level.
    \end{tablenotes}
    \end{threeparttable}
    \end{table}
    
        As before, the coefficient on Avg. anomalies is negative and statistically significant at 5\% level and the coefficient on Avg. anomalies squared is positive and statistically significant at 5\% level, thereby providing evidence that the climatic experiences of ancestors feed into the attention parameter in the same way and increase the stock of environmental folklore for either extremely stable or extremely unstable ancestral climates, with a dip in the intermediate range. 
    	An increase of 0.01 ($\sim$ 10 \%) in average deviations from the typical climate conditions in the most stable environments leads to a $\sim$3.6\% \textit{decrease} in the proportion of environmentally related folklore and an increase of 0.01 in average deviations in the most unstable environments lead to a $\sim$6.3\% \emph{increase} in the proportion of environmentally related folklore.
    	
    	To provide further evidence on this relationship, we additionally evaluate the effects of ancestral climatic anomalies on the prevalence of environmental themes in descriptive narratives drawn from historical records of ethnic groups available on English Wikipedia and the Joshua Project. We first use the composite index capturing environmental themes constructed from English Wikipedia text as our main dependent variable,\footnote{That is, $\mathfrak{F}_{ec}$ now denotes the normalized score derived from Wikipedia text capturing the level of environmental attention reflected in written records for each ethnic group.} and re-estimate \Cref{folklore_eq}. As before, we control for historical ethnic-group-level characteristics and include country fixed effects. To account for variation in the quality and granularity of available information, we additionally control for the length of the associated English Wikipedia page and the number of distinct language editions in which the ethnic group has a corresponding page. \Cref{table:wiki-reg} reports the results. Across all specifications, we observe a statistically significant U-shaped relationship between environmental content in written narratives and the degree of ancestral climatic anomalies.
    
        We replicate this exercise using the index constructed using written narratives of ethnic groups available in the Joshua Project. We estimate the same specification as in \Cref{folklore_eq}, where $\mathfrak{F}_{ec}$ now denotes the normalized score derived from Joshua Project text capturing the level of environmental attention reflected in written records for each ethnic group. The vector $\mathbf{X}_{ec}$ includes ethnic-group-level characteristics available in the Joshua Project dataset. Specifically, we control for the group’s specified religion, whether it is classified as least reached or frontier (i.e., variables capturing the difficulty of reaching these groups as reported by Joshua Project contributors), and whether the group is classified as indigenous. In addition, we include geographic characteristics at the ethnic-group level, such as distance to the nearest coast, distance to the equator, and the Köppen climate classification of the spatial grid corresponding to the ethnic group's location.
    	Country fixed effects are captured by $\alpha_c$, and standard errors are clustered at the people group level as specified in the Joshua Project. Results from this specification are reported in \Cref{table:reg-joshua}.
        Across all specifications, from Columns 1 through 4, where we sequentially add ethnic-group-level controls and country fixed effects, we again find a statistically significant U-shaped relationship between environmental content in written narratives and the degree of ancestral climatic anomalies.

\section{An evolutionary theory of environmental attention}\label{sec:theory}

    We propose a general theoretical framework to analyze the formation of perceptions about the value of environmental attention through cultural transmission.
    The theory has two main distinct goals.
    First, we characterize why and how the specific distribution of conditions faced by ancestors leaves a persistent imprint.
    This establishes that descendants' perception generally reflects the \emph{average} underlying true value of attention under ancestral climate distribution.
    Second, we show that if true value is driven by the dual motive structure, this transfers through evolution and aggregation to generate the U-shape pattern in average ancestral anomalies.

    \subsection{Model}\label{subsec:theory-model}
    
        \paragraph{Setup}
        Take the point of view of a single lineage of ancestral generations and a fixed descendant.
        Each ancestral generation, indexed by $g \in G$, contains a population of individuals which each live for $T$ (non-overlapping) periods indexed $t=1,...,T$.
        In each period $t$ of generation $g$'s lifetime, climate condition $z_{g,t}$ is realized. 
        We represent climate conditions by \emph{intensity of deviation} from a well-understood "normal" range, so that $z \in Z := [0,\overline{z}]$ measures how far realized conditions depart from typical levels in a given period. 
        Ancestral generations, living in a fixed geographical area, face a common distribution $z_{g,t} \sim q \in \Delta(Z)$ which is known (draws are iid); this captures that a fixed geographical area faces a relatively stable \emph{distribution} of conditions (though random \emph{realizations} in any given generation's lifespan).
        The descendant faces potentially different conditions $\tilde{q} \in \Delta(Z)$ (due to e.g. migration or climate change).
        In each period of an individual's life, they face $n=1,...,N$ independent "situations" in which they must decide on some course of action.
        We index a generic situation by $s=(g,t,n)$.

        \paragraph{Environmental concerns and its stakes}
        Coming to a decision in any specific situation requires making costly effort to endogenize the relevant environmental concerns: acquiring and processing information, thinking about all relevant tradeoffs, evaluating consequences of one's action, etc. 
        To model this sparsely, represent each problem via the \emph{loss from imprecise attention to environmental aspects}.
        In any particular situation $s=(g,t,n)$, there is an unknown (situation-specific) state of the world $\omega_s \sim \mathcal{N}(0,\sigma^2)$; the agent's chosen "action" $a_s \in \reals$ induces losses $v_s(z_{g,t}) (\omega_s - a_s)^2$, where $v_s(z) > 0$ is a situation-specific function of \emph{realized climate conditions} $z_{g,t}$.
        
        This is a flexible modeling approach.  
        One natural interpretation is to view $\omega_s$ as the action that would be optimal under perfect assessment of environmental consequences, normalized so that action zero is chosen when environmental considerations are ignored. 
        Alternatively, this can be seen as a reduced-form representation of losses arising from uncertainty; in that case $\omega_s$ represents an abstract state of knowledge, the agent’s action corresponds to their best estimate of that state, and losses reflect the expected mean-squared error in estimating $\omega_s$.
 
        The climate-sensitive stakes function $v_s(z_{g,t})$ captures both heterogeneity of situations (dependence on $s$: environmental aspects may matter more in some problems than others) and variation in realized losses with climate conditions (dependence on $z_{t,g}$). 
        The \emph{true value} of attention $v_{g,t,n}$ in problem $(g,t,n)$ is a random function drawn iid from a given distribution over a suitable class with mean $v^\star$, taking values bounded away from zero and infinity.
        The function $v^\star: Z \rightarrow \reals_+^*$ is the underlying \textbf{true average stakes function} across situations.
        The \emph{subjective perception of the stakes function} is the key object we seek to microfound via cultural transmission.
        This first requires mapping perceptions into attention choices.

        \paragraph{Optimal attention}
        In each situation, the agent chooses how much costly attention to devote to environmental consequences, given their subjective perceived stakes $\hat{v}$. We model this as a rational inattention problem: the agent pays a cost proportional to how well informed they are about $\omega_s$ (measured by Shannon mutual information). Given quadratic losses and Gaussian priors, the problem summarizes without loss as a choice of posterior uncertainty $\xi \le \sigma^2$:
        \[
        \min_{\xi \le \sigma^2} \; \Esp_q[\hat v] \cdot \xi 
        \;+\; \kappa\log\!\left(\frac{\sigma^2}{\xi}\right),
        \]
        where $\Esp_q[\hat v] > 0$ denotes the ex ante expected perceived stakes at the time attention is chosen.
        The derivation of this form of the problem is standard in the rational inattention literature -- see \Cref{subsec:rational-inattention-details} for details and \citet{MackowiakMatejkaWiederholt2023} for a review.
        Assuming interior solutions for expositional simplicity gives
        \(
        \xi^* = \frac{\kappa}{\Esp_q[\hat v]}.
        \)
        
        Higher perceived value of attention induces strictly greater attention (uncertainty reduction). 
        Perceived expected stakes $\Esp_q[\hat v]$ is the sufficient statistic governing attention effort, while \emph{true stakes} govern realized losses. 
        The key object which will condense these ideas is the \emph{reduced loss function} $\mathcal{L}(\mu|v_s)$ which describes realized losses when attention is chosen according to perceived ex-ante \emph{expected} stakes $\mu > 0$ (the $\Esp_q[\hat{v}]$ term) and ex-post \emph{realized} situation stakes are $\nu > 0$ (the $v_s(z_{t,g})$ term). Formally, let:
        \[
            \mathcal{L}( \mu | \nu) := \nu \times \frac{\kappa}{\mu} + \kappa \log \left( \frac{\sigma^2}{\kappa}   \mu \right) \text{ for any } \mu,\nu \in \reals_+^*.
        \]

        \paragraph{Cultural types and realized losses}
        A \textbf{cultural type} is described by a \emph{perceived average stakes function} $f:Z \rightarrow \reals_+^*$ (the $\hat{v}$ in the previous discussion).
        Let $\mathcal{F}=\mathcal{C}^0(Z)$ the class of admissible cultural types, which are continuous functions on the domain $Z$ of climate-anomaly intensities, containing the true average $v^\star$ and any situation-specific value $v_s$.
        An individual with cultural type $f$ chooses their optimal attention in problem $(g,t,n)$ according to the RI problem using perceived stakes $f(z)$, with $z \sim q$. 
        Define optimal posterior variance as a function of cultural type,\footnote{
         This assumes for simplicity that agents have no problem-specific information and choose attention based on aggregate perceived value; a natural extension can be written where agents have some information about the problem specific-noise.
        }
        and the induced \emph{realised losses}  $\ell_{g,t,n}(f)$  of a cultural type $f$ in problem $(t,g,n)$:
        \begin{align*}
         \xi_q(f):= \frac{\kappa}{\Esp_q[f]}, \quad \ell_{g,t,n}(f):= \mathcal{L} \bigl( \Esp_q[f] \bigm| v_{g,t,n}(z_{g,t}) \bigr),
        \end{align*} 
        
        which depend on the attention choice of the agent (hence their cultural type) and the realized climate condition $z_{g,t}$, which enters via the true stakes function.
        Define the lifetime average realized losses of a type $f$ in generation $g$:
        \(
        L_g(f):= \frac{1}{T} \frac{1}{N} \sum_{t=1}^T \sum_{n=1}^N \ell_{g,t,n}(f).
        \)
        Average realized losses $L_g(f)$ can be rewritten as a simple function of \emph{ex-ante expected stakes} $\Esp_q[f]$ and \emph{empirical average stakes} $\hat{\overline{v}}_g$:
        \begin{align*}
        L_g(f) = \mathcal{L} \bigl( \Esp_q[f] \bigm| \hat{\overline{v}}_g \bigr); \quad
         \text{ where  } & \;  \hat{\overline{v}}_g := \frac{1}{T} \frac{1}{N} \sum_{t=1}^T \sum_{n=1}^N v_{g,t,n}(z_{g,t})
            && \text{ \color{gray} (empirical gen. $g$ mean value).}
        \end{align*}
          
        \paragraph{Transmission mechanism: evolutionary dynamics}
        In each generation, a plurality of types coexist and each behaves optimally according to their belief.
        The next generation's population is generated from the previous one by the interaction of four forces. 
        Transmission first operates by \emph{blending}: each new type combines the types of two cultural models (for instance two parents, or a parent and another member of the community).
        \emph{Selection} then captures that types which fare empirically better are more likely to serve as models (they transmit more successfully). 
        \emph{Random mutations} represent exogenous changes in cultural types, on top of blending and selection.
        \emph{Default-reversion} captures that cultural types tend to revert to a default perception, absent other forces.
        
        Formally, let $P_g \in \Delta(\mathcal{F})$ be the population distribution over cultural types in generation $g$. 
        Define the \emph{replicator-mutator update} as:
        \begin{align*}\tag{RM}\label{eq:replicator-mutator-dynamics}
            P_{g+1}(df'&)
            =
            \int_\mathcal{F} \int_\mathcal{F}
            \,
            \textcolor{gray}{
                \underbrace{
                    \textcolor{black}{ M\bigl(df' \bigm| \tfrac{f_1+f_2}{2} \bigr)}
                    \vphantom{\Phi_g(f)}
                }_{\text{mutation}}
            }
            \;
            \textcolor{gray}{
                \underbrace{
                    \textcolor{black}{\tilde{P}_g(df_1) \, \tilde{P}_g(df_2)}
                }_{\text{selection}}
            },
            \\[.5em]
            & \text{where } \tilde{P}_g(df) := \frac{\Phi_g(f) \, P_g(df)}{\int_\mathcal{F} \Phi_g \, dP_g}, \quad P_0 = \delta_{f_0}.
        \end{align*}
        The fitness function $\Phi_g$ captures the relationship between a type's ``success'' or ``fit'' to the environment they face and reproduction. 
        The mutation kernel $M( \cdot | h ) \in \Delta(\mathcal{F})$ captures exogenous changes across generations, applied to the combined type $h=\frac{f_1+f_2}{2}$ of two cultural models drawn independently from the post-selection law $\tilde{P}_g$.
        Overall, a better relative fit means more reproduction (overweighting) of that type in the next generation, through more frequent use as a cultural model.
        We initialize dynamics at some default belief $f_0$, interpreted as a primitive starting point absent any evolutionary adaptation; we assume $0 \leq f_0 \leq v^\star$ (the default does not already overvalue attention).
        The dynamics encode how repeated selection and mutation perturb this baseline. 
        We make two standard assumptions.
        
        \begin{assumption}[Exponential selection]\label{assumption:exponential-selection}
          Fitness is a decreasing exponential function of average losses: $\Phi_g(f):= \varphi \bigl( L_g(f) \bigr)$, where $\varphi(L):=\exp(- \tau L)$ and $\tau > 0$ is the strength of evolutionary pressure.  
        \end{assumption}
        
        \begin{assumption}[Mean-reverting mutations]\label{assumption:mutation}
            Given cultural models $f_1,f_2$, new mutated type $f'$ is:
            \begin{align*}
                f' := (1-\rho \lambda) \frac{f_1+f_2}{2} + \rho \lambda f_0 + \sqrt{\rho} \, \varepsilon
            \end{align*}
            where the \emph{noisy mutation term} $\varepsilon$ is a centered Gaussian process with continuous sample paths such that $\varepsilon(z) \sim N(0,1)$ and $\mathrm{Cov}(\varepsilon(z),\varepsilon(x)) = e^{-|z-x|}$ for any $z,x \in Z$; $\rho > 0$ is the \emph{mutation scale} and $\lambda \in (0,1)$ is the relative strength of the pull to the default (on the mutation scale).
        \end{assumption}


        Absent the force of selection, mutation on its own tends to revert to a ``default'' value $f_0$.
        A natural choice is to take $f_0 \equiv 0$, so that if evolution does not actively select individuals who do well by paying optimal attention, the population reverts to a perception that there is \emph{no value}.
        The parameter $\lambda$ captures the relative strength of the pull towards the default versus persistence.
        The random component extends Gaussian noise on a function space, where parameter $\rho$ controls the mutation strength.\footnote{
            The additive expression ignores the positivity constraint for simplicity, but this is not an issue since we consider the case where $\rho$ is small: probabilities of negative values are vanishing if $f$ is strictly positive. Formally, \Cref{sec:fullmodel} handles non-positive values with a technical extension of the loss function below an arbitrarily small threshold, which plays no role in the results.
            }
        Economically, the assumptions capture that: (i) the structure of random mutations does not depend on the current type; (ii) variance of mutation noise is uniform (and normalized) along the domain: $\Var(\varepsilon(z))=1$ for all $z$; (iii) correlation between mutation at different climate conditions decays exponentially, capturing the idea that perceptions are continuous so mutations of perceived stakes at nearby conditions are coupled while distant ones are (essentially) independent.
        Exponential selection is a standard choice which gives convenient representations, but the functional form plays no decisive role beyond tractability.

        An object which will play an important formal role is the mutation covariance kernel $K(x, z) := \mathrm{Cov}(\varepsilon(x),\varepsilon(z))$ and the associated covariance operator $C$, which maps finite measure $q$ to the function obtained by integrating the kernel $Cq(z):=\int K(z,x)dq(x)$. 
        The kernel $K$ characterizes the geometry mutation imposes on cultural variations: which directions $f$ are easier for mutations to explore. It will determine the relevant notion of distance between cultural types in our main result.

    \subsection{Characterization of evolutionary biases}\label{theory:general-characterization}


        We study whether biases persist in the limit outcome of the evolutionary process \eqref{eq:replicator-mutator-dynamics}, i.e. how accurately $v^\star$ is learned through evolutionary selection.
        To do so, we focus on the case when the mutation noise $\rho$ is small and the number of situations per generation $N$ is large.
        This gives a precise characterization of the long-run behaviour of the \emph{representative cultural type}: $f_g := \Esp_{f \sim P_g}[f]$.
        Importantly, such a focus amounts to stacking the deck \emph{in favor} of evolution, which strengthens our results. 
        Our model provides an idealized identification of persistent biases induced by the distribution of conditions $q$, which remain even under a noiseless benchmark. This gives a generous upper bound on how well evolution can  identify the true stakes function $v^\star$.
        We now state the main result.
       
        \begin{theorem}\label{thm:convergence}
            Fix ancestral distribution of conditions $q$. 
            In the limit of many periods and situations per generation $(N,T \uparrow \infty)$, as noise becomes small $(\rho \downarrow 0)$, the average cultural type $f_g$ converges in the long run $(g \uparrow \infty)$ to the unique minimizer $f^\star$ of the following ``penalized fit objective'':
            \begin{align*}
                f^\star = \argmin J, \quad \text{ with } \; \; J(f):= 2\tau \mathcal{L} \bigl( \Esp_q[f] \bigm| \Esp_q[v^\star] \bigr) + \frac{\lambda}{2} \Vert f - f_0 \Vert_C^2,
            \end{align*}
            where $\Vert f - f_0 \Vert_C^2$ denotes the mutation weighted distance to the default:
            \[
                \Vert f - f_0 \Vert_C^2 := \inf_{ h \in \mathcal{M}(Z) \, | \, Ch = f-f_0} \int (f-f_0) dh
                \quad \text{ ($+\infty$ if no such $h$ exists) }
            \]

            Further, as the default weight $\lambda$ vanishes, $f^\star$ approaches the mutation-weighted-closest function to the default which correctly identifies average stakes $\Esp_q[v^\star]$:
            \[
                f^\star \xrightarrow[\lambda \rightarrow 0]{} \argmin_{ f \; | \; \Esp_q[f]=\Esp_q[v^\star]} \Vert f - f_0 \Vert_C^2 = f_0 + \frac{\Esp_q [v^\star] - \Esp_q[f_0]}{\Esp_{q \otimes q}[K]} Cq
            \]
        \end{theorem}

        \input{run_062/plot_config.tex}
        \begin{figure}
            \centering
            \begin{tikzpicture}
              \pgfplotstableread[col sep=comma]{run_062/solutions.csv}\datatable
              \begin{groupplot}[
                group style={group size=1 by 2, vertical sep=.2cm},
                width=.9\textwidth,
                legend style={font=\footnotesize},
                label style={font=\footnotesize},
              ]
    
              \nextgroupplot[
                ylabel={value},
                ymin=0,
                xmin=0, xmax=5,
                xtick = \empty,
                ytick distance = 2,
                ticklabel style = {font=\tiny},
                legend columns=2,
                height=6.5cm,
              ]
              \addplot+  [no marks, smooth,very thick,black] table[x=z, y=v, col sep=comma] {\datatable};
              \addlegendentry{$v^\star$}
    
              \addplot+ [no marks, smooth,ultra thick,gray] table[x=z, y=f0, col sep=comma]{\datatable};
              \addlegendentry{$f_0$}
    
                \addplot+ [dashed, no marks, thick, plotcolor1, domain=\pgfkeysvalueof{/pgfplots/xmin}:\pgfkeysvalueof{/pgfplots/xmax}] { \EqvA };
                \addlegendentry{$E_{q}[v]$}
    
              \addplot+ [no marks, smooth,very thick,plotcolor2] table[x=z, y=fstarA, col sep=comma]{\datatable};
              \addlegendentry{$f^\star$}

              \def\i{90}
              \def\j{70}
    
              \pgfplotstablegetelem{\i}{z}\of{\datatable}\let\zUp\pgfplotsretval
              \pgfplotstablegetelem{\i}{fstarA}\of{\datatable}\let\fsUp\pgfplotsretval
    
              \pgfplotstablegetelem{\j}{z}\of{\datatable}\let\zDown\pgfplotsretval
              \pgfplotstablegetelem{\j}{fstarA}\of{\datatable}\let\fsDown\pgfplotsretval
              \pgfplotstablegetelem{\j}{f0}\of{\datatable}\let\fzeroDown\pgfplotsretval
    
              \draw[-{Stealth[length=2mm]}, thick,dotted,plotcolor2!70!white]
                (axis cs:\zUp,\fsUp) -- (axis cs:\zUp,\EqvA)
                node[pos=0.8, right] {\scriptsize selection};
    
              \draw[-{Stealth[length=2mm]}, thick, dotted,plotcolor2!70!white]
                (axis cs:\zDown,\fsDown) -- (axis cs:\zDown,\fzeroDown)
                node[pos=0.75, left] {\scriptsize mutation};
    
              \nextgroupplot[
                xlabel={z},
                ylabel={scaled $q$},
                xmin=0, xmax=6,
                ymin=0, ymax=1.0,
                xtick distance = 2,
                ytick = {0,.5,1},
                ticklabel style = {font=\tiny},
                legend columns=2,
                height=3cm,
              ]
              
    
              \addplot[name path=p, draw=none,forget plot] table[x=z, y=qA] {\datatable};
              \addplot[name path=zero, draw=none, forget plot] coordinates {(0,0) (6,0)};
    
              \addplot[fill=plotcolor1, draw=none] fill between[of=p and zero];
    
              \end{groupplot}
            \end{tikzpicture}
            \caption{Illustration for the limit outcome of evolutionary dynamics}
            \label{fig:thm1-illustration}
        \end{figure}

        One essential observation underpins the intuition for the theorem. 
        Since attention choice optimally depends on the \emph{expected value} $\Esp_q[f]$, because specific climate conditions are not known ex ante, realized losses and hence fitness too only depends on the expected value induced by a cultural type.
        This means that, in a given generation, evolutionary dynamics can only reward \emph{fitness to the mean} and do not distinguish further properties of the function $f$.
        In other words, if two cultural types $f_1$ and $f_2$ induce the same mean value under the given conditions i.e. $\int f_1 dq = \int f_2 dq$, they induce the exact same attention choice $\nu(f_1)=\nu(f_2)$, losses $\ell_{t,g,n}(f_1)=\ell_{t,g,n}(f_2)$, and fitness $\Phi_g(f_1)=\Phi_g(f_2)$.
        In that sense, two cultural types which induce the same mean in a given generation are completely \emph{evolutionary equivalent}: they will have the same reproductive advantage in the next generation.
        
        Crucially, this wedge is introduced by the interplay of attention choices being made \emph{ex ante} with evolutionary dynamics selecting for \emph{ex post} losses.
        As a benchmark, consider a simple alternative model where agents observe the current $z$ \emph{before} choosing attention.
        In that case, losses would penalize for the fit to the \emph{entire function} $v^\star$; as the weight of the default $\lambda$ vanishes, we would then obtain that the average cultural type converges to $v^\star$ itself.
        This hints at a general phenomenon caused by the combination of decision under uncertainty (or information acquisition) across rich situations and evolutionary selection.
        The proof of \Cref{thm:convergence} relies on a pivotal result which translates this observation into dynamics, using empirically realized losses and dispersion of types.
        \begin{proposition}\label{prop:kernel-gradient-form}
            The population mean evolves according to:
            \begin{align*}
            f_{g+1}= f_g - \rho \lambda (f_g - f_0) - \tau \mathcal{L}' \bigl( \Esp_q[f_g]  \bigm| \hat{\overline{v}}_g \bigr) \Sigma_g q  + o(\rho)
            \end{align*} 
            where $\Sigma_g$ denotes the covariance operator for the population distribution $P_g$, and $o(\rho)$ is uniform in $g$.
        \end{proposition}
        This is \emph{almost} a gradient descent step on the penalized fit objective $J$. The first difference is that this features the random sample average $\hat{\overline{v}}_g$ instead of the theoretical expectation $\Esp_q[v^\star]$. 
        Note, however, that the former is an unbiased estimator of the latter -- so the expectation obtains from taking a limit when the number of situations is large.
        The second difference is that this involves the \emph{population covariance} $\Sigma_g$ of cultural types instead of mutation covariance $C$, but because each new type combines two cultural models, the dispersion of types is renewed in every generation, and in the long run $\Sigma_g q = 2\rho \, Cq + o(\rho)$.
        The last technical difference is that this yields a $o(\rho)$-order approximation for the limit, which becomes an exact first-order condition of $J$ when $\rho$ is small.
        The complete proofs of \Cref{thm:convergence} and \Cref{prop:kernel-gradient-form} can be found in \Cref{sec:fullmodel}.
        We further provide an explicit form for the solution, which is tractable but omitted in the main text for conciseness.

    \subsection{From ancestral anomalies to descendants' beliefs}\label{theory:U-shape}

        We characterize the implications of \Cref{thm:convergence} for the effect of \emph{ancestral heterogeneity}. Under a primitive "dual motive" for environmental attention, this predicts that descendants' perceived overall value of attention is approximately U-shaped in the average intensity of ancestral anomalies.

        \paragraph{Ancestral heterogeneity and descendants' beliefs}
        The effect of ancestral conditions is cleanly captured by considering the effect of varying the distribution $q$ faced by ancestors.
        Define $B(\tilde{q}|q)$ to be the subjectively perceived value of attention for a descendant now facing climate $\tilde{q}$ when their ancestor faced climate $q$.
        This is exactly the object we measure in our data. 
        Formally:
        \begin{align*}
            B(\tilde{q}|q) := \Esp_{z \sim \tilde{q}}[\hat{v}(z|q)] = \int \hat{v}(z|q) d\tilde{q}(z),
        \end{align*}
        where $\hat{v}(\cdot | q)$ denotes the limit average cultural type (stakes function), characterized in \Cref{thm:convergence}, given ancestral conditions $q$.     
        The expression for $B(\tilde{q}|q)$ highlights the bias introduced by climate distribution $q$.
        \emph{If} evolutionary dynamics induced effective pointwise learning of $v^\star$, then we should see \emph{no effect} of descending from a particular group of ancestors: we would have $\hat{v} \approx v^\star$ and $B(\tilde{q}|q) \approx \Esp_{\tilde{q}}[v^\star]$ (correct perceptions, regardless of ancestral climate).
        But since evolutionary dynamics only fit to the mean, the climate faced by one's ancestor $q$ distorts today's perceived value: $\hat{v}$ depends on $q$.
        Hence, $B(\tilde{q}|q)$ is generically biased away from $\Esp_{\tilde{q}}[v^\star]$.
        Further, 
        evolutionary dynamics push $\hat{v}_q$ towards overweighting the relative value of conditions that are frequent under $q$, causing potentially severe mismatch of shape and magnitude between $\hat{v}(\cdot | q)$ and $v^\star$. 
        This can be seen formally from the penalized fit and is clearest in the case when $\lambda$ vanishes (see \Cref{fig:thm1-illustration} or \Cref{fig:three-panels-comparative-statics} for illustrations).
        \Cref{thm:convergence} provides a toolbox which reduces studying the effect of ancestral experiences on descendants' attention to a comparative statics exercise (of $B$, in $q$). 
        From now on, focus on the solution with vanishing default weight ($\lambda \downarrow 0$) and take $f_0\equiv 0$ for simplicity.
        \begin{proposition}\label{prop:decomposition}
            The contemporaneous average belief about the importance of environmental concerns decomposes into:
             \(
                 B(\tilde{q}|q) = V(q) \times S(\tilde{q}|q)
             \),
             where $V(q):=\Esp_{q}[v^\star]$ is average ancestral value in conditions $q$ and $S$ is a mutation-weighted climate similarity measure: $S(\tilde{q}|q):= \frac{\Esp_{q \otimes \tilde{q}}[K]}{\Esp_{q \otimes q}[K]}$.
        \end{proposition}
        This formalizes the intuition that the average value among descendants is essentially the average value among ancestors, modulated by something which depends on how different a climate they face.
        In particular, $S(q|q)=1$, by construction.
        An explicit derivation is provided in \Cref{sec:fullmodel}.
        To tie back to our empirical exercise, assume that the possible distributions of climate anomalies faced by different ancestral ethnic groups are parameterized by a scale family, $\{q_\theta\}$:
        \begin{align*}
        z \sim q_{\theta} \; \Longleftrightarrow \; z \sim \theta \tilde{z}\text{, where } \tilde{z} \sim \bar{q},
        \end{align*}
        for $\theta \in [\underline{\theta},\overline{\theta}]$.
        We normalize without loss in terms of the maximal scale distribution $\bar{q} \in \Delta(Z)$: $\overline{\theta}=1$, $\bar{q}=q_1$.
        The scale parameter $\theta$ controls ``how stretched out to the right'' the distribution is, so it parametrizes the average intensity of deviations from the typical range.
        \Cref{prop:decomposition} can be written with all quantities as function as $\theta,\tilde{\theta}$ for any two distributions $q_\theta,q_{\tilde{\theta}}$; taking log derivatives in $\theta$ for fixed $\tilde{\theta}$ implies that the marginal effect of ancestral climates decomposes neatly into a \emph{transmission effect} and a \emph{dissimilarity effect}:
            \begin{align*}
             \frac{
                    \frac{\partial}{\partial{\theta}} B
                }{
                    B
                } (q_{\tilde{\theta}}|q_\theta)
            = 
            \underbrace{
                \frac{
                        \frac{\partial}{\partial{\theta}} V
                    }{
                        V
                    } (q_\theta)
            }_{\text{transmission effect}}
            + 
            \underbrace{
                \frac{
                        \frac{\partial}{\partial{\theta}} S
                    }{
                        S
                    } (q_{\tilde{\theta}}|q_\theta)
            }_{\text{dissimilarity effect}}.
            \end{align*}
        If the dissimilarity effect is "not too strong" relative to the transmission effect, then the descendant average value $B$, as a function of ancestral conditions, will retain the structural properties of ancestral average value $V$ -- although the dissimilarity effect still alters the strength of the effect at the margin.
        To make this operational, we anchor to a simple micro-foundation of the value of environmental attention.

        \paragraph{From primitive motives to observable patterns}
        Assume that in any given situation, the true value of attention derives from a (possibly problem-specific) mix of two motives.
        The first is an \emph{exploitation motive}, which induces losses that are decreasing in deviations from normal conditions (there is more to gain from attention when conditions are typical).
        The second is a \emph{protection motive}, which induces losses that are increasing in deviations from normal conditions (there is more to lose from inattention when conditions are unusual).
        Formally, in any situation $s$:
        \begin{align*}
        v_s(z) = \alpha_s w(z) + \beta_s r(z), 
        \end{align*}
        where $w(z)$ represents the gains from adaption to typical conditions, assumed strictly convex decreasing and with vanishing slope at the right boundary of $Z$ and $r(z)$ represents the reduction of losses from extreme events, assumed strictly convex increasing with vanishing slope at the left boundary; $\alpha_s,\beta_s \geq 0$ are problem specific weights.
        Assume we can average across the range of possible situations to give the representation:
        \begin{align*}
        \Esp[v_s] = v^\star \text{, where: } v^\star(z) = \alpha \; w(z) + \beta \; r(z) 
        \end{align*}
        where $\alpha,\beta >0$ represent the fundamental average weights of the exploitation and protection motives across the situations that one might encounter.
        This delivers our second main theoretical result.
        \begin{theorem}\label{thm:U-shape}
            Denote by $\bar{\mu}_k$ the $k$-th uncentered moment of $\bar{q}$.
            \begin{enumerate}[label=\roman*.]
            \item \emph{(U-Shape of Ancestral Value)} There exists a unique $\theta^\dagger$ such that $V$ is strictly decreasing if $\theta<\theta^\dagger$ and strictly increasing if $\theta>\theta^\dagger$.
            \item \emph{(Approximate U-Shape of Descendants' Value -- tube bounds)} For all $\tilde{\theta}$ the mapping $\theta \mapsto B(\tilde{\theta}|\theta)$ is within a $\delta$-sized multiplicative tube around $V$, i.e.
               \begin{align*}
                \forall \theta, (1-\delta) V(\theta) \leq B(\tilde{\theta}|\theta) \leq (1+\delta)V(\theta) \quad \text{ with }  \delta := \frac{\bar{\mu}_1}{e^{-\overline{z}}} |\overline{\theta} - \underline{\theta}|
               \end{align*} 

               \item \emph{(Approximate U-Shape of Descendants' Value -- trough displacement)} If $v^\star$ is strongly convex with $(v^\star)'' \geq c>0$, any minimizer  $\theta^*_B$ of $B$ satisfies:
               \begin{align*}
                |\theta^*_B - \theta^\dagger| \leq \sqrt{\frac{4\delta}{c\bar{\mu}_2(1-\delta)} V(\theta^\dagger)}
               \end{align*}
            \end{enumerate}
            \end{theorem}

        \begin{figure}
            \centering
            \input{run_062/plot_config.tex}
            \begin{tikzpicture}
                \pgfplotstableread[col sep=comma]{run_062/solutions.csv}\datatable
                
                \begin{groupplot}[
                  group style={group size=3 by 2, vertical sep=.5cm},
                  width=.36\textwidth,
                  legend style={font=\footnotesize},
                  label style={font=\footnotesize},
                ]
        
                \nextgroupplot[
                  title={$\theta=\thetaA$},
                  ylabel={value},
                  ymin=0,
                  xmin=0, xmax=6,
                  xtick distance = 2,
                  ytick distance = 2,
                  ticklabel style = {font=\tiny},
                  legend columns=2,
                  height=5.5cm,
                ]
                \addplot+  [no marks, smooth,very thick, black] table[x=z, y=v, col sep=comma] {\datatable};
        
                \addplot+ [no marks, smooth,very thick, gray] table[x=z, y=f0, col sep=comma]{\datatable};
                
                \addplot+ [no marks, smooth,very thick,plotcolor2] table[x=z, y=fstarA, col sep=comma]{\datatable};
        
                  \addplot+ [dashed, no marks, plotcolor1, domain=\pgfkeysvalueof{/pgfplots/xmin}:\pgfkeysvalueof{/pgfplots/xmax}] { \EqvA };

                \nextgroupplot[
                  title={$\theta=\thetaB$},
                  ymin=0,
                  xmin=0, xmax=6,
                  xtick distance = 2,
                  ytick distance = 2,
                  ticklabel style = {font=\tiny},
                  legend columns=2,
                  height=5.5cm,
                ]
                \addplot+  [no marks, smooth,very thick, black] table[x=z, y=v, col sep=comma] {\datatable};
        
                \addplot+ [no marks, smooth,very thick, gray] table[x=z, y=f0, col sep=comma]{\datatable};
        
                \addplot+ [no marks, smooth,very thick,plotcolor2] table[x=z, y=fstarB, col sep=comma]{\datatable};
        
                  \addplot+ [dashed, no marks, plotcolor1, domain=\pgfkeysvalueof{/pgfplots/xmin}:\pgfkeysvalueof{/pgfplots/xmax}] { \EqvB };
        
                \nextgroupplot[
                  title={$\theta=\thetaC$},
                  ymin=0,
                  xmin=0, xmax=6,
                  xtick distance = 2,
                  ytick distance = 2,
                  ticklabel style = {font=\tiny},
                  legend columns=2,
                  height=5.5cm,
                ]
                \addplot+  [no marks, smooth,very thick, black] table[x=z, y=v, col sep=comma] {\datatable};
        
                \addplot+ [no marks, smooth,very thick, gray] table[x=z, y=f0, col sep=comma]{\datatable};
        
                \addplot+ [no marks, smooth,very thick,plotcolor2] table[x=z, y=fstarC, col sep=comma]{\datatable};
        
                  \addplot+ [dashed, no marks, plotcolor1, domain=\pgfkeysvalueof{/pgfplots/xmin}:\pgfkeysvalueof{/pgfplots/xmax}] { \EqvC };

                \nextgroupplot[
                  xlabel={z},
                  ylabel={scaled density},
                  xmin=0, xmax=6,
                  ymin=0, ymax=1.0,
                  xtick distance = 2,
                  ytick = {0,.5,1},
                  ticklabel style = {font=\tiny},
                  legend columns=2,
                  height=2.5cm,
                ]
        
                  \addplot[name path=p, draw=none,forget plot] table[x=z, y=qA] {\datatable};
                  \addplot[name path=zero, draw=none, forget plot] coordinates {(0,0) (6,0)};
                  \addplot[fill=plotcolor1, draw=none] fill between[of=p and zero];
        
                \nextgroupplot[
                  xlabel={z},
                  xmin=0, xmax=6,
                  ymin=0, ymax=1.0,
                  xtick distance = 2,
                  ytick = {0,.5,1},
                  ticklabel style = {font=\tiny},
                  legend columns=2,
                  height=2.5cm,
                ]
        
                  \addplot[name path=p, draw=none,forget plot] table[x=z, y=qB] {\datatable};
                  \addplot[name path=zero, draw=none, forget plot] coordinates {(0,0) (6,0)};
                  \addplot[fill=plotcolor1, draw=none] fill between[of=p and zero];
        
                \nextgroupplot[
                  xlabel={z},
                  xmin=0, xmax=6,
                  ymin=0, ymax=1.0,
                  xtick distance = 2,
                  ytick = {0,.5,1},
                  ticklabel style = {font=\tiny},
                  legend columns=2,
                  height=2.5cm,
                ]
        
                  \addplot[name path=p, draw=none,forget plot] table[x=z, y=qC] {\datatable};
                  \addplot[name path=zero, draw=none, forget plot] coordinates {(0,0) (6,0)};
                  \addplot[fill=plotcolor1, draw=none] fill between[of=p and zero];
        
                \end{groupplot}
        
            \end{tikzpicture}
            \caption{Evolutionary outcome: comparative statics when increasing the scale of climate anomalies}
            {\setstretch{.9} \footnotesize Scaled densities are given in the lower panel for three values of $\theta$; $v^\star$ is the black curve (identical across all panels); the dashed line represents $\Esp_{q_\theta}[v^\star]$, which decreases then increases, showing the U-shape of ancestral value; the evolutionary outcome $\hat{v}(\cdot|q_\theta)$ is represented by the pink curve. This illustrates that $\hat{v}(\cdot|q_\theta)$ tends to \emph{overweight} the climate realizations which are most common under $q_\theta$ at the cost of a potentially severe shape mismatch with the true value $v^\star$.}
            \label{fig:three-panels-comparative-statics}
        \end{figure}    

        A first step towards the result is the elementary observation that under our exploitation-protection assumption, the primitive $v^\star$ is U-shaped \emph{in climate conditions $z$}. 
        However, this is misleading: it establishes a U-shaped property, but not the one we observe (only of the true value in terms of realized climate conditions).
        How does this transfer to descendants' perception, and average values in terms of the whole distribution?
        To see the intuition behind \Cref{thm:U-shape}, first consider an individual facing the same anomaly distribution $q_\theta$ as their ancestors.
        Then average realized $z_{t,g}$ (which is exactly our explanatory variable in the data) is an unbiased estimator of the scale parameter $\theta$, so for enough generations:
        \(
        \sum_g \frac{1}{G} \sum_t \frac{1}{T} z_{t,g} \approx \theta
        \).
        The perceived value of attention is approximately equal to the mean value of attention under ancestral conditions via evolutionary convergence:
        \begin{align*}
        \Esp_{q_\theta}[\hat{v}_{q_\theta}] \approx \int v^\star(z) dq_\theta(z) = \alpha W(\theta) + \beta R(\theta) \text{ where } W(\theta)=\Esp_{q_\theta}[w], R(\theta)=\Esp_{q_\theta}[r]
        \end{align*}
        which leads to the first point in the theorem: ancestral value $V$ is U-shaped in $\theta$. 
        However, this approximate equality masks the modulating effect of the descendants' climate distribution $q_{\tilde{\theta}}$.
        While \Cref{prop:decomposition} gives a general intuition why the U-shape will be preserved provided that climate dissimilarity effects are not too strong, the effects remain very complex.
        Indeed, there are many possible ways to produce a precise ``U-shape preservation'' statement, that each require taking a stand on fine properties of $v^\star$ and $\bar{q}$. 

        \begin{figure}
            \centering
            \input{sweep_001/plot_config.tex}
              \begin{tikzpicture}
                \pgfmathsetmacro{\onepluseps}{1.05}
                \pgfmathsetmacro{\oneminuseps}{.95}
                \pgfplotstableread[col sep=comma]{run_062/solutions.csv}\datatableind
                \pgfplotstableread[col sep=comma]{sweep_001/expectation_sweep.csv}\datatableagg
                \begin{groupplot}[
                  group style={group size=2 by 1, vertical sep=.2cm},
                  width=.45\textwidth,
                  legend style={font=\footnotesize},
                  label style={font=\footnotesize},
                ]
        
                \nextgroupplot[
                  xlabel={z},
                  ylabel={value},
                  ymin=0,
                  xmin=0, xmax=6,
                  xtick distance = 1,
                  ytick distance = 2,
                  ticklabel style = {font=\tiny},
                  legend columns=2,
                  height=6cm,
                ]
        
                \addplot+  [no marks, smooth,very thick,black] table[x=z, y=v] {\datatableind};
                \addlegendentry{$v^\star$}

                 \nextgroupplot[
                  xlabel={$\theta$},
                  ylabel={expected value},
                  ymin=5.5, ymax=10,
                  xmin=.4, xmax=1.5,
                  xtick distance = .5,
                  ytick distance = 1,
                  ticklabel style = {font=\tiny},
                  legend columns=2,
                  height=6cm,
                  extra x ticks={\thetamin},
                  extra x tick labels={$\theta^\dagger$},
                ]
            
                \addplot+  [no marks, smooth,very thick,plotcolor3] table[x=theta, y=Eq_v] {\datatableagg};
                \addlegendentry{$V(\theta)=\Esp_{q_\theta}[v]$}
        
                \draw[plotcolor3!50!black] (axis cs:\thetamin,0) -- (axis cs:\thetamin,\Vmin);
        
                  \addplot[name path=p,dashed,plotcolor3!50!white,forget plot] table[x=theta, y expr=\thisrow{Eq_v}*\oneminuseps] {\datatableagg};
                  \addplot[name path=zero,dashed,plotcolor3!50!white,forget plot] table[x=theta, y expr=\thisrow{Eq_v}*\onepluseps] {\datatableagg};
        
                  \addplot[fill=plotcolor3!30!white, draw=none] fill between[of=p and zero];
        
                  \node[plotcolor3,above] at (axis cs:\thetamin,\Vmin) {};
                  \node[plotcolor3,below] at (axis cs:\thetamin,5.5) {$\theta^\dagger$};
        
                  \draw[dotted, plotcolor3!50!black, thick] (axis cs:{\thetamin-0.07},0) -- (axis cs:{\thetamin-0.07},10);
                  \draw[dotted, plotcolor3!50!black, thick] (axis cs:{\thetamin+0.07},0) -- (axis cs:{\thetamin+0.07},10);
                \end{groupplot}
              \end{tikzpicture}
            \caption{Illustration of \Cref{thm:U-shape} U-shape and approximate U-shape.}
            \label{fig:U-shape}
        \end{figure}
        
        \Cref{thm:U-shape}, focuses on two precise ways to quantify the U-shape preservation.
        First, how far $B$ can stray away from the shape of $V$ is controlled by the dispersion term $\delta$, which is itself proportional to the maximal dispersion of possible condition distributions $\overline{\theta}-\underline{\theta}$.
        Intuitively, climate anomalies $z$ are already tail realizations, so $\overline{\theta}-\underline{\theta}$ controls the spread of the tail of a tail. 
        This is likely to be small, since little variations in $\theta$ correspond to sizable variations in the concrete weather distribution.
        In our sample, we indeed find that variables for average anomaly scales are quite concentrated in scale (.01 to .09, see \Cref{fig1}).
        Second, the convexity of $v^\star$ (the sensitivity of protection and exploitation marginal effects to variation in conditions), together with the same distributional parameters, controls the displacement of the trough of $\theta \mapsto B(\tilde{\theta}|\theta)$.
        This gives a robust sense in which higher and lower levels of anomalies lead to higher perceived value of attention.
        See \Cref{sec:fullmodel} for the complete proof.

        Together, \Cref{thm:convergence} and \Cref{thm:U-shape} give a precise description of how descendants' average perception of the value of environmental attention is shaped by the distribution of climate conditions their ancestors faced.
        We have focused on the U-shape, which is our central prediction. 
        Further interesting comparative statics exercises can be carried out, using this theoretical machinery, to analyze the effect of various parameters -- we discuss some in \Cref{sec:robustness-checks} and \Cref{sec:heterogeneity}.
        The core of the theory is more general than our application; we hope it may serve as a building block.

\section{Robustness}\label{sec:robustness-checks}
	
	We verify the robustness of the U-shaped relationship between average ancestral climate anomalies and the perceived value of environmental attention. 
	We run multiple alternate specifications to establish that our findings are not driven by the idiosyncrasies of the sample, our choice of lifespan, range of typical conditions or misspecification of the functional form in the regression equation.
    We further highlight that the core theoretical results hold under natural variants of the model, and discuss extensions.
	
	\paragraph{Choice of language sample} 
	Matching individuals to their corresponding ethnic ancestors through the use of spoken language could introduce systematic noise. 
	First, even though our approach is agnostic on the channel of association apart from existence of a link between individuals and some of their ancestors through shared language, the mapping of descendants to their ancestors is inherently more imprecise for languages with a larger group of speakers (e.g. English, Spanish, Arabic).
	Further, if a large part of our sample is composed of individuals belonging to a particular language group, it is natural to test the sensitivity of our results to the inclusion of such groups.  
    We assess sensitivity to large language groups by successively excluding English, Spanish, and Arabic speakers (three of the largest language groups from our sample) and by weighting observations inversely by ethnic / language-group frequency. We also exclude respondents whose home language is proxied by the interview language (approximately 7\% of the sample). The estimated U-shape remains statistically significant across these specifications (see \Cref{robust}).
	
	\paragraph{Contemporaneous region confounds} 
	Individuals belonging to a specific ethnic group may decide to migrate to specific regions whose characteristics might influence environmental attitudes. 
	Our original empirical specification tackles this through the inclusion of respondent's current country-year fixed effects, making identification of coefficients $\beta_1$ and $\beta_2$ in our sample independent of any variation that would have jointly explained the responses of two individuals from the same ethnic group residing in the same country at the time of survey. 
    Within-country sorting may remain a concern as different regions face differential contemporaneous climatic shocks.
	To robustly address this concern, we replace country-year fixed effects with region-year fixed effects (\Cref{robust}, column 6),\footnote{Where the region is a sub-territory within the country. See IVS documentation for further details.} absorbing contemporaneous shocks and other common factors at a finer geographic level.
    The estimated coefficients retain their signs and statistical significance.

    \paragraph{Construction of ancestral climatic anomalies} 
	We also test robustness of the results to our choice of measure of average climatic anomalies across ancestral generations. 
	We vary our choice of lifespan for each generation from 20 (baseline) to 50 years and our definition of the range of typical conditions: $10^\text{th}$--$90^\text{th}$, $20^\text{th}$--$80^\text{th}$ (baseline), $30^\text{th}$--$70^\text{th}$ percentiles of temperatures within the lifetime. 
    We also replace average absolute excursions with squared excursions or within-generation temperature standard deviations.
    Almost across all cases, the signs and the significance of the coefficients are retained (see \Cref{method_var}).\footnote{
        A similar exercise with our group level variable of attention to environment (environmental themes in ethnic folklore) shows broadly the same results.
        }

    \paragraph{Other characteristics of temperature distributions and model fit} 
    We further assess functional-form sensitivity and potential confounding by other characteristics of historical temperature distributions (\Cref{highermom}).
	We augment the baseline specification by including a cubic anomaly term, across-generation variation in temperature \citep[as in][]{nunn_giuliano}, and the first four moments of the historical temperature distribution.
    We also compare the baseline quadratic specification with a linear alternative using the Akaike Information Criterion (AIC) and the Bayesian Information Criterion (BIC).
    The quadratic specification improves both criteria by approximately 200 points relative to the linear model, while additional terms do not substantially improve fit.
    Across all specifications, both the signs and statistical significance of our coefficients of interest remain robust.
    This addresses concerns that (a) our baseline measure may proxy for dissimilarities in ancestral experiences across generations \citep[consistent with issues raised in the original approach of][]{nunn_giuliano} or (b) our measure of climatic anomalies may capture idiosyncratic geographic characteristics of group locations beyond the topographic and climate-region controls already included in the baseline specification. It also robustly supports the quadratic relationship.
    
	

    \paragraph{Alternate functional forms for folklore}
	We test the sensitivity of the folklore results to alternative outcome constructions and estimation methods (\Cref{folklore-altspec}).
    These include the manual SVO-based motif classification introduced in \Cref{sec:data}, untransformed folklore shares (addressing potential concerns associated with log transformations), inverse-hyperbolic-sine-transformed shares, and a Poisson specification with a log link. 
    We consistently find a statistically significant U-shaped relationship between the prevalence of environmentally related folklore at the group level and average ancestral climatic anomalies.

    \paragraph{Adjustments for Spatial Non-Independence}
    Finally, to ensure that our results are not biased by spatial correlation in the standard errors, we re-estimate all specifications allowing for spatial dependence in the error structure within radii of 200 km and 500 km (\Cref{conley-table}). 
    Across all specifications and samples, the statistical significance of our estimates largely remains intact.

    \paragraph{Variants and generalizations of the model}
    The core predictions of the model are robust to variants, and direct generalizations have been omitted for clarity of exposition. Natural generalizations include noisy information about problem-specific stakes, corner solutions in attention choices, and alternative mutation kernels. While these changes affect the precise analytical characterization, the central distinction between average and pointwise learning remains. In particular, The decomposition in \Cref{prop:decomposition} generalizes with an appropriately modified similarity kernel, and analogous bounds to those in \Cref{thm:U-shape} can be derived under suitable assumptions.
    
    The tractability of our model lends itself to many further exercises that are beyond the scope of this paper.
    For instance, we may probe the effect of the default on the shape of beliefs.
    Our main U-shape result was derived under the assumption that the default is $f_0 \equiv 0$ (no value), which is both natural and helpful for tractability, but different assumptions will have implications for the shape of generated perceptions and how sensitive they are to underlying distributions.
    As a simple illustration, \Cref{fig:default-effect-comparative-statics} reproduces the exercise from \Cref{fig:three-panels-comparative-statics} (varying the underlying distribution), with a different default.

    \paragraph{Extensions of the theory}
    More substantive extensions are possible. 
    When considering heterogeneous climate distributions among ancestors, the main results are largely preserved if more nuanced: instead of fitting to \emph{one} moment, the evolutionary process will fit to multiple -- but as long as the diversity of experienced climates remain small, evolutionary bias will be induced.
    The mean-reversion to the default in the mutation structure can be removed; this creates a technical difficulty: the evolutionary process is now dominated by inertia and noise, and no longer guaranteed to converge to a unique point. Nonetheless, the core evolutionary bias remains and this gives additional scope for even stronger forms of path dependency.    
    The dual motive assumption suggests another natural assumption in the form of assuming that agents learn separately the value of exploitation and the value of protection; i.e a cultural type is a \emph{pair} of functions $(\hat{w},\hat{r})$. 
    This makes the analysis more intricate, but the machinery can plausibly be adapted, and this gives scope for possibly richer descriptions, allowing to study entanglement and separation between the two channels.

\section{Heterogeneity Analysis}\label{sec:heterogeneity}
	

    If cultural transmission across generations is driven by evolutionary forces, then any social or technological factor which influences the true value of adaptation to one's environment should also influence resulting perceptions.
    In the language of our model: if different groups face have different "true value" $v^\star$, the evolutionary learning process will induce different descendant beliefs.
    This leads us to examine, theoretically and empirically, \emph{historical economic development} as a mitigating factor for both the sensitivity to climate shocks and payoffs from adaptation (\Cref{sec:heterogeneity-development}).
    Similarly, the current climate conditions faced by descendants ($\tilde{q}$ in the model) was explicitly shown in \Cref{prop:decomposition} and \Cref{thm:U-shape} to have a mitigating effect on the resulting beliefs. 
    We investigate this effect empirically in \Cref{sec:heterogeneity-contemporaneous-conditions}.
    Finally, we further consider whether \emph{migration} acts as an alternative channel. 
    Since our framework rests on cultural transmission about \emph{intrinsic importance of climate concerns}, rather than about the climate itself, it predicts that migration itself should not matter except through the climate distributions (of descendants and ancestors).
    \Cref{sec:heterogeneity-migration} verifies that the U-shape pattern is insensitive to splitting the sample by migration status, supporting the independence of transmission at the language group level.

    \subsection{Historical economic development}\label{sec:heterogeneity-development}

        It is reasonable to hypothesize that heterogeneity in social, economic, or technological development, because it affects the degree of reliance on natural resources and resilience to climate shocks, would lead to a true value $v^\star$ which may differ in \emph{sensitivity} but not in \emph{shape}.
        In other words, the same exploitation/protection motive is at play across different groups, attenuated or accentuated.
        If so, we should expect to see a corresponding \emph{attenuation} or \emph{accentuation} of the U-shape result as a function of group-level development characteristics.
        We provide a direct comparative statics result below which formalizes this idea.
        We then investigate this idea empirically by carrying out a series of heterogeneity analyses, splitting our World Values Survey sample by measures of historical development of ancestor ethnic groups.
    
        \paragraph{Comparative statics}
        The decomposition in \Cref{prop:decomposition} together with \Cref{thm:U-shape} immediately yields a comparative statics result on the primitive value of attention which makes the above idea precise. 
        Fix the relative weights $\alpha,\beta>0$ of the exploitation and protection motives, and consider two primitives $v^\star = \alpha w + \beta r$ and $\tilde{v}^\star = \alpha \tilde{w} + \beta \tilde{r}$ both satisfying the assumptions of \Cref{thm:U-shape}. 
        We say that $\tilde{v}^\star$ is \emph{more sensitive} than $v^\star$ if both $\tilde{w}-w$ and $\tilde{r}-r$ are convex on $Z$, i.e. the marginal effects of climate conditions under $\tilde{v}^\star$ are sharper than under $v^\star$, in both motives. 
        \begin{proposition}\label{prop:het-convexity}
            If $\tilde{v}^\star$ is more sensitive than $v^\star$, then $\tilde{V}''(\theta) \geq V''(\theta)$ for all $\theta \in [\underline{\theta},\overline{\theta}]$, and the tube width $\delta$ in \Cref{thm:U-shape}(ii) is the same for $\tilde{B}$ and $B$.
        \end{proposition}
        The proof is direct: $\tilde{v}^\star - v^\star = \alpha(\tilde{w}-w) + \beta(\tilde{r}-r)$ is convex as a positive combination of convex functions, so $(\tilde{V}-V)''(\theta) = \Esp_{\bar{q}}[\bar{z}^2 (\tilde{v}^\star-v^\star)''(\theta\bar{z})] \geq 0$; and $\delta$ depends only on $\bar{q}$ and $Z$, not on the primitive $v^\star$. 
        Hence \emph{a more sensitive exploitation--protection structure produces a more pronounced U-shape in ancestral value $V$, and this curvature ordering transmits to descendants' value $B$} up to a primitive-invariant margin induced by contemporaneous conditions. 
        The condition that $\tilde{w}-w$ and $\tilde{r}-r$ be convex is a natural partial order.
        This formalizes the structural prediction we now take to the data: groups with sharper underlying exploitation--protection sensitivities (i.e. less developed) should display a more accentuated U-shape, and conversely the effect should be muted by higher development.
    	
    	\paragraph{Empirical results} 
        We provide suggestive evidence of a muted effect by rerunning \Cref{main_eq} separately on sub-samples split by level of economic and institutional development at the ethnic group level. 
    	The sample is divided along four characteristics, depending on whether the ethnic ancestors of an individual (1) were hunter / gatherers or non-hunter / gatherers, (2) had low / high level of economic complexity, (3) had low / high level of local jurisdiction and (4) had low / high level of global jurisdiction.\footnote{Economic complexity is captured by the size of local communities \citep[similar to][]{nunn_giuliano}. Communities of size above 400 are considered as those with high economic complexity. If a group had a local jurisdiction of at least 4 levels or global jurisdiction of at least one level, they are considered as communities with high local (global) jurisdiction. See \citet{nunn_giu_database} for more details.} 
    	
    	\begin{figure}[t!]
    		{\centering
    		\includegraphics[scale=0.3]{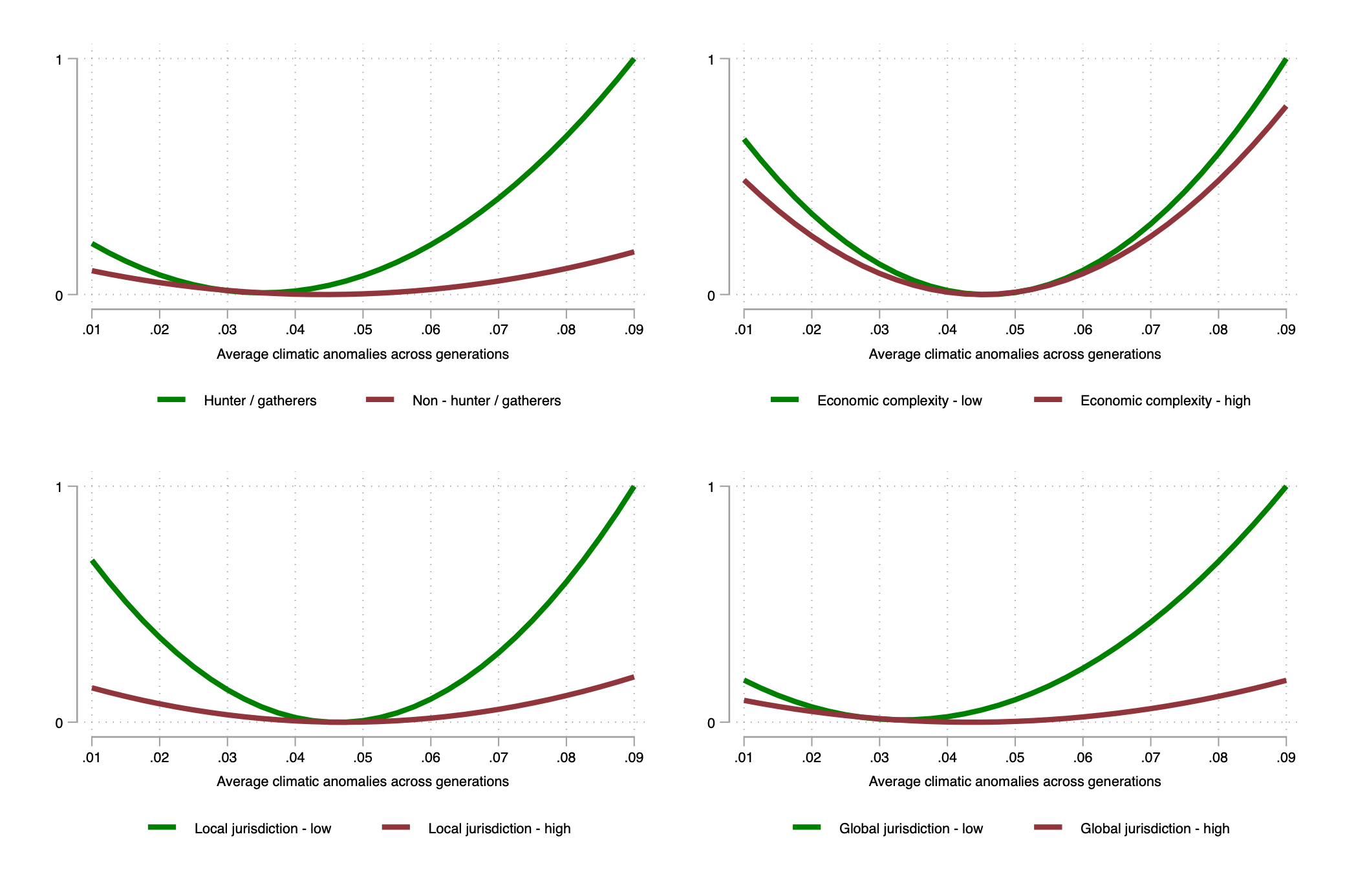}
    		\caption{Margins plots - Relative slopes categorized by the development level of the ethnic group} \label{fig_het}
            }
    		{\setstretch{.9} \footnotesize \textbf{Note:} Margins corresponding to \Cref{main_eq}, plotted for samples split by the level of ethnic group's historical development. All values are normalized to facilitate comparison of slopes, i.e. the minimum point of margin plots across the sample split are made equal to each other and all other margin values are normalized to lie between 0 and 1. Controls are fixed at their mean values from the original sample. Top left sub-figure corresponds to margins obtained from hunter / gatherers vs non-hunter / gatherers. Top right (Bottom left / Bottom right) figure corresponds to margins obtained from low vs high level of economic complexity (local / global jurisdiction) respectively.\par}
    	\end{figure} 
    		
    	\Cref{reg-IVS-het_eth} reports the results from our original specification for each sub-sample. Note that, irrespective of how the sample is spliced, the U-shaped relationship is maintained. However, the main purpose of this exercise is to compare the slopes across the two samples, not the levels. 
    	Indeed, historical economic development may independently impact other factors such as the level of education or awareness towards climate issues in general.
    	In this exercise, we analyze how transmission through the climatic variability at the ancestral level gets muted or amplified in the presence or absence of historical economic protection. 
    	If ancestors had a higher stock of economic and institutional protection then the actual temperature variability would result in a less than one to one transition into climate related ancestral experiences. 
    	Therefore, to facilitate comparison, we normalize the margin plots we obtain from our separate regressions, so that the minimum point of the predicted level of attention function from both the sub-samples are equalized and values are normalized to be in $[0,1]$.
    	
    	\Cref{fig_het} plots predicted attention functions for each sub-sample under the four different types of sample splits. 
    	In all cases, we see that the U-shape of the attention function is more pronounced when the ancestral level of economic and institutional development is low, i.e. for individuals coming from more economically developed ethnic groups, the U-shaped curves are pointwise flatter.
        This supports the idea that cultural transmission is driven by the realized payoffs to paying attention (or not) to environmental issues, which is ultimately determined by the interaction of the conditions a group faces with their sensitivity to both the exploitation and protection motives.

    \subsection{Contemporaneous environmental conditions}\label{sec:heterogeneity-contemporaneous-conditions}
    
        So far, we have focused on characterizing how descendants' beliefs about the value of environmental attention is shaped by ancestral experiences. 
        However, this is fundamentally intertwined with a contemporaneous condition effect. 
        We now examine when and how the relationship between ancestral experiences and current environmental attention becomes more or less pronounced by studying their interaction with contemporaneous factors arising from an individual’s current environment.

        \paragraph{Disentangling transmission from current conditions}
        Recall from the model's formalization that our central object of interest is the expected value under today's condition, of the stakes function which is determined by ancestral conditions: $B(\tilde{q}|q):= \Esp_{\tilde{q}}[\hat{v}(\cdot|q)]$.
        Since the transmitted object is a \emph{function} of which the resulting average belief is an expectation, this allows for many complex effects and interplay between the shapes of $\tilde{q}$ and $\hat{v}$.
        Disciplining this richness was the main object of \Cref{prop:decomposition}: we can decompose the current belief into an ancestral value component $V(q)$ and a climate similarity factor $S(\tilde{q}|q)$: $B(\tilde{q}|q)=V(q)S(\tilde{q}|q)$. 
        Although \Cref{thm:U-shape} establishes that $B$ approximately retains the shape of $V$, under specific distributions at the individual level this still leaves room for complex mitigating effects in \emph{either direction} in general -- exhaustively disciplining all such effects theoretically is not possible.
        Nonetheless, it is natural to expect a more disciplined interplay between the U-shape and contemporaneous conditions under specific forms of climate heterogeneity.
        We consider two specific mechanisms below, one of which is directly anchored in our decomposition while the other is a natural but informal extension.
        

        \paragraph{Effect of (dis)similarity} 
        When current climate conditions more closely resemble ancestral ones, the dissimilarity factor $S(\tilde{q}|q_\theta)$ in \Cref{prop:decomposition} is closer to one; its variation in $\theta$ is correspondingly muted, so the U-shape of $V$ is transmitted to $B$ with less distortion. Conversely, larger climate dissimilarity introduces a non-trivial $\theta$-dependence in $S$ that partially offsets the transmission effect, attenuating the U-shape in $B$.

\begin{figure}[t!]
    		{\centering
    		\includegraphics[scale=0.3]{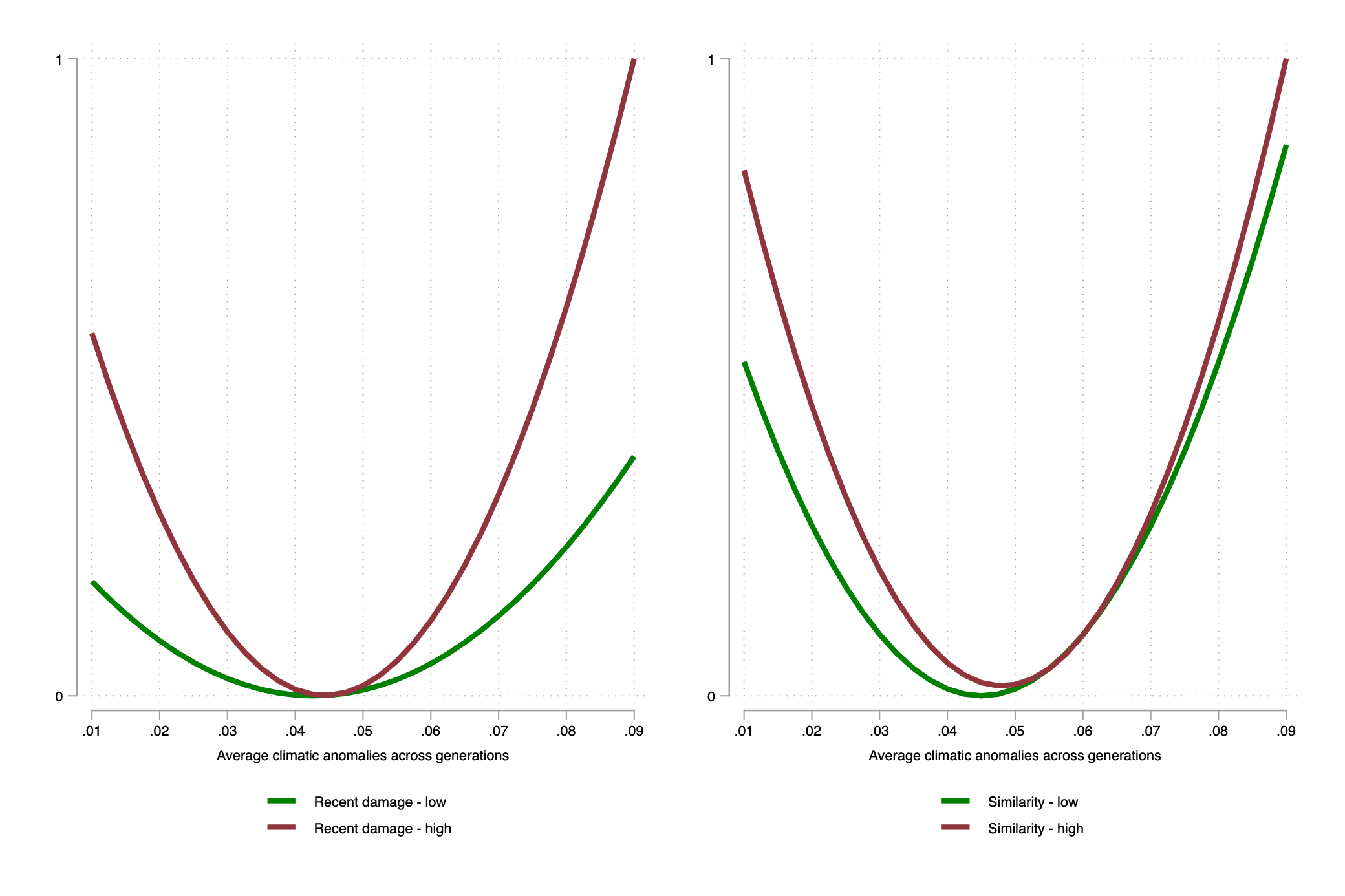}
    		\caption{Margins plots - Relative slopes categorized by current climatic conditions faced by the respondents.} \label{fig_current_char}
    		}
            {\setstretch{.9} \footnotesize \textbf{Note:} Margins corresponding to \Cref{main_eq}, plotted for samples split by the current level of damages and similarity in country's weather and ancestral weather. All values are normalized to facilitate comparison of slopes, i.e. the minimum point of margin plots across the sample split are made equal to each other and all other margin values are normalized to lie between 0 and 1. Controls are fixed at their mean values from the original sample. Left sub-figure corresponds to margins obtained from low vs high level of recent disaster related damages. Right sub-figure corresponds to margins obtained from low vs high level of similarity between current environment and ancestral environment.
            }
    	\end{figure}

        To formally investigate this mechanisms, we divide respondents in the World Values Survey by the median similarity between current climatic variability and ancestral climatic variability.\footnote{This measure is constructed as the absolute difference between ancestral climatic anomalies and the current climatic anomalies experienced in the respondent's country of residence.}
        \Cref{reg-IVS-het_char} reports estimates of our baseline specification for each subsample. 
        A first observation is that the U-shaped relationship between ancestral climatic variability and the perceived value of environmental attention is preserved in both subsamples. 
        However, as before, our primary interest lies in comparing slopes rather than levels across subsamples. To facilitate this comparison, we normalize the predicted margins from each subsample so that the minimum of the predicted perceived importance of climate issue is equalized across groups, and all values are scaled to lie in the interval $[0,1]$.
    	\Cref{fig_current_char} plots the resulting predicted outcome, which shows that when current climatic conditions more closely resemble ancestral environments, the influence of ancestral experiences is stronger.
        
        \paragraph{Damage exposure} 
        Exposure to damage caused by contemporaneous disasters should plausibly heighten the salience of inherited beliefs about the value of attention in unusual conditions, i.e. the part of $v^\star$ driven by the protection motive. We do not formalize this second mechanism as it would require additional structure beyond what is delivered by our evolutionary learning model;\footnote{A relatively basic issue is since evolutionary dynamics identify only the average $\Esp_q[v^\star]$ and not the shape of $v^\star$ separately, the descendant value $B(\tilde{q}|q) = \Esp_q[v^\star] \cdot S(\tilde{q}|q)$ does not decompose into separately re-weighted exploitation and protection components driven by $\tilde{q}$. A more complex issue is that what we measure technically combines anomalies and the damage they cause, which depends on other technological factors as previously argued. Lastly, it is plausible that many other channels (e.g. salience or awareness) combine with those in our model. The mechanism we describe is therefore an extension of the model's intuition rather than a formal implication.} we view it as a natural extension consistent with the empirical pattern.
        To test this empirically, we split the sample by the median level of recent natural disaster impacts at the country level.\footnote{We rely on disaster-impact data \citep{disasters} to compute country-level averages over the past 20 years of (i) the share of GDP affected by natural disasters, (ii) the proportion of the population that died due to natural disasters, and (iii) the proportion of the population affected. To reduce dimensionality, we perform a principal component analysis over these variables and use the first component as our measure of disaster exposure.}
        
        Results are also collected in \Cref{reg-IVS-het_char}, for each subsample. 
        Once again, the U-shaped relationship is found in both subsamples. 
        Using the same normalization as before, the predicted perceived importance of climate issue in each sample is plotted in \Cref{fig_current_char}.	
    	The U-shaped curves corresponding to individuals facing fewer recent disaster impacts are pointwise flatter.
        In other words, greater exposure to recent natural disasters heightens the marginal impact of ancestral climatic variability on current environmental attention.

    \subsection{Migration and language as a channel of transmission}\label{sec:heterogeneity-migration}
    
        A potential concern is that our estimates may be biased if descendants migrate far from their ancestral climate zones or if ancestral climatic conditions are correlated with other long-run historical traits—such as institutions or colonial legacies—that also shape contemporary environmental attitudes. While we already control for a rich set of historical ethnic-group-level characteristics to mitigate these concerns, this concern is further attenuated in our setting because individuals are linked to ancestral climatic experiences through language spoken, rather than through country of origin or current location. This mapping implies that the relevant channel of transmission operates through linguistic and cultural narratives, such as family values, stories, and cultural practices, that are transmitted across generations and can persist irrespective of individuals’ current place of residence. This interpretation is consistent with our emphasis on folklore as a cumulative stock of knowledge and beliefs embodied in language, and underpins the structure of our cultural transmission model, which operates only through realized gains and losses from chosen attention.
    	
    	Nevertheless, to directly assess whether historical migration weakens the relationship between ancestral climate and contemporary environmental attention, we split the sample based on whether at least one parent of the respondent was an immigrant to the country of residence.\footnote{Directly identifying first-generation immigrants is not feasible across all waves of the World Values Survey, as this information is not consistently collected, which would lead to substantial sample loss.} Columns 7 and 8 of \Cref{reg-IVS-het_char} report the results for children of immigrants and non-immigrants, respectively. We find that the estimated relationships are strikingly similar across the two subsamples: the U-shaped pattern is preserved, and both the magnitudes and confidence intervals of the coefficients almost completely overlap. This evidence suggests that the transmission of ancestral climatic experiences operates primarily through language-based and cultural channels, rather than through country-specific institutions or persistent exposure to ancestral locations.

\section{Conclusion}\label{sec:conclusion}

    The starting point of this paper is a simple intuition: attention to climate-related issues is influenced by ancestors' climatic experiences through cultural transmission. 
    Our empirical results show a significant and nonmonotonic impact and suggest that cultural narratives play a driving role, while our theory provides a toolbox to characterize the biases induced by ancestors' experienced climate on descendants' perceptions.
    Beliefs about environmental issues are of critical importance in the age of climate change. 
    By highlighting the role of culture and socialization, our work complements prior work in ecological anthropology and contributes to the broader study of cultural transmission and its impact on climate preferences, showing the insights generalize to a global scale.
    We hope that this project provides clear evidence and a simple analytical framework, laying groundwork for further research on the cultural and ancestral roots of environmental concerns.

\printbibliography

\vspace{1em}
\hrule
\vspace{1em}

\appendix
\renewcommand{\thesection}{\Alph{section}}
\renewcommand{\thesubsection}{\Alph{section}.\arabic{subsection}}
\renewcommand{\thetable}{\Alph{section}.\arabic{table}}
\renewcommand{\thefigure}{\Alph{section}.\arabic{figure}}
\setcounter{table}{0}
\setcounter{figure}{0}

\titleformat{\section}{\centering\scshape\large}{Appendix \thesection}{0.7em}{}

\setcounter{equation}{0}


\section{Additional Tables}\label{sec:additional-tables}


Throughout, Avg. anomalies denotes average ancestral climate anomalies, and Avg. anomalies sq. its square. Unless otherwise indicated, controls follow the baseline specifications in  \Cref{sec:empirical-results}. Table-specific variable definitions and clustering conventions are reported in the notes.


\begin{table}[!htbp]\centering
    \caption{
    Ancestral climate anomalies and environmental attitudes --- IVS
    \label{table:ivs_reg}}
    \footnotesize
    \begin{threeparttable}
    \begin{tabular}{lcccc}
    \hline
    \multicolumn{5}{c}{Dependent variable: Pro-environment index (IVS)}\\
                        &         (1)   &         (2)   &         (3)   &         (4)   \\
    \hline
    \hline
    Avg. Anomalies    &      -2.975** &      -2.980** &      -3.119*  &      -3.087** \\
                        &     (1.317)   &     (1.408)   &     (1.756)   &     (1.374)   \\
    Avg. Anomalies sq.&      35.442** &      35.555** &      37.140*  &      34.611** \\
                        &    (14.899)   &    (15.734)   &    (19.527)   &    (14.808)   \\
    Demographic Controls & Y & Y & Y & Y \\
    Historical ethnic group characteristics & Y & Y & Y & Y \\
    Historical topographic characteristics & Y & Y & Y & Y \\
    Country-year Fixed effects & Y & Y & Y & Y \\
    WVS Sample only & N & Y & N & N \\
    Exc. Env. Confidence from PCA & N & N & Y & N \\
    Exc. Env. vs Growth from PCA & N & N & N & Y \\
    \hline
    Mean of dep var     &       0.443   &       0.450   &       0.334   &       0.365   \\
    St. Dev. of dep var &       0.230   &       0.234   &       0.310   &       0.237   \\
    Min value Avg. Anomalies&       0.015   &       0.015   &       0.015   &       0.015   \\
    Max value Avg. Anomalies&       0.093   &       0.093   &       0.093   &       0.093   \\
    R-sq                &       0.124   &       0.121   &       0.118   &       0.146   \\
    Adj. R-sq           &       0.123   &       0.120   &       0.117   &       0.145   \\
    N                   &      219900   &      180293   &      238141   &      295507   \\
    \hline
    \multicolumn{5}{c}{ {*}{*}{*} p$<$0.01, {*}{*}p$<$0.05, {*} p$<$0.1, {+} p$<$0.15}\\
    \hline
    \hline
    \end{tabular}
    \end{threeparttable}
    \\[.2em]
    {\setstretch{.9} \footnotesize \justifying
        \textbf{Notes on \Cref{table:ivs_reg}:}
        Unit of observation is an individual. The dependent variable is constructed by rescaling the first PCA from answers to pro-environment actions and support for environmental causes (in particular: voluntary membership in environmental groups, confidence in and approval of environmental movements, willingness to choose environmental protection over growth), ranging between 0 and 1 and increasing with reported level of environmental engagement and action. WVS Sample only excludes respondents from EVS. Columns 3 and 4 recalculate the PCA by removing individual responses to either confidence in and approval of environmental movement or choice between environment vs growth. 
         Standard errors are clustered at the ethnicity level.
        \par
    }
\end{table}


\begin{table}[!htbp]\centering
    \caption{
        Ancestral climate anomalies and environmental attitudes --- ESS
        \label{table:ess-reg}}
    \footnotesize
    \begin{threeparttable}
    \begin{tabular}{lcccc}
    \hline
    \multicolumn{5}{c}{Dependent variable: Pro-environment index (ESS)}\\
                        &         (1)   &         (2)   &         (3)   &         (4)   \\
    \hline
    \hline
    Avg. Anomalies    &      -2.391***&      -4.369***&      -4.700***&      -2.018***\\
                        &     (0.841)   &     (1.030)   &     (0.963)   &     (0.773)   \\
    Avg. Anomalies sq.&      17.951***&      27.973***&      29.219***&      14.816***\\
                        &     (6.511)   &     (6.252)   &     (4.952)   &     (4.496)   \\
    Demographic Controls & Y & Y & Y & Y \\
    Historical ethnic group characteristics & Y & Y & Y & Y \\
    Historical topographic characteristics & Y & Y & Y & Y \\
    Country Fixed effects & Y & Y & Y & Y \\
    PCA inc. renewable resources & N & Y & Y & N \\
    PCA inc. climate change impact & N & N & Y & N \\
    PCA w. climate change opinions only & N & N & N & Y \\
    \hline
    Mean of dep var     &       0.580   &       0.552   &       0.557   &       0.589   \\
    St. Dev. of dep var &       0.136   &       0.148   &       0.143   &       0.172   \\
    Min value Avg. Anomalies&       0.022   &       0.022   &       0.022   &       0.019   \\
    Max value Avg. Anomalies&       0.119   &       0.119   &       0.119   &       0.119   \\
    R-sq                &       0.162   &       0.158   &       0.161   &       0.124   \\
    Adj. R-sq           &       0.160   &       0.155   &       0.156   &       0.123   \\
    N                   &       23787   &       11696   &        9241   &       33055   \\
    \hline
    \multicolumn{5}{c}{ {*}{*}{*} p$<$0.01, {*}{*}p$<$0.05, {*} p$<$0.1, {+} p$<$0.15}\\
    \hline
    \hline
    \end{tabular}
    \end{threeparttable}
    \\[.2em]
    {\setstretch{.9} \footnotesize \justifying
        \textbf{Notes on \Cref{table:ess-reg}:}
        Unit of observation is an individual. Dependent variable is the first component of the PCA carried over pro-environment actions and support for environmental causes (detailed in \Cref{subsec:ESS-details}). 
        \par
        }
\end{table}

\begin{table}[!h]\centering
    \caption{
        Ancestral climate anomalies and environmental themes: Wikipedia
        \label{table:wiki-reg}}
    \footnotesize
    \begin{threeparttable}
    \begin{tabular}{lccc}
    \hline
    \multicolumn{4}{c}{Dependent variable: Environment score for description on Wikipedia}\\
                        &         (1)   &         (2)   &         (3)   \\
    \hline
    \hline
    Avg. Anomalies    &      -4.088** &      -3.623*  &      -3.295*  \\
                        &     (2.013)   &     (1.955)   &     (1.913)   \\
    Avg. Anomalies sq.&      42.798** &      40.107** &      38.738** \\
                        &    (19.281)   &    (18.744)   &    (18.501)   \\
    Wikipedia Entry characteristics & Y & Y & Y \\
    Historical ethnic group characteristics & Y & Y & Y \\
    Historical topographic characteristics & Y & Y & Y \\
    Country Fixed effects & Y & Y & Y \\
    Language pages included & N & N & Y \\
    Geographical / Political Entity pages included & N & Y & Y \\
    \hline
    Mean of dep var     &       0.508   &       0.502   &       0.494   \\
    St. Dev. of dep var &       0.212   &       0.214   &       0.218   \\
    Min value Avg. Anomalies&       0.010   &       0.010   &       0.010   \\
    Max value Avg. Anomalies&       0.097   &       0.097   &       0.097   \\
    R-sq                &       0.325   &       0.336   &       0.336   \\
    Adj. R-sq           &       0.161   &       0.180   &       0.188   \\
    N                   &        1031   &        1081   &        1137   \\
    \hline
    \multicolumn{4}{c}{ {*}{*}{*} p$<$0.01, {*}{*}p$<$0.05, {*} p$<$0.1, {+} p$<$0.15}\\
    \hline
    \hline
    \end{tabular}
    \end{threeparttable}
    \\[.2em]
    {\setstretch{.9} \footnotesize \justifying
        \textbf{Notes on \Cref{table:wiki-reg}:}
        Unit of observation is an ethnic group. 
        Wikipedia Entry characteristics include page length (number of words in the Wikipedia biography for the ethnic group) and number of editions (number of Wikipedia language editions the ethnic group has a biography in). Language pages refer to biographies that primarily capture information on the language spoken by the ethnic group instead of the group itself. Geographical / Political Entity pages refer to biographies that primarily capture information on the geographical or political entity that the ethnic group resided in instead of the group itself. 
        Column 1 presents estimates from specifications that include only pages directly describing the ethnic group, excluding pages related to the geographic or political entity in which the group is located or the language it speaks. Column 2 expands the sample to include ethnic groups for which only pages related to geographic or political entities are available, while Column 3 further includes groups for which only language-based pages are available.
        \par
    }
\end{table}


\begin{table}[!htbp]\centering
    \footnotesize
    \begin{threeparttable}
    \caption{
        Ancestral climate anomalies and environmental themes: Joshua Project
        \label{table:reg-joshua}}
    \begin{tabular}{lccccc}
    \hline
    \multicolumn{5}{c}{Dependent variable: Environment score for description in Joshua Project}\\
                        &         (1)   &         (2)   &         (3)   &         (4)   \\
    \hline
    \hline
    Avg. Anomalies    &      -3.734***&      -3.365***&      -2.871***&      -1.390** \\
                        &     (0.455)   &     (0.452)   &     (0.469)   &     (0.569)   \\
    Avg. Anomalies sq.&      33.568***&      30.165***&      26.856***&      13.163** \\
                        &     (4.525)   &     (4.465)   &     (4.565)   &     (5.539)   \\
    Group characteristics & N & Y & Y & Y \\
    Topographic characteristics & N & N & Y & Y \\
    Country Fixed effects & N & N & N & Y \\
    \hline
    Mean of dep var     &       0.236   &       0.238   &       0.238   &       0.238   \\
    St. Dev. of dep var &       0.154   &       0.154   &       0.154   &       0.154   \\
    Min value Avg. Anomalies&       0.010   &       0.010   &       0.010   &       0.010   \\
    Max value Avg. Anomalies&       0.097   &       0.097   &       0.097   &       0.097   \\
    R-sq                &       0.008   &       0.065   &       0.080   &       0.149   \\
    Adj. R-sq           &       0.007   &       0.064   &       0.079   &       0.131   \\
    N                   &       12898   &       11873   &       11873   &       11866   \\
    \hline
    \multicolumn{5}{c}{ {*}{*}{*} p$<$0.01, {*}{*}p$<$0.05, {*} p$<$0.1, {+} p$<$0.15}\\
    \hline
    \hline
    \end{tabular}
    \end{threeparttable}
    \\[.2em]
    {\setstretch{.9} \footnotesize \justifying
        \textbf{Notes on \Cref{table:reg-joshua}:}
        The unit of observation is an ethnic group. Average anomalies ranges between 0.01 and 0.097 within the sample. Group characteristics include dummies for group's religion, whether the group is categorized as least reached, is at the frontier and is an indigenous group. Least Reached and Frontier categories capture the difficulty highlighted by the Joshua Project in terms of reaching the group in consideration and the success they had in converting individuals to Christianity.\par
    }
\end{table}



\begin{table}[!h]\centering
    \caption{Heterogeneity by ancestral economic development \label{reg-IVS-het_eth}}
    \resizebox{\linewidth}{!}{%
    \begin{threeparttable}
    \begin{tabular}{lcc|cc|cc|cc}
    \hline
    \multicolumn{9}{c}{Dependent variable: Self reported measure of attention to environment}\\
    & \multicolumn{2}{c}{\underline{Hunter-Gatherers}} & \multicolumn{2}{c}{\underline{Economic Complexity}} & \multicolumn{2}{c}{\underline{Local Jurisdiction}} & \multicolumn{2}{c}{\underline{Global Jurisdiction}}\\
    & Yes & No & Low & High & Low & High & Low & High\\
                        &         (1)   &         (2)   &         (3)   &         (4)   &         (5)   &         (6)   &         (7)   &         (8)   \\
    \hline
    \hline
    Avg. Anomalies    &     -29.650***&      -9.741***&     -10.085***&      -7.634***&     -20.416***&      -4.190*  &     -25.868*  &      -8.989***\\
                        &     (0.568)   &     (2.063)   &     (3.750)   &     (2.772)   &     (4.160)   &     (2.204)   &    (13.129)   &     (2.307)   \\
    Avg. Anomalies sq.&     421.557***&     110.145***&     109.993** &      84.733***&     220.699***&      44.355*  &     387.463** &     103.220***\\
                        &     (7.796)   &    (22.096)   &    (41.799)   &    (29.555)   &    (39.900)   &    (23.527)   &   (180.468)   &    (25.330)   \\
    Demographic Controls & Y & Y & Y & Y & Y & Y & Y & Y \\
    Historical ethnic group characteristics & Y & Y & Y & Y & Y & Y & Y & Y \\
    Historical topographic characteristics & Y & Y & Y & Y & Y & Y & Y & Y \\
    Country-year Fixed effects & Y & Y & Y & Y & Y & Y & Y & Y \\
    \hline
    Mean of dep var     &       0.683   &       0.704   &       0.703   &       0.704   &       0.710   &       0.695   &       0.721   &       0.702   \\
    St. Dev. of dep var &       0.256   &       0.250   &       0.252   &       0.249   &       0.243   &       0.258   &       0.246   &       0.250   \\
    Min value Avg. Anomalies&       0.015   &       0.015   &       0.015   &       0.015   &       0.018   &       0.015   &       0.015   &       0.015   \\
    Max value Avg. Anomalies&       0.084   &       0.093   &       0.080   &       0.093   &       0.084   &       0.093   &       0.074   &       0.093   \\
    R-sq                &       0.061   &       0.115   &       0.075   &       0.127   &       0.118   &       0.114   &       0.097   &       0.117   \\
    Adj. R-sq           &       0.054   &       0.115   &       0.072   &       0.126   &       0.117   &       0.112   &       0.092   &       0.116   \\
    N                   &        3925   &      134140   &       25801   &      112262   &       76519   &       61543   &       12076   &      125981   \\
    \hline
    \multicolumn{9}{c}{ {*}{*}{*} p$<$0.01, {*}{*}p$<$0.05, {*} p$<$0.1, {+} p$<$0.15}\\
    \hline
    \hline
    \end{tabular}
    \end{threeparttable}
    }
    \\[.2em]
    {\setstretch{.9} \footnotesize 
        \textbf{Notes on \Cref{reg-IVS-het_eth}:}
         Specification is as in \Cref{table:main-regression}.
    }
\end{table}


\FloatBarrier

\begin{table}[!htb]\centering
    \caption{
        Heterogeneity by contemporary exposure and migration
        \label{reg-IVS-het_char}}
    \resizebox{\linewidth}{!}{%
    \begin{threeparttable}
    \begin{tabular}{lcc|cc|cc}
    \hline
    \multicolumn{7}{c}{Dependent variable: Self reported measure of attention to environment}\\
    & \multicolumn{2}{c}{\underline{Recent Disaster Levels}} & \multicolumn{2}{c}{\underline{Similarity in Environments}} & \multicolumn{2}{c}{\underline{Children of Immigrants}}\\
    & Low & High & Low & High & Yes & No\\
                        &         (1)   &         (2)   &         (3)   &         (4)   &         (5)   &         (6)   \\
    \hline
    \hline
    Avg. Anomalies    &      -4.317** &     -12.341***&     -10.368***&     -14.070***&      -9.291***&      -8.662***\\
                        &     (2.163)   &     (2.566)   &     (2.360)   &     (3.119)   &     (3.492)   &     (2.077)   \\
    Avg. Anomalies sq.&      50.572** &     139.489***&     123.843***&     146.427***&      99.097***&      98.057***\\
                        &    (24.278)   &    (27.289)   &    (25.435)   &    (31.526)   &    (36.977)   &    (22.082)   \\
    Demographic Controls & Y & Y & Y & Y & Y & Y \\
    Historical ethnic group characteristics & Y & Y & Y & Y & Y & Y \\
    Historical topographic characteristics & Y & Y & Y & Y & Y & Y \\
    Country-year Fixed effects & Y & Y & Y & Y & Y & Y \\
    \hline
    Mean of dep var     &       0.692   &       0.712   &       0.698   &       0.716   &       0.708   &       0.707   \\
    St. Dev. of dep var &       0.248   &       0.251   &       0.250   &       0.248   &       0.252   &       0.251   \\
    Min value Avg. Anomalies&       0.021   &       0.015   &       0.015   &       0.018   &       0.015   &       0.015   \\
    Max value Avg. Anomalies&       0.093   &       0.093   &       0.093   &       0.093   &       0.085   &       0.093   \\
    R-sq                &       0.120   &       0.110   &       0.120   &       0.101   &       0.117   &       0.113   \\
    Adj. R-sq           &       0.119   &       0.109   &       0.119   &       0.100   &       0.107   &       0.112   \\
    N                   &       59581   &       78486   &       97432   &       40632   &       10966   &      111600   \\
    \hline
    \multicolumn{7}{c}{ {*}{*}{*} p$<$0.01, {*}{*}p$<$0.05, {*} p$<$0.1, {+} p$<$0.15}\\
    \hline
    \hline
    \end{tabular}
    \end{threeparttable}
    }
    \\
    {\setstretch{.9} \footnotesize \justifying
        \textbf{Notes on \Cref{reg-IVS-het_char}:}
         The unit of observation is an individual. 
         Recent Disaster levels are captured by the impact of natural disasters on GDP and population affected. Similarity of Environments captures how similar the stock of contemporaneous climate anomalies is to the average ancestral climatic anomalies faced by the ancestors. Children of immigrants are respondents for whom either parent was an immigrant to the country where the individual is currently located. Standard errors are clustered at the ethnicity level.\par
    }
\end{table}

\section[Model]{Model: additional details and omitted proofs}\label{sec:fullmodel}
    
    
        Throughout, consider the ambient space of continuous functions $\mathcal{F} = \mathcal{C}^0(Z)$ as the class of admissible cultural types, equipped with the sup norm $\Vert \cdot \Vert_\infty$.
        All integrals over functions are understood in the Bochner sense.
        $\mathcal{C}^0(Z)$ is considered with its natural duality pairing with $\mathcal{M}(Z)$ the space of finite signed measures and we denote by $\langle \cdot, \cdot \rangle$ the duality bracket $\langle q, f \rangle := \int f dq$. 
        The mutation noise term $\varepsilon$ introduced in \Cref{assumption:mutation} is a canonical Gaussian random variable on $\mathcal{C}^0(Z)$ with covariance operator $C:\mathcal{M}(Z) \rightarrow \mathcal{C}^0(Z)$ defined by:
        \(
            C q(z) := \int_Z K(z,x) dq(x) \text{ for any $q \in \mathcal{M}(Z)$, where } K(z,x):= e^{-|z-x|}.
        \)
        This is a positive bounded operator on $\mathcal{C}^0(Z)$; this generates paths that are Hölder continuous hence in $\mathcal{C}^0$ a.s.
        One can verify that this is well-defined directly from standard references, or observe that this corresponds to the truncation on $Z$ of an Ornstein-Uhlenbeck process on $\reals$.

        Since mutations can produce types with non-positive values, for which the attention problem is not defined, we use a technical device which plays no role in the results.
        Fix an arbitrarily small threshold $\epsilon > 0$, below the lower bound on realized stakes, and for any $\nu > 0$ extend the reduced loss $\mathcal{L}(\cdot|\nu)$ below $\epsilon$ to a twice continuously differentiable function on $\reals$ which is convex and decreasing on $(-\infty,\epsilon)$ with bounded slope; for instance, take its derivative to be $\mathcal{L}'(\epsilon|\nu) + \mathcal{L}''(\epsilon|\nu) \tanh(\mu - \epsilon)$ for $\mu < \epsilon$.
        Fitness is then given by \Cref{assumption:exponential-selection} for all types, using the extended loss, which we still denote $\mathcal{L}$.
        In particular, this makes the dynamics well defined for the default $f_0 \equiv 0$.
        Since realized stakes are bounded away from zero and infinity, $\hat{\overline{v}}_g$ takes values in a compact subset of $\reals_+^*$.
        Using $\mathcal{L}'(\mu|\nu) = \kappa(\mu-\nu)/\mu^2$ and $\mathcal{L}''(\mu|\nu) = \kappa(2\nu - \mu)/\mu^3$ on $[\epsilon,\infty)$, it follows that the loss $\mathcal{L}(\cdot|\hat{\overline{v}}_g)$ has first and second derivatives bounded uniformly in $g$; its second derivative is bounded below by $-\kappa/(27 \hat{\overline{v}}_g^2)$, the minimum of $\mathcal{L}''(\cdot|\hat{\overline{v}}_g)$ (attained at $3\hat{\overline{v}}_g$); the extension is convex.
        These are the only properties of the loss used in the proofs below.
        Throughout, constants implicit in $O(\cdot)$ depend only on $\tau$, $\kappa$, $\lambda$, $\langle q, Cq \rangle$, $\epsilon$ and the bounds on realized stakes; in particular, they do not depend on $g$ or $\rho$.
        


        \paragraph{Exact recursion on population averages}
        To characterize average type dynamics, we first derive an exact recursion for $f_g$.
        By using Fubini-Tonelli in the replicator-mutator dynamics:
        \begin{align*}
            f_{g+1}
            &=\int_{\mathcal F} \int_{\mathcal F} \Esp_{f'\sim M(\cdot\mid \frac{f_1+f_2}{2})}[f'] \, \tilde{P}_g(df_1) \, \tilde{P}_g(df_2).
            \end{align*}
        By assumption $\Esp_{f'\sim M(\cdot\mid h)}[f']=(1-\rho\lambda)h + \rho\lambda f_0$ so we can replace and simplify, using that $\Esp_{\tilde P_g}[\frac{f_1+f_2}{2}] = \Esp_{\tilde P_g}[f]$, to get that:
        $f_{g+1} = (1-\rho\lambda) \Esp_{f \sim \tilde{P}_g}[f] + \rho\lambda f_0$,
        where $\tilde{P}_g$ is the selection-tilted (or post-selection, pre-mutation) law, i.e. $d\tilde{P}_g(f):=\frac{\Phi_g(f)}{\Esp_{P_g}[\Phi_g]}dP_g(f)$.
        The selection shift $\Esp_{f \sim \tilde{P}_g}[f]-f_g$ can be rewritten into the ``Price decomposition'': 
        \begin{align*}
        \Esp_{\tilde{P}_g}[f]-f_g
        =\frac{\Esp_{P_g}[(f-f_g)\Phi_g(f)]}{\Esp_{P_g}[\Phi_g(f)]}.
        \end{align*}
        This is a canonical result in the evolution literature  \citep[see e.g.][]{page2002unifying}.

        \paragraph{Reduction to one dimension}
        Fitness depends on a type $f$ only through $\Esp_q[f]$; we show that the dynamics reduce exactly to a one-dimensional problem for this quantity.
        Denote by $\Sigma_g$ the covariance operator of $P_g$, defined on $\mathcal{M}(Z)$ as $\Sigma_g : q \mapsto \Esp_{f \sim P_g}[(f-f_g) \langle q, f- f_g \rangle]$.
        Let $m_g := \Esp_q[f_g]$ and $V_g := \langle q, \Sigma_g q \rangle$ be the mean and variance of $\Esp_q[f]$ under $P_g$, and $\tilde{m}_g$, $\tilde{V}_g$ the same quantities under $\tilde{P}_g$.
        For any type $f$, call $f - f_0 - \bigl(\Esp_q[f]-\Esp_q[f_0]\bigr) \frac{Cq}{\langle q, Cq \rangle}$ its \emph{residual}.

        \begin{lemma}[Reduction to one dimension]\label{lem:reduction}
        For every generation $g$:
        \begin{enumerate}[label=(\roman*)]
            \item under $P_g$ and under $\tilde{P}_g$, the residual of $f$ has mean zero and is independent of $\Esp_q[f]$;
            \item $f_g = f_0 + (m_g - \Esp_q[f_0]) \frac{Cq}{\langle q, Cq \rangle}$, $\Esp_{\tilde{P}_g}[f] = f_0 + (\tilde{m}_g - \Esp_q[f_0]) \frac{Cq}{\langle q, Cq \rangle}$, and $\Sigma_g q = V_g \frac{Cq}{\langle q, Cq \rangle}$;
            \item the law of $\Esp_q[f]$ under $P_{g+1}$ is that of $(1-\rho\lambda)\frac{\Esp_q[f_1]+\Esp_q[f_2]}{2} + \rho\lambda \Esp_q[f_0] + \sqrt{\rho} \, \langle q, \varepsilon \rangle$, where $f_1, f_2$ are independent draws from $\tilde{P}_g$ and $\langle q, \varepsilon \rangle \sim N(0, \langle q, Cq \rangle)$ is independent of them; the law of $\Esp_q[f]$ under $\tilde{P}_g$ is its law under $P_g$ reweighted by $\varphi\bigl(\mathcal{L}(\cdot|\hat{\overline{v}}_g)\bigr)$.
        \end{enumerate}
        \end{lemma}
        In words: selection only sees $\Esp_q[f]$, and the rest of the type is determined by mutation geometry, so that all movement of the population mean is along the single direction $Cq$.
        The proof of this technical lemma (and other technical results below) is in \Cref{online-app:technical-lemmas}.

        \paragraph{Laws with Gaussian-like spread}
        It remains to control the law of $\Esp_q[f]$, which is not Gaussian since selection distorts it in every generation.
        We use the following notion.
        A probability law on $\reals$ with a positive, twice differentiable density has \emph{spread at most $s$}, for $s > 0$, if minus its log-density has second derivative at least $1/s$ everywhere: its log-density is at least as curved as that of a Gaussian with variance $s$ (which has spread exactly $s$). 
        Equivalently, its density is the product of a log-concave function and of the Gaussian density with variance $s$: this is the class of strongly log-concave laws, see \citet{saumard2014log}.
        Such laws behave like Gaussians in all the bounds we need, and the class is stable under the three operations of evolution --- i.e. selection, combination and mutation and taking variance bounds --- which is all we use below.
        Background details on these stability properties, which follow from standard facts, are in \Cref{online-app:technical-lemmas}.

        \paragraph{One round of selection}
        Selection moves the mean in proportion to the variance and to the local slope of fitness, and only changes the variance at higher order.
        For a Gaussian law and a quadratic loss everything is explicit: $\tilde{m}_g - m_g = - \tau \mathcal{L}'(m_g|\hat{\overline{v}}_g) V_g / \bigl(1 + \tau \mathcal{L}''(m_g|\hat{\overline{v}}_g) V_g\bigr)$.
        The following technical lemma (whose proof is in \Cref{online-app:technical-lemmas}) is essentially the Price equation with an explicit error.
        \begin{lemma}\label{lem:one-step}
        If the law of $\Esp_q[f]$ under $P_g$ has spread at most $s$, with $s$ small enough, then:
        \begin{align*}
            \tilde{m}_g - m_g = O(s),
            \qquad
            \tilde{m}_g - m_g = - \tau \mathcal{L}'\bigl( m_g \bigm| \hat{\overline{v}}_g \bigr) V_g + O(s^{3/2}),
            \qquad
            \tilde{V}_g - V_g = O(s^{3/2}).
        \end{align*}
        \end{lemma}
        
        \paragraph{Controlling population spread}
         In each generation, selection widens the spread slightly, combining two cultural models halves it, and mutation adds $\rho \langle q, Cq \rangle$. The following lemma (proved in \Cref{online-app:technical-lemmas}) shows that the variance therefore settles approximately at $2\rho \langle q, Cq \rangle$.
        \begin{lemma}[Population spread]\label{lem:spread}
        For $\rho$ small enough and every $g \geq 1$:
        \begin{enumerate}[label=(\roman*)]
            \item the law of $\Esp_q[f]$ under $P_g$ has spread at most $3 \rho \langle q, Cq \rangle$;
            \item $|m_g - \Esp_q[f_0]|$ is bounded uniformly in $g$ and $\rho$;
            \item $|V_g - 2\rho \langle q, Cq \rangle| \leq 2^{1-g} \rho \langle q, Cq \rangle + O(\rho^{3/2})$.
        \end{enumerate}
        \end{lemma}
        
       \paragraph{Proof of \Cref{prop:kernel-gradient-form}}
       Using the exact recursion and \Cref{lem:reduction}(ii), exactly:
       \begin{align*}
           f_{g+1} = f_g - \rho\lambda(f_g - f_0) + (1-\rho\lambda)\bigl(\Esp_{\tilde{P}_g}[f] - f_g\bigr) = f_g - \rho\lambda(f_g - f_0) + (1-\rho\lambda)(\tilde{m}_g - m_g)\frac{Cq}{\langle q, Cq \rangle}.
       \end{align*}
       By \Cref{lem:one-step} and \Cref{lem:spread}(i), $\tilde{m}_g - m_g = -\tau \mathcal{L}'\bigl(m_g \bigm| \hat{\overline{v}}_g\bigr) V_g + O(\rho^{3/2})$ uniformly in $g$, where $V_g = O(\rho)$ and $V_g \frac{Cq}{\langle q, Cq \rangle} = \Sigma_g q$ by \Cref{lem:reduction}(ii).
       This gives the result of \Cref{prop:kernel-gradient-form}, with a remainder $O(\rho^{3/2})$ uniform in $g$.
        
       \paragraph{Proof of \Cref{thm:convergence}}
        By \Cref{lem:spread}(iii) and \Cref{lem:reduction}(ii), $\Sigma_g q = V_g \frac{Cq}{\langle q, Cq \rangle} = 2 \rho \, Cq + O(\rho^{3/2})$ for all $g$ beyond a large enough number of generations.
        In the large $N$,$T$ limit, the law of large numbers gives $\hat{\overline{v}}_g = \Esp_q[v^\star]$, so that the recursion of \Cref{prop:kernel-gradient-form} becomes:
        \begin{equation}\label{eq:approx-mean-dynamics}
           f_{g+1}
          = f_g - \rho \biggl( \lambda (f_g-f_0)  + 2 \tau \mathcal{L}'\bigl( \Esp_q[f_g] \bigm| \Esp_q[v^\star] \bigr) C q \biggr) + o(\rho).
        \end{equation}
        This is a gradient step on the \emph{theoretical} penalized fit objective $J$, in the geometry induced by $C$.
        By \Cref{lem:reduction}(ii), $f_g - f_0$ is proportional to $Cq$ in every generation, so \eqref{eq:approx-mean-dynamics} is equivalent to:
        \begin{equation}\label{eq:scalar-dynamics}
            m_{g+1} = m_g - \rho \Bigl( \lambda \bigl(m_g - \Esp_q[f_0]\bigr) + 2\tau \langle q, Cq \rangle \mathcal{L}'\bigl( m_g \bigm| \Esp_q[v^\star] \bigr) \Bigr) + o(\rho).
        \end{equation}
        The rest of the proof (i) characterizes the unique zero of the term in parentheses, (ii) shows that $m_g$ converges to it, and (iii) identifies the corresponding type as the minimizer of $J$.

        \emph{Rest point.}
        With the loss derived from the rational inattention problem, $\mathcal{L}'(a|\nu) = \frac{\kappa}{a}\bigl(1 - \frac{\nu}{a}\bigr)$ for $a \geq \epsilon$, so the term in parentheses in \eqref{eq:scalar-dynamics}, evaluated at $m_g = a$, equals $\lambda/a^2$ times the cubic:
        \[
            a^2 \bigl(a - \Esp_q[f_0]\bigr) + \gamma \bigl(a - \Esp_q[v^\star]\bigr), \quad \text{ where } \gamma := \frac{2\tau \kappa \langle q, Cq \rangle}{\lambda}.
        \]
        Since $0 \leq \Esp_q[f_0] \leq \Esp_q[v^\star]$, this cubic has a unique positive root $a^\star$, which lies in $[\Esp_q[f_0], \Esp_q[v^\star]]$: both terms are positive for $a > \Esp_q[v^\star]$ and negative for $0 < a < \Esp_q[f_0]$, while on $[\Esp_q[f_0], \Esp_q[v^\star]]$ the cubic is increasing (its derivative is $a(3a - 2\Esp_q[f_0]) + \gamma > 0$), non-positive at $\Esp_q[f_0]$ and non-negative at $\Esp_q[v^\star]$.\footnote{
            The root can be computed explicitly. In the case where $f_0 \equiv 0$, this is directly a depressed cubic so we can apply Cardano's formula to get:
            \(
            a^\star = \sqrt[3]{\frac{\gamma \Esp_q[v^\star]}{2} + \sqrt{\frac{\gamma^2 \Esp_q[v^\star]^2}{4} + \frac{\gamma^3}{27}}} + \sqrt[3]{\frac{\gamma \Esp_q[v^\star]}{2} - \sqrt{\frac{\gamma^2 \Esp_q[v^\star]^2}{4} + \frac{\gamma^3}{27}}}.
            \)
            In the general case, first reduce it to a depressed cubic, then apply Cardano's formula.
        }
        Hence the term in parentheses is negative below $a^\star$ and positive above it.
        This holds on $(0,\infty)$ for the original loss, and on all of $\reals$ for the extended loss provided $\epsilon < a^\star$ (recall that $\epsilon$ can be taken arbitrarily small), since the extended loss has slope at most $\mathcal{L}'(\epsilon|\Esp_q[v^\star])$ below $\epsilon$.

        \emph{Convergence.}
        Fix any neighbourhood of $a^\star$.
        Since $m_g$ remains in a bounded interval (\Cref{lem:spread}(ii)), the term in parentheses is bounded, and bounded away from zero outside that neighbourhood, with the sign of $m_g - a^\star$.
        Squaring \eqref{eq:scalar-dynamics}, $(m_{g+1} - a^\star)^2$ equals $(m_g - a^\star)^2$ minus $2\rho(m_g - a^\star)$ times the term in parentheses, up to $o(\rho)$.
        So for $\rho$ small enough, the squared distance to $a^\star$ decreases by at least a fixed multiple of $\rho$ in every generation in which $m_g$ lies outside the neighbourhood, while a single generation moves $m_g$ by $O(\rho)$ only.
        Hence $m_g$ eventually enters the neighbourhood and never moves further than $O(\rho)$ beyond.
        As the neighbourhood is arbitrary, $\lim_{\rho \rightarrow 0} \limsup_{g \rightarrow \infty} |m_g - a^\star| = 0$.
        By \Cref{lem:reduction}(ii), $f_g - f^\star = (m_g - a^\star) \frac{Cq}{\langle q, Cq \rangle}$ with $f^\star := f_0 + \frac{a^\star - \Esp_q[f_0]}{\langle q, Cq \rangle} Cq$, so the same holds for $\Vert f_g - f^\star \Vert_\infty$.

        \emph{Characterization as the minimizer of $J$.}
        If $\Vert f - f_0 \Vert_C < \infty$, write $f - f_0 = Ch$ with $h \in \mathcal{M}(Z)$ (all such $h$ give the same value $\langle h, Ch \rangle = \int (f-f_0) dh = \Vert f - f_0 \Vert_C^2$).
        Since $K$ is a positive definite kernel, the Cauchy-Schwarz inequality gives $(\Esp_q[f] - \Esp_q[f_0])^2 = \langle q, Ch \rangle^2 \leq \langle q, Cq \rangle \Vert f - f_0 \Vert_C^2$, with equality if and only if $f - f_0$ is proportional to $Cq$ (equality forces $h$ to differ from a multiple of $q$ by a measure in the kernel of $C$).
        Hence:
        \begin{align*}
            J(f) \geq 2\tau \, \mathcal{L}\bigl( \Esp_q[f] \bigm| \Esp_q[v^\star] \bigr) + \frac{\lambda}{2 \langle q, Cq \rangle} \bigl(\Esp_q[f] - \Esp_q[f_0]\bigr)^2,
        \end{align*}
        with equality if and only if $f = f_0 + (\Esp_q[f] - \Esp_q[f_0]) \frac{Cq}{\langle q, Cq \rangle}$.
        As a function of $\Esp_q[f] > 0$, the right-hand side has derivative equal to the term in parentheses in \eqref{eq:scalar-dynamics} divided by $\langle q, Cq \rangle$, hence is strictly decreasing on $(0, a^\star)$ and strictly increasing on $(a^\star, \infty)$.
        Hence $f^\star$ is the unique minimizer of $J$.

        \emph{Vanishing default weight.}
        Rearranging the cubic at $a^\star$ gives $\Esp_q[v^\star] - a^\star = \frac{\lambda \, a^{\star2} (a^\star - \Esp_q[f_0])}{2\tau\kappa\langle q, Cq \rangle} \leq \frac{\lambda \, \Esp_q[v^\star]^3}{2\tau\kappa\langle q, Cq \rangle}$, so $a^\star \rightarrow \Esp_q[v^\star]$ as $\lambda \downarrow 0$ and:
        \(
        f^\star \xrightarrow[\lambda \downarrow 0]{} f_0 + \frac{\Esp_q[v^\star] - \Esp_q[f_0]}{\langle q, Cq \rangle} Cq = \argmin_{ f | \langle q,f \rangle = \Esp_q[v^\star] } \Vert f-f_0 \Vert_{C}^2,
        \)
        where the last equality is the equality case of the Cauchy-Schwarz inequality above.
        This concludes the proof of \Cref{thm:convergence}.
    
    \paragraph{Proof of \Cref{prop:decomposition}}
    Recall that by definition for any distribution $p$, $Cp(z) = \int K(x,z) dp(x)$, so that in particular: 
        $\langle p, Cq \rangle = \Esp_p[Cq] = \int \int  K(x,z) dq(x) dp(z) = \Esp_{p \otimes q }[K]$.
    Directly applying the characterization of \Cref{thm:convergence} with $f_0 \equiv 0$ and using Fubini gives:
    \begin{align*}
        B(\tilde{q}|q) & = \Esp_{\tilde{q}} [\hat{v}(\cdot|q)]
        = \Esp_{\tilde{q}} \biggl[ \frac{\Esp_q[v^\star] }{\langle q, Cq \rangle} Cq \biggr]
        = \Esp_q[v^\star] \frac{\Esp_{\tilde{q}}[Cq] }{ \Esp_{q} [Cq] }
        = \Esp_q[v^\star] \frac{\Esp_{\tilde{q} \otimes q}[K] }{ \Esp_{q \otimes q} [K] }
    \end{align*}
    
    \paragraph{Proof of \Cref{thm:U-shape}} To prove the first part of the theorem (core U-shape), define $W(\theta):= \int_{\reals_+} w(z) dq_\theta(z)$ and $R(\theta) := \int_{\reals_+} r(z) dq_\theta(z)$.
    It is straightforward to observe that $W$ and $R$ inherit the properties of $w$ and $r$ almost directly. 
    For instance, a change of variable on $W$ gives:
    \begin{align*}
    W(\theta) := \int_{Z} w(z) \frac{1}{\theta} d\bar{q} \left( \frac{z}{\theta} \right) =   \int_{Z} w(\theta z) d\overline{q}(z)
    \end{align*}
    Hence, assuming differentiability only for simplicity of exposition $W'(\theta) = \int_{Z} z w'(\theta z) d\bar{q}(z) \leq 0$ and $W''(\theta)= \int_{Z} z^2 w''(\theta z) d\bar{q}(z) \geq 0$. 
    $W' \xrightarrow[\bar{z}]{} 0$ follows from dominated convergence. Similarly, we get $R \geq 0$, $R' \geq 0$, $R'' \geq 0$ and $R'(0)=0$.
    
    Tube bounds are obtained via:
    \(
      \biggl| \frac{B(q|\tilde{q})}{V(q)} - 1\biggr| = \bigl| S(q|\tilde{q})-1 \bigr| = \frac{|E_{q \otimes \tilde{q}}[K] - E_{q \otimes q}[K]|}{E_{q \otimes q}[K]} \leq \frac{W_1(q|\tilde{q})}{e^{-\overline{z}}},
    \)
    where the Wasserstein bound comes from Kantorovitch-Rubinstein duality with the Lipschitz constant $|\partial_y K| \leq 1$; the denominator bound is immediate since $K(x,z) \geq e^{-\overline{z}}$ as $x,z \in [0,\overline{z}]$. The standard Wasserstein identity $W_1(q_\theta,q_{\tilde{\theta}}) = \bar{\mu}_1 |\theta - \tilde{\theta} |$ gives the final form of the bound.
    
    For the last statement (minimizer displacement), let $\theta^*_B$ a minimizer of $B$. Then, using the tube bounds:
    \(
    B(\theta^*_B) \leq B(\theta^\dagger) \Longrightarrow (1-\delta) V(\theta^*_B) \leq (1+\delta) V(\theta^\dagger) \Longrightarrow V(\theta^*_B)-V(\theta^\dagger) \leq \frac{2 \delta}{1-\delta} V(\theta^\dagger)
    \).
    Quadratic growth implies $V(\theta^*_B)-V(\theta^\dagger) \geq \frac{c}{2} \bar{\mu}_2 |\theta^*_B - \theta^\dagger|^2$, which yields the result.

\FloatBarrier

\newpage

\begin{center}
    \textbf{\LARGE \scshape Online Appendix}
\end{center}

\section{Data: more details on construction and descriptions}

    \subsection{Integrated Value Survey}\label{subsec:IVS-details}

    The Integrated Values Surveys (IVS) is a harmonized global dataset that integrates and standardizes responses from the World Values Survey (WVS) and the European Values Study (EVS), and is designed to facilitate cross-country and within-country analyses over time of human values, beliefs, and preferences using consistent measures. WVS covers nationally representative samples from 104 countries between 1981 and 2022. Similarly, EVS covers nationally representative samples from 49 European nations between 1981 and 2022. 
    Individuals in each survey are asked questions along many thematic categories (perceptions of life, environment, work, family, politics, society, religion and national identity). 
    Each survey captures respondent-level information on demographic characteristics such as age, education level, occupational category, income level (ten standardized categories) and gender, which we use as controls. Respondents also report the primary language that they speak at home, which allows us to match them to the corresponding ethnic ancestors from the Ancestral Characteristics database \citep{nunn_giu_database}.
    
    For our lead question of interest, \emph{"How important is it to you to take care of the environment?"}, we have a total of 157,142 respondents from 78 countries in waves 5 and 6 (2005-2014). In marginal cases (7\% of the total sample), where we do not have information on the language spoken at home, we exploit the language of the interview as a proxy for the language spoken at home. In cases where an individual reports multiple languages spoken at home or the same language is spoken by multiple ethnic groups, we link the individual to all the potential historical ethnic groups and assign equal weight to them. Our sample of interest for main variable of interest from World Values Surveys contains 202 different language / ethnic groups. 
    Less than 5\% of the individuals in the main sample of consideration have multiple ethnic groups associated to them.

    \begin{table}[!h]\centering
		\resizebox{\linewidth}{!}{%
			\begin{threeparttable}
				\caption{Variable source (Wave and Study) - stated environmental preferences}
                \label{env_descstat}
				\begin{tabular}{l|ccccccc|ccccc}\hline\hline
					\multicolumn{1}{c}{\textbf{Variable}} & \multicolumn{7}{c}{\textbf{WVS}} & \multicolumn{5}{c}{\textbf{EVS}} \\
					\textit{Wave} & (1) & (2) & (3) & (4) & (5) & (6) & (7) & (1) & (2) & (3) & (4) & (5) \\
					\textit{Starting year} & 1981 & 1989 & 1994 & 1999 & 2005 & 2010 & 2017 & 1981 & 1990 & 1999 & 2008 & 2017 \\
					\hline
					Membership in Env.\ Groups & \checkmark & \checkmark & \checkmark & \checkmark & \checkmark & \checkmark & \checkmark & \checkmark & \checkmark & \checkmark & \checkmark & \checkmark \\
					Voluntary Work for Env.\ Groups &  & \checkmark &  & \checkmark &  &  &  & \checkmark & \checkmark & \checkmark & \checkmark &  \\
					Participated in and Provided Money for Env.\ Actions &  &  & \checkmark &  &  & \checkmark &  &  &  &  &  &  \\
					\textbf{Env.\ Participation} & \checkmark & \checkmark & \checkmark & \checkmark & \checkmark & \checkmark & \checkmark & \checkmark & \checkmark & \checkmark & \checkmark & \checkmark \\
					\hline
					Approval:\ Ecology movement or nature protection &  & \checkmark &  &  &  &  &  &  & \checkmark &  &  &  \\
					Confidence:\ Environmental protection movement &  &  & \checkmark & \checkmark & \checkmark & \checkmark & \checkmark &  &  &  & \checkmark & \checkmark \\
					\textbf{Env.\ Support} &  & \checkmark & \checkmark & \checkmark & \checkmark & \checkmark & \checkmark &  & \checkmark &  & \checkmark & \checkmark \\
					\hline
					\textbf{Env.\ over Growth} &  &  & \checkmark & \checkmark & \checkmark & \checkmark & \checkmark &  &  & \checkmark &  & \checkmark \\
					\hline
					\hline\end{tabular}
				\begin{tablenotes}
					\small
					\item \textbf{Note: } Source for each stated environment preference measure described in section \ref{env_pref}. Environment is abbreviated as Env.
				\end{tablenotes}
		\end{threeparttable}}
	\end{table}

    For the PCA exercises conducted to construct alternate proxies capturing pro-environmental sentiment and actions, we rely on measures capturing (i) voluntary membership in environmentally related organizations (WVS Waves 1–7; EVS Waves 1–5), (ii) approval of and confidence in country-specific environmental and nature movements (WVS Waves 2–7; EVS Waves 2, 4, and 5), and (iii) willingness to prioritize environmental protection over economic growth (WVS Waves 3–7; EVS Waves 3 and 5) -- see  \Cref{env_descstat}. All variables are normalized so that higher numbers indicate greater subjective value, and the first PCA component is scaled to lie between 0 and 1. Depending on the specification,\footnote{Our main results use an index constructed as the first principal component from a PCA over individual responses to all the questions described above. To assess the sensitivity of our findings to the specific inputs used in the PCA, we construct alternative indices based on narrower subsets of variables. In particular, we retain voluntary environmental group membership (which is available across most survey waves) and, in turn, combine it with either (i) willingness to prioritize environmental conservation over economic growth or (ii) approval of and confidence in national environmental movements, but not both simultaneously.} information on the constructed proxy is available for approximately 180,000 to 300,000 respondents across 67 to 93 countries.

    \begin{figure}[h!]
        \caption{Self-reported mother tongue (language spoken at home)} \label{lang_at_home}
        \centering
        \includegraphics[scale=0.3]{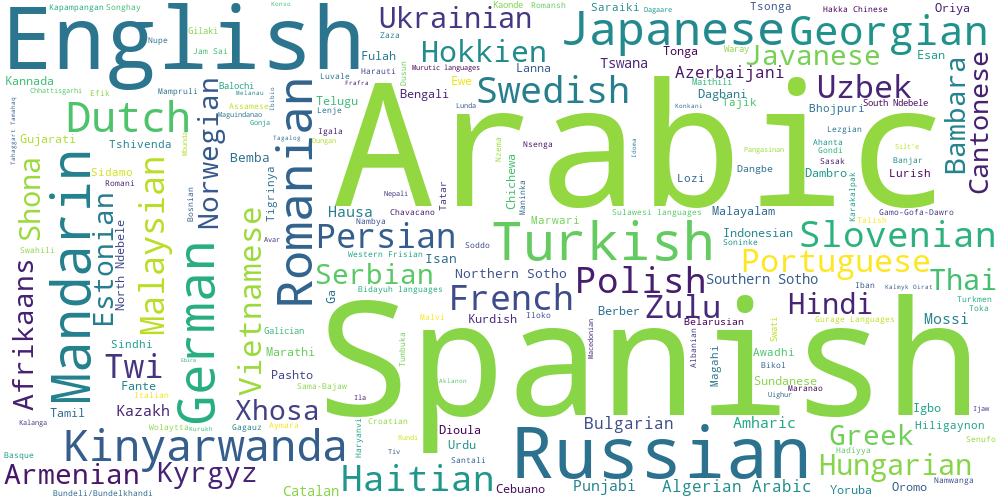}
        \fnote{\textbf{Note:} Word cloud of self-reported languages spoken at home which allow us to match individuals to the historical ethnic groups. Word size is weighted by the proportion of individuals in the sample of WVS that reported the language as their primary language and answered the environmental question of interest.}
    \end{figure}

    \subsection{European Social Survey}\label{subsec:ESS-details}

    The European Social Survey is a cross-national repeated social survey that measures public attitudes, beliefs, values, and behaviors across a broad set of countries, primarily in Europe. Each round includes rotating modules focused on specific thematic domains. The 8\textsuperscript{th} round contains a dedicated climate module that elicits respondents’ views and behaviors related to climate change, energy use, and environmental policy.

    Using responses from this module, we conduct four PCA exercises:
    \begin{enumerate}
        \item In the first PCA, we use responses to questions capturing: (i) confidence in one’s ability to reduce energy consumption, (ii) frequency of actions taken to reduce energy use, (iii) concern about national dependence on fossil fuels, (iv) prior consideration of climate change, and (v) support for climate-related policies, including increasing taxes on fossil fuels, subsidizing renewable energy, and banning the sale of the least energy-efficient household appliances.\footnote{The exact survey questions include: (1) confidence in using less energy; (2) frequency of actions to reduce energy use; (3) worry about national dependence on fossil fuels; (4) extent of prior thought about climate change; (5) belief about whether climate change is driven by human activity (coded higher when attributed mainly to human causes); (6) worry about climate change; (7) support for increasing fossil fuel taxes; (8) support for subsidizing renewable energy; and (9) support for banning the least energy-efficient household appliances.}
        \item In the second PCA exercise, we augment this set with questions, asked to a smaller subsample, regarding respondents’ views on the appropriate share of energy that should be generated from renewable sources such as hydroelectric, nuclear, solar, wind, and bioenergy. 
        \item In the third PCA, we further expand the variable set by including questions, asked to an even smaller subsample, on the likelihood of purchasing highly energy-efficient household appliances and beliefs about whether climate change will have negative global consequences. 
        \item In the fourth PCA, we solely rely on beliefs related to climate-change. We use responses to questions on prior consideration of climate change, belief in its existence, concern about its impacts, and attribution of climate change primarily to human activity.
    \end{enumerate}

    To conduct the PCA, as with IVS, we first normalize all variables so that higher numbers indicate greater subjective value, and the first PCA components which we use as our indices is scaled to lie between 0 and 1. ESS also provides rich demographic information on respondents. We use information on the language spoken at home to link individuals to their ethnic ancestry through the language–ethnicity mappings available in the ancestral characteristics database. Additional control variables, which we use on our specifications, include age, gender, education level, income deciles (standardized within country), and occupation categories based on ISCO-08 classifications.

    \subsection{Folklore catalog}\label{subsec:folklore-details}

        Berezkin's original catalog consists of motifs related to the mythology, folklore and oral traditions for 958 groups worldwide. 
		After parsing over 6,239 books and journal articles (documenting oral traditions) from 4,041 authors edited by 4,932 publishing houses in 32 different languages, Yuri Berezkin categorized 2,564 motifs and linked them to the ethnolinguistic groups. 
		As per the original catalog, a motif reflected a combination of images, episodes, or structural elements found in two or more texts, including sacred and profane ones. 
		The median group in Berezkin's catalog has 62 motifs.
		Each motif is accompanied with a title and a short description of an image or an episode in the group's oral tradition.

        We used two different methods to classify environmental themes in folklore.
        \begin{enumerate}
            \item \textbf{ConceptNet Classification:} ConceptNet \citep{conceptnet} is a large, multilingual semantic knowledge graph that encodes general human common-sense knowledge in the form of structured relationships between concepts. It represents information as nodes (words or short phrases) connected by labeled edges that describe relations such as IsA, PartOf, UsedFor, CapableOf, Causes, and RelatedTo. The knowledge in ConceptNet is aggregated from a variety of sources, including crowdsourced contributions (notably the Open Mind Common Sense project), lexical resources, and structured data repositories. Each connection is weighted to reflect confidence and frequency. Designed to support common-sense reasoning in natural language processing systems, ConceptNet provides a structured representation of everyday knowledge that can be used for semantic similarity, reasoning, and knowledge-augmented modeling. As mentioned before, we construct a ConceptNet list using the seed words: weather, climate, temperature, environment and natural disaster. We then retrieve all associated words and phrases for each seed that have a relevance score of greater than 1 and then manually verify the relevance of each item. We finally end up with a list of 228 words which we employ to classify motifs as environmental vs non-environmental related using a dictionary based approach like \citet{folklore}, i.e. we check the occurrence of the words from this list in the description of the motifs to classify whether a motif is environmentally oriented or not.\footnote{For more details on ConceptNet, see \citep{conceptnet}.}
            \item \textbf{Classification procedure based on syntactic structure.} To reduce dimensionality, we extract subject–verb–object (SVO) triplets from each motif using spaCy’s dependency parser. Subject–verb–object (SVO) triplets are a structured way of representing the basic relational content of text by identifying who is performing an action, what the action is, and who or what is affected by it. This structure captures the core meaning of a sentence in a compact and interpretable form.\footnote{By reducing unstructured text to SVO triplets, complex narratives can be transformed into standardized relational units that are comparable across documents, speakers, or contexts. SVO triplets are widely used in text analysis because they preserve directional and role-based information that is lost in bag-of-words or topic-based approaches. Unlike word counts or embeddings alone, SVO representations explicitly encode agency and responsibility by distinguishing actors from recipients of actions, making them particularly useful for analyzing narratives.} Each motif text is first parsed to identify its grammatical structure. For every verb within the motif, we then identify the corresponding subject and object based on their syntactic relationships to the verb, expanding these arguments to full noun phrases when relevant. We retain the set of unique subjects and objects obtained from the SVO triplets, which yields a substantially smaller and more interpretable vocabulary. We manually classify these terms as environmental or non-environmental (e.g., whether they refer to plants, animals, or other natural entities). Using this classification, we label a motif as environmentally related if it contains at least one subject or object classified as environmental. We use this alternative procedure to construct an additional measure of environmental focus in folklore, which serves as a robustness check for our main classification approach.
            \end{enumerate}

    \subsection{The Joshua Project}\label{subsec:Joshua-details}

    	We construct an alternative group-level measure of environmental attention using textual descriptions from the Joshua Project. Similar in spirit to the folklore-based measure, this approach treats written ethnographic descriptions as reflecting the cumulative stock of experiences, beliefs, and knowledge embodied within an ethnic group over time.
    	
    	The Joshua Project dataset aggregates information from a wide range of secondary sources - including global, regional, and national researchers, mission organizations, census materials, and other published records - which are integrated to provide a consistent global picture of ethnic groups, including their geographic distribution, population estimates, and descriptive narratives of social organization and cultural practices. While the project originated in a religious and missionary context, its dataset draws on diverse ethnographic, linguistic, and historical sources and has been used in academic research as a practical resource for harmonized ethnic group identifiers and descriptive cultural information, particularly in settings where official census data on ethnicity are sparse or inconsistent.
    	
    	We begin by downloading the full Joshua Project dataset and identifying all ethnic groups for which descriptive information is available. For each group, we scrape textual content from the project’s official website, which provides narrative descriptions of ethnic groups in a relatively standardized format. These descriptions are generally organized into key sections such as a brief introduction, accounts of daily life and subsistence, and discussions of beliefs and worldviews.\footnote{The website also includes a separate section focused on religious needs, which is explicitly oriented toward Christian missionary activity. Because this content reflects an external normative framing rather than endogenous cultural beliefs, we exclude it from our analysis.}
    	
    	To understand how environmental attention is expressed in these texts, we manually processed a random subset of 100 group descriptions with the help of trained research assistants. This exercise was used to identify recurring thematic structures and linguistic patterns through which environmental conditions, ecological constraints, and nature-related beliefs are discussed. In particular, this manual review informed our understanding of how environmental attention appears in narratives related to subsistence activities, exposure to climate and weather shocks, natural resource dependence, environmental degradation, and spirituality linked to nature.
    	
    	Guided by this analysis, we then apply natural language processing tools to the full corpus of Joshua Project texts. Specifically, we use a BART-based zero-shot text classification approach, fine-tuned over MultiNLI \citep{williams2018broad}.\footnote{BART is a transformer-based sequence-to-sequence model that can be adapted for classification tasks using a zero-shot framework grounded in natural language inference. Specifically, we use a version of BART fine-tuned on the MultiNLI corpus, which trains the model to determine whether a premise sentence entails, contradicts, or is neutral with respect to a hypothesis. In a zero-shot setting, classification is implemented by converting each candidate thematic category into a short hypothesis statement and scoring the degree to which the input text entails that hypothesis. The model assigns the category with the highest entailment probability, without any task-specific retraining or fine-tuning on our data.}\textsuperscript{,}\footnote{The use of a MultiNLI-trained model is particularly well-suited for descriptive classifications of ethnographic and encyclopedic text—such as content drawn from the Joshua Project website or Wikipedia—because MultiNLI exposes the model to a wide range of genres (including fiction, government documents, spoken dialogue, and written nonfiction) during training. This broad genre coverage improves the model’s ability to generalize to diverse narrative and descriptive styles. Importantly, the entailment-based formulation aligns closely with our conceptual objective: we are not asking the model to learn dataset-specific labels, but rather to evaluate whether a given description semantically implies the presence of particular thematic content. This approach allows for flexible, transparent, and reproducible categorization of text into conceptually meaningful themes while maintaining interpretability and eliminating researcher-driven retraining choices.} We classify each group’s textual description along several predefined dimensions capturing environmental relevance: (i) references to environment and ecology, (ii) discussion of climate or weather shocks as central features of group life, (iii) embedding of natural resources and ecological constraints in subsistence activities, (iv) resistance to or concern about environmental degradation and threats, and (v) expressions of spirituality or belief systems explicitly linked to nature. We aggregate these classification scores to construct a composite index, which we normalize to lie between 0 and 1 and interpret as our second group-level measure of attention to environmental issues.
    	
    	The Joshua Project database also provides information on the geographic location of ethnic groups, which allows us to match groups to surface grid cells in the temperature anomaly dataset constructed by \citet{mann_2009}. Additional control variables include indicators for whether a group is classified as indigenous, measures of religious composition (which also proxy for the extent and quality of available descriptive information), and classifications identifying frontier or least-reached groups. Our final sample includes 12,988 ethnic groups distributed globally.

    \subsection{Wikipedia}\label{subsec:Wikipedia-details}

        To construct the final group-level measure of environmental attention we rely on textual descriptions scraped from English Wikipedia pages on ethnic groups. Similar to the Joshua Project–based measure, this approach treats encyclopedic written records as reflecting a cumulative stock of knowledge, experiences, and cultural narratives associated with a group, while recognizing that such descriptions are curated and externally authored.
	
    	We first manually compile a list of English Wikipedia pages for the corresponding ethnic groups in the ancestral characteristics database. We then scrape the full text of the corresponding Wikipedia page which typically provide narrative content summarizing a group’s livelihood strategies, history, social organization, and interactions with external actors. In a small number of cases, information on the ethnic group itself is unavailable; in such instances, we rely on pages describing the language spoken by the group or the political or geographic entity with which the group is associated, which often contain brief ethnographic descriptions of the group.
    	
    	To identify how environmental attention is expressed in these narratives, we first manually reviewed a random subset of 50 Wikipedia pages with the assistance of trained research assistants. This manual review was used to identify recurring themes, narrative structures, and linguistic patterns through which environmental conditions, ecological knowledge, and nature-centered beliefs are discussed. Based on this exercise, we defined four thematic dimensions capturing environment-centered cultural narratives: (i) nature's central to group identity and livelihood; (ii) indigenous religious or spiritual systems that link ancestors or spirits directly to specific natural entities; (iii) detailed traditional ecological knowledge related to desert, marine, forest, or island environments; and (iv) narratives of cultural resistance to external actors, including corporations, missionaries, or states, that are perceived as a threat to environmental factors.
    	
    	Guided by these themes, we apply a natural language classification approach to the full corpus of Wikipedia texts using a BART-based zero-shot classification framework, fine-tuned over MultiNLI. In this zero-shot setting, the model evaluates each group’s textual description against the four predefined themes without requiring task-specific retraining, allowing for flexible and interpretable classification of ethnographic narratives. For each group, we aggregate classification scores across the four dimensions to construct a composite index, which we normalize to lie between 0 and 1 and interpret as a measure of environmental relevance embedded in encyclopedic descriptions of ethnic groups.
    	
    	In addition, following \citet{wiki2022}, we collect data on the total number of Wikipedia language editions in which a biography or description of the ethnic group exists, which we use as a proxy for the breadth of documentation available for each group. Finally, we match the resulting environmental attention scores back to the ancestral characteristics database, which allows us to link ethnic groups to surface grid cells in the temperature anomaly dataset constructed by \citet{mann_2009}.

    \subsection{Ancestral Ethnic Groups}\label{subsec:ancestors-details}

        \citet{nunn_giu_database} database on Ancestral Characteristics of Modern Populations combines the pre-industrial characteristics of ethnic groups available in Murdock's Ethnographic Atlas, \citet{Bondarenko} dataset on People of Easternmost Europe and \citet{Murdock} World Ethnographic sample. with the 1265 groups available from the Ethnographic Atlas. \citet{nunn_giu_database} manually match over 7,000 different dialects and languages from \posscite{gordon} Ethnologue: Languages of the World to the ethnic groups from the ethnographic data sources.
        
        Information on the pre-industrial characteristics of the ethnic groups in the ethnic atlases has been coded for the earliest period for which satisfactory ethnographic data is available or can be reconstructed. 
		In total, 23 ethnicities are observed during the seventeenth century or earlier, 16 during the eighteenth century, 310 during the nineteenth century, 876 between 1900 and 1950 and 31 after 1950. See \citet{bahrami2021tabulated} for the relevance of the \textit{Ethnographic Atlas} in capturing traditional practices.
		The other two samples that \citet{nunn_giu_database} use, \citet{Bondarenko} and  \posscite{Murdock} World Ethnographic Sample, add 27 more ethnic groups to the 1265 groups available from the Ethnographic Atlas. The availability of the latitude and longitude of the ethnic groups location in the ethnographic datasets allows us to match the ethnographic groups to the surface grids in the \citet{mann_2009} temperature anomalies database. 

        Historical pre-industrial characteristics of ethnic group controls that we inclue in our regressions include:
        \begin{itemize}
            \item Primary mode of subsistence; Fishing, Hunting, Gathering, Animal Husbandry and Pastoralism and Agriculture.
            \item complexity of settlement: values go from 1 to 8. 1: fully-nomadic, 2: semi-nomadic, 3: semisedentary, 4: compact but impermenant settlements, 5: neighborhoods of dispersed family homesteads, 6: separated hamlets forming a single community, 7: compact and relatively permanent, 8: complex settlements.
            \item Size of local community: Ranging from 1 to 8 where 1: fewer than 50, 2: 50-99, 3: 100-199, 4: 200-399, 5: 400-1,000, 6: 1,000-4,999, 7: 5,000-50,000 and 8: $> 50,000$.
            \item Intensity of agriculture: ranging from 1 to 6 where 1 corresponds to no agriculture, 2 to casual agriculture, ... , 6 to intensive irrigated agriculture.
            \item Level of local and global jurisdictional development: ranging from 1 to 5, denoting the number of levels within a jurisdiction.
            \item Ethnicity level geographical variables such as distance to closest coast and equator and K\"{o}ppen climate classification of the spatial grid the ethnic group was historically located in.
        \end{itemize}

    \subsection{Climate conditions}\label{sec:mann}

\begin{figure}[h!]
                \centering
    			\includegraphics[scale=0.4]{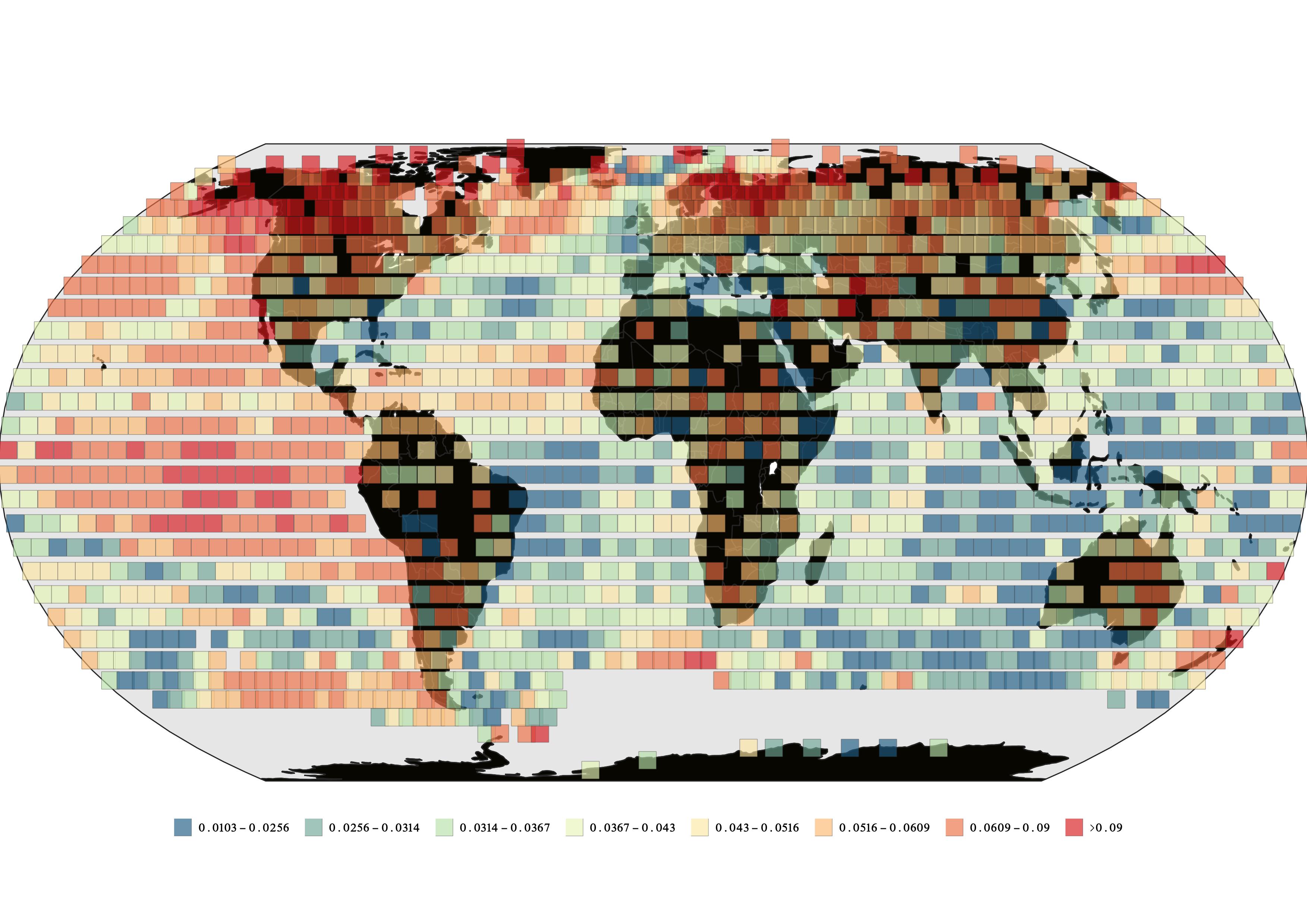}
                \caption{Spatial distribution of Average Ancestral Climatic Variability}      
                \label{temp_map}
    			\fnote{\textbf{Note:} Distribution of average ancestral climatic variability constructed from \citet{mann_2009}. The variable reports the average intensity of deviations from a defined range of normal temperatures specific to a generation. Generation lifespan is assumed to be 20 years and the normal range of temperature are the temperatures within the 20\textsuperscript{th} and 80\textsuperscript{th} percentile of temperatures within the lifespan of a generation. Variable is constructed at the 5 by 5 degree grid level. For more details refer to \Cref{sec:data}{Section}.}
    		\end{figure}

        We construct measures of ancestral climatic exposure using the climate field reconstruction (CFR) data from \citet{mann_2009}. The CFR approach reconstructs past surface temperature fields by combining instrumental temperature records with a large network of climate proxy data, allowing for spatially resolved estimates of temperature variability over long historical horizons. The reconstruction relies on a multiproxy dataset with broad global coverage, comprising 1,036 tree-ring series, 32 ice core series, 15 marine coral series, 19 documentary series, 14 speleothem series, 19 lacustrine (lake) sediment series, and 3 marine sediment series \citep{mann2}. These proxy records are statistically calibrated against observed temperature data over the instrumental period and then used to infer historical temperature patterns at the grid-cell level. Details on the spatial distribution of climate conditions are provided in \Cref{temp_map}.

\section{Robustness: Tables and Figures}

The following section presents detailed results for the robustness checks in \Cref{sec:robustness-checks}. 

\subsection{Choice of language sample and Contemporaneous Region Confounds}

    {\setstretch{1} \footnotesize \textbf{Notes on \Cref{robust}:}
    The unit of observation is an individual. The dependent variable is the individual's level of attention paid to environment. The dependent variable ranges between 0 and 1 and increases with the reported level of attention. The variable is constructed by rescaling the answer to the prompt: On a scale of 1 to 6, how important is it for this individual to take care of the environment. Average anomalies refers to the average intensity of deviations from the typical climate conditions (\textit{specific to each ancestral generation}) across generations. Average anomalies sq. refers to the square of the average anomalies term. Average anomalies ranges between 0.015 and 0.093 within the sample. Region-year fixed effects control for geography specific year fixed effects at a more granular level than the country. Historical ethnic group characteristics include measure of development such as agricultural intensity, complexity of settlement, level of political hierarchies, size of the local community and main source of subsistence. Demographic controls include dummies for income deciles, occupation categories, gender, education level and age. Historical topographic characteristics include controls for distance to the equator, distance to the closest coast and geographical K\"{o}ppen climate classification for the location of the ethnic group obtained from Ethnographic Atlas. Lang. at home only refers to the case where we remove all individuals for whom we proxied the language spoken at home by the language of interview. Standard errors are clustered at the ethnicity level.\par}

    \begin{table}[!h]\centering
    \resizebox{\linewidth}{!}{%
\begin{threeparttable}
\caption{Robustness checks: Coefficients for the impact of average ancestral climatic anomalies on individual's self-reported attention to the environment \label{robust} }
\begin{tabular}{lcccccc}
\hline
\multicolumn{7}{c}{Dependent variable: Self reported measure of attention to environment}\\
                    &         (1)   &         (2)   &         (3)   &         (4)   &         (5)   &         (6)   \\
\hline
\hline
Avg. Anomalies    &      -9.877***&      -9.243***&      -8.903***&     -14.751***&      -9.897***&      -6.649***\\
                    &     (2.003)   &     (1.988)   &     (2.206)   &     (2.647)   &     (2.011)   &     (2.023)   \\
Avg. Anomalies sq.&     111.767***&     107.503***&     103.956***&     163.717***&     111.862***&      71.929***\\
                    &    (21.346)   &    (21.557)   &    (25.184)   &    (27.890)   &    (21.505)   &    (22.402)   \\
Demographic Controls & Y & Y & Y & Y & Y & Y \\
Historical ethnic group characteristics & Y & Y & Y & Y & Y & Y \\
Historical topographic characteristics & Y & Y & Y & Y & Y & Y \\
Country-year Fixed effects & Y & Y & Y & Y & Y & N \\
Region-year Fixed effects & N & N & N & N & N & Y \\
\multicolumn{7}{l}{\textit{\underline{Sample restrictions}}}\\
English speakers excluded & N & Y & Y & N & N & N \\
Spanish speakers excluded & N & Y & Y & N & N & N \\
Arabic speakers excluded & N & N & Y & N & N & N \\
Weights at ethnicity level & N & N & N & Y & N & N \\
Lang. at home only & N & N & N & N & Y & N \\
\hline
Mean of dep var     &       0.704   &       0.700   &       0.689   &       0.704   &       0.703   &       0.705   \\
St. Dev. of dep var &       0.250   &       0.251   &       0.249   &       0.250   &       0.251   &       0.250   \\
Min value Avg. Anomalies&       0.015   &       0.015   &       0.015   &       0.015   &       0.015   &       0.015   \\
Max value Avg. Anomalies&       0.093   &       0.093   &       0.093   &       0.093   &       0.093   &       0.093   \\
R-sq                &       0.113   &       0.122   &       0.118   &       0.117   &       0.114   &       0.160   \\
Adj. R-sq           &       0.113   &       0.121   &       0.117   &       0.117   &       0.113   &       0.151   \\
N                   &      138067   &      110484   &       93184   &    138067   &      128532   &      135228   \\
\hline
\multicolumn{7}{c}{ {*}{*}{*} p$<$0.01, {*}{*}p$<$0.05, {*} p$<$0.1, {+} p$<$0.15}\\
\hline
\hline
\end{tabular}
\end{threeparttable}}
\end{table}

    \subsection{Construction of Ancestral Climatic Anomalies}

    {\setstretch{1} \footnotesize
        \textbf{Notes on \Cref{method_var}:} The unit of observation is an individual. The dependent variable is the individual's level of attention paid to environment. The dependent variable ranges between 0 and 1 and increases with the reported level of attention. The variable is constructed by rescaling the answer to the prompt: On a scale of 1 to 6, how important is it for this individual to take care of the environment. Avg. refers to the average intensity of deviations from the typical climate conditions (\textit{specific to each ancestral generation}) across generations. Avg. sq. refers to the square of the average anomalies term. Every regression includes controls for demographic characteristics, ethnicity level group and topographic characteristics and country-year fixed effects. Each column varies the lifespan of one ancestral generation in our method of construction of the average anomalies term. Each row varies the choice of the typical range. Dev. from 20-80 (our main specification) refers to the deviations from the cut points of the range of typical temperatures for each generation, where typical temperatures lie between $20^\text{th}$ and $80^\text{th}$ percentile of generation specific temperatures. Dev. from 10-90 and 30-70 are defined analogously. Dev. (sq.) from 20-80 instead squares the deviations from the cut points before taking the average. Std. Dev. instead of defining the deviations from the typical range just takes the Standard deviation of temperatures within a generation as a measure of climatic shocks.  {*}{*}{*}p$<$0.01, {*}{*}p$<$0.05, {*}p$<$0.1, {+} p$<$0.15\par}
    
    \begin{table}[h!]
         \centering
            \begin{threeparttable}
                \caption{Coefficient on average ancestral climatic anomalies - Robustness by method of variable construction \label{method_var}}
                \centering
                \begin{tabular}{llcccc}\hline\hline
                \multicolumn{6}{c}{Dependent variable: Self reported measure of attention to environment}\\
                \hline
                \textbf{Lifespan of generation:} & & \underline{20 years} & \underline{30 years} & \underline{40 years} & \underline{50 years} \\
                \multirow{4}{*}{Dev. from 10-90} & Avg. & -14.742*** & -13.926*** & -10.453*** & -5.884*** \\
                 &  & (4.739) & (2.433) & (2.067) & (1.770) \\
                 & Avg. sq. & 316.651*** & 235.773*** & 143.257*** & 64.708*** \\
                 &  & (103.216) & (40.585) & (25.865) & (18.135) \\
                 \hline
                \multirow{4}{*}{Dev. from 20-80} & Avg. & -9.877*** & -5.064*** & -5.618*** & -4.423*** \\
                 &  & (2.003) & (1.770) & (1.159) & (1.256) \\
                 & Avg. sq. & 111.767*** & 44.420*** & 44.969*** & 31.565*** \\
                 &  & (21.346) & (16.391) & (8.496) & (8.888) \\
                 \hline
                \multirow{4}{*}{Dev. from 30-70} & Avg. & -7.082*** & -2.055+ & -3.529*** & -3.402*** \\
                 &  & (1.602) & (1.399) & (1.095) & (1.132) \\
                 & Avg. sq. & 57.857*** & 12.856 & 23.773*** & 19.033*** \\
                 &  & (12.367) & (9.672) & (6.999) & (6.150) \\
                 \hline
                \multirow{4}{*}{Dev. (sq.) from 20-80} & Avg. & -9.064* & -13.880*** & -7.059** & -5.105** \\
                 &  & (4.686) & (4.641) & (2.750) & (2.144) \\
                 & Avg. sq. & 1076.265*** & 1016.342*** & 400.162*** & 238.084*** \\
                 &  & (377.298) & (342.086) & (127.251) & (74.105) \\
                 \hline
                \multirow{4}{*}{Standard Deviation} & Avg. & -4.131*** & -1.675* & -1.529* & -1.707** \\
                 &  & (0.993) & (0.993) & (0.922) & (0.799) \\
                 & Avg. sq. & 21.710*** & 6.709+ & 6.903+ & 6.353** \\
                 &  & (5.015) & (4.347) & (4.316) & (3.039) \\
                \hline
                \hline
                \end{tabular}
            \end{threeparttable}
        \end{table}

    \subsection{Other characteristics of temperature distribution and model fit}

     {\setstretch{1} \footnotesize \textbf{Notes on \Cref{highermom}:}
     The unit of observation is an individual. The dependent variable is the individual's level of attention paid to environment. The dependent variable ranges between 0 and 1 and increases with the reported level of attention. The variable is constructed by rescaling the answer to the prompt: On a scale of 1 to 6, how important is it for this individual to take care of the environment. Average anomalies refers to the average intensity of deviations from the typical climate conditions (\textit{specific to each ancestral generation}) across generations. Average anomalies sq. (cub.) refers to the square (cube) of the average anomalies term. Variability across generations captures the standard deviations in the average temperatures across generations, i.e. how distinct the overall climate conditions were between generations. Moments of the full temperature distribution include the mean, standard deviation, skewness and kurtosis of the whole temperature distribution associated to an ethnic group in the 320 years (1600-1920) used to construct the per-generation anomalies index. Average anomalies ranges between 0.015 and 0.093 within the sample. Demographic controls include dummies for income deciles, occupation categories, gender, education level and age. Historical ethnic group characteristics include measure of development such as agricultural intensity, complexity of settlement, level of political hierarchies, size of the local community and main source of subsistence. Historical topographic characteristics include controls for distance to the equator, distance to the closest coast and geographical K\"{o}ppen climate classification for the location of the ethnic group obtained from Ethnographic Atlas. Standard errors are clustered at the ethnicity level.\par }
    
    \begin{table}[h!]
        \centering
         \caption{Coefficients for the impact of average ancestral climatic anomalies on individual's self-reported attention to the environment \label{highermom}}
            \begin{threeparttable}
                \centering
                \begin{tabular}{lccccc}
                \hline
                \multicolumn{6}{c}{Dependent variable: Self reported measure of attention to environment}\\
                                    &         (1)   &         (2)   &         (3)   &         (4)   &         (5)   \\
                \hline
                \hline
                Avg. Anomalies    &       0.177   &      -9.877***&     -17.402** &      -9.055***&      -9.677***\\
                                    &     (0.428)   &     (2.003)   &     (7.035)   &     (2.108)   &     (2.135)   \\
                Avg. Anomalies sq.&               &     111.767***&     276.862*  &     105.537***&     109.681***\\
                                    &               &    (21.346)   &   (145.560)   &    (21.700)   &    (21.829)   \\
                Avg. Anomalies cub.&               &               &   -1130.897   &               &               \\
                                    &               &               &   (951.849)   &               &               \\
                Variability across generations&               &               &               &      -0.166   &               \\
                                    &               &               &               &     (0.123)   &               \\
                Demographic Controls & Y & Y & Y & Y & Y \\
                Historical ethnic group characteristics & Y & Y & Y & Y & Y \\
                Historical topographic characteristics & Y & Y & Y & Y & Y \\
                Moments of full temp. distribution & N & N & N & N & Y \\
                Country-year Fixed effects & Y & Y & Y & Y & Y\\
                \hline
                Mean of dep var     &       0.704   &       0.704   &       0.704   &       0.704   &       0.704   \\
                St. Dev. of dep var &       0.250   &       0.250   &       0.250   &       0.250   &       0.250   \\
                Min value Avg. Anomalies&       0.015   &       0.015   &       0.015   &       0.015   &       0.015   \\
                Max value Avg. Anomalies&       0.093   &       0.093   &       0.093   &       0.093   &       0.093   \\
                R-sq                &       0.112   &       0.113   &       0.114   &       0.114   &       0.114   \\
                Adj. R-sq           &       0.111   &       0.113   &       0.113   &       0.113   &       0.113   \\
                N                   &      138067   &      138067   &      138067   &      138067   &      138067   \\
                AIC                 &    -7632.03   &    -7822.44   &    -7827.27   &    -7832.44   &    -7868.15   \\
                BIC                 &    -7346.80   &    -7527.37   &    -7522.37   &    -7527.54   &    -7533.75   \\
                \hline
                \multicolumn{6}{c}{ {*}{*}{*} p$<$0.01, {*}{*}p$<$0.05, {*} p$<$0.1, {+} p$<$0.15}\\
                \hline
                \hline
                \end{tabular}
            \end{threeparttable}
        \end{table}

    \subsection{Alternate functional forms for Folklore}

    {\setstretch{1} \footnotesize \textbf{Notes on \Cref{folklore-altspec}:}
     The unit of observation is an ethnic group. Average anomalies refers to the average intensity of deviations from the typical climate conditions (\textit{specific to each ancestral generation}) across generations. Average anomalies sq. refers to the square of the average anomalies term. Average anomalies ranges between 0.01 and 0.097 within the sample. Folklore document characteristics include the first year of publication, number of authors that have covered the motifs in their works, number of languages, publishers and publications that contain the motif records. Historical ethnic group characteristics include measure of development such as agricultural intensity, complexity of settlement, level of political hierarchies and main source of subsistence. Historical topographic characteristics include controls for distance to the equator, distance to the closest coast and geographical K\"{o}ppen climate classification for the location of the ethnic group obtained from Ethnographic Atlas. Standard errors are clustered at language group level.\par}

    \begin{table}[!h]\centering
        \caption{Coefficients for the impact of average ancestral climate anomalies on occurrences of environment-related ethnic folklore (Different Specifications)  \label{folklore-altspec}}
        \resizebox{\linewidth}{!}{%
        \begin{threeparttable}
        \begin{tabular}{lc|cccc}
        \hline
         & \multicolumn{1}{c}{\underline{SVO Triplets}} & \multicolumn{4}{c}{\underline{ConceptNet}} \\
                            &log(1+Share)   &log(1+Share)   &       Share   &Asinh(Share)   & Log-Poisson   \\
        \hline
        \hline
        Avg. Anomalies    &      -1.524***&      -1.548** &      -1.943*  &      -1.884*  &      -5.527*  \\
                            &     (0.561)   &     (0.764)   &     (1.061)   &     (0.975)   &     (3.017)   \\
        Avg. Anomalies sq.&      16.219***&      17.237** &      22.423** &      21.324** &      63.612** \\
                            &     (5.881)   &     (7.112)   &     (9.964)   &     (9.116)   &    (27.998)   \\
        Folklore document characteristics & Y & Y & Y & Y & Y \\
        Historical ethnic group characteristics & Y & Y & Y & Y & Y \\
        Historical topographic characteristics & Y & Y & Y & Y & Y \\
        Country Fixed effects & Y & Y & Y & Y & Y \\
        \hline
        Mean of dep var     &       0.434   &       0.277   &       0.325   &       0.317   &       0.325   \\
        St. Dev. of dep var &       0.068   &       0.096   &       0.133   &       0.122   &       0.133   \\
        Min value Avg. Anomalies&       0.010   &       0.010   &       0.010   &       0.010   &       0.010   \\
        Max value Avg. Anomalies&       0.097   &       0.097   &       0.097   &       0.097   &       0.097   \\
        R-sq                &       0.323   &       0.439   &       0.417   &       0.432   &               \\
        Adj. R-sq           &       0.235   &       0.366   &       0.341   &       0.358   &               \\
        N                   &         982   &         982   &         982   &         982   &         982   \\
        \hline
        \multicolumn{6}{c}{ {*}{*}{*} p$<$0.01, {*}{*}p$<$0.05, {*} p$<$0.1, {+} p$<$0.15}\\
        \hline
        \hline
        \end{tabular}
        \end{threeparttable}}
        \end{table}

    \subsection{Conley standard errors to account for spatial correlation}

    {\setstretch{1} \footnotesize \textbf{Notes on \Cref{conley-table}:}
     The unit of observation is an individual for columns (1), (2) and (3). The unit of observation is an ethnic group for columns (4), (5) and (6). Average anomalies refers to the average intensity of deviations from the typical climate conditions (\textit{specific to each ancestral generation}) across generations. Average anomalies sq. refers to the square of the average anomalies term. Each regression includes the specified controls used in our primary analysis. Individual level controls include dummies for gender, occupation categories, age, income and education level available for individuals in the respective datasets. Historical ethnic group characteristics include measure of development such as agricultural intensity, complexity of settlement, level of political hierarchies and main source of subsistence. Historical topographic characteristics include controls for distance to the equator, distance to the closest coast and geographical K\"{o}ppen climate classification for the location of the ethnic group. Conley Standard Errors which account for spatial correlation within the range of 200 kms and 500 kms are reported in curly and square brackets respectively.\par}

    \begin{table}[!h]\centering
        \caption{Coefficients for average ancestral climatic Anomalies with Conley Standard Errors \label{conley-table}}
        \resizebox{\linewidth}{!}{
        \begin{threeparttable}
        \begin{tabular}{lcccccc}
        \hline
        & \multicolumn{3}{c}{\textbf{\underline{Individual Level Variables}}} & \multicolumn{3}{c}{\textbf{\underline{Group Level Variables}}} \\
        & \textbf{WVS} & \textbf{IVS} & \textbf{ESS} & \textbf{Folklore} & \textbf{Wikipedia} & \textbf{Joshua} \\
                            &         (1)   &         (2)   &         (3)   &         (4)   &         (5)   &         (6)   \\
        \hline
        \hline
        Avg. Anomalies    &      -9.877***&      -2.975** &      -2.391***&      -1.548** &      -4.088** &      -1.390** \\
                            &     (2.003)   &     (1.317)   &     (0.841)   &     (0.764)   &     (2.013)   &     (0.569)   \\
                            &     \{1.929\}   &     \{1.279\}   &     \{0.830\}   &     \{0.852\}   &     \{1.975\}   &     \{0.660\}   \\
                            &     [1.852]   &     [1.174]   &     [0.802]   &     [1.036]   &     [1.924]   &     [0.653]   \\
        Avg. Anomalies sq.&     111.767***&      35.442** &      17.951***&      17.237** &      42.798** &      13.163** \\
                            &    (21.346)   &    (14.899)   &     (6.511)   &     (7.112)   &    (19.281)   &     (5.539)   \\
                            &    \{20.717\}   &    \{14.512\}   &     \{6.455\}   &     \{8.858\}   &    \{18.998\}   &     \{6.105\}   \\
                            &    [19.685]   &    [12.956]   &     [6.259]   &     [9.906]   &    [18.829]   &     [6.072]   \\
        Relevant individual controls & Y & Y & Y & N & N & N \\
        Relevant ethnic group controls & Y & Y & Y & Y & Y & Y \\
        Country Fixed effects & Y & Y & Y & Y & Y & Y \\
        \hline
        Mean of dep var     &       0.704   &       0.443   &       0.580   &       0.277   &       0.508   &       0.238   \\
        St. Dev. of dep var &       0.250   &       0.230   &       0.136   &       0.096   &       0.212   &       0.154   \\
        Min value Avg. Anomalies&       0.015   &       0.015   &       0.022   &       0.010   &       0.010   &       0.010   \\
        Max value Avg. Anomalies&       0.093   &       0.093   &       0.119   &       0.097   &       0.097   &       0.097   \\
        R-sq                &       0.113   &       0.124   &       0.162   &       0.439   &       0.325   &       0.149   \\
        Adj. R-sq           &       0.113   &       0.123   &       0.160   &       0.366   &       0.161   &       0.131   \\
        N                   &      138067   &      219900   &       23787   &         982   &        1031   &       11866   \\
        \hline
        \multicolumn{7}{c}{ {*}{*}{*} p$<$0.01, {*}{*}p$<$0.05, {*} p$<$0.1, {+} p$<$0.15}\\
        \hline
        \hline
        \end{tabular}
        \end{threeparttable}
        }
    \end{table}

    \begin{figure}[t]
        \centering
        \input{run_061/plot_config.tex}
            \begin{tikzpicture}
             \pgfplotstableread[col sep=comma]{run_061/solutions.csv}\datatable
              \begin{groupplot}[
                group style={group size=3 by 2, vertical sep=.5cm},
                width=.36\textwidth,
                legend style={font=\footnotesize},
                label style={font=\footnotesize},
              ]
    
            \nextgroupplot[
              title={$\theta=\thetaA$},
              ylabel={value},
              ymin=0,
              xmin=0, xmax=6,
              xtick distance = 2,
              ytick distance = 2,
              ticklabel style = {font=\tiny},
              legend columns=2,
              height=5.5cm,
            ]
            \addplot+  [no marks, smooth,very thick, black] table[x=z, y=v, col sep=comma] {\datatable};
    
            \addplot+ [no marks, smooth,very thick, gray] table[x=z, y=f0, col sep=comma]{\datatable};
            
            \addplot+ [no marks, smooth,very thick,plotcolor2] table[x=z, y=fstarA, col sep=comma]{\datatable};
    
              \addplot+ [dashed, no marks, plotcolor1, domain=\pgfkeysvalueof{/pgfplots/xmin}:\pgfkeysvalueof{/pgfplots/xmax}] { \EqvA };

            \nextgroupplot[
              title={$\theta=\thetaB$},
              ymin=0,
              xmin=0, xmax=6,
              xtick distance = 2,
              ytick distance = 2,
              ticklabel style = {font=\tiny},
              legend columns=2,
              height=5.5cm,
            ]
            \addplot+  [no marks, smooth,very thick, black] table[x=z, y=v, col sep=comma] {\datatable};
    
            \addplot+ [no marks, smooth,very thick, gray] table[x=z, y=f0, col sep=comma]{\datatable};
    
            \addplot+ [no marks, smooth,very thick,plotcolor2] table[x=z, y=fstarB, col sep=comma]{\datatable};
    
              \addplot+ [dashed, no marks, plotcolor1, domain=\pgfkeysvalueof{/pgfplots/xmin}:\pgfkeysvalueof{/pgfplots/xmax}] { \EqvB };
    
            \nextgroupplot[
              title={$\theta=\thetaC$},
              ymin=0,
              xmin=0, xmax=6,
              xtick distance = 2,
              ytick distance = 2,
              ticklabel style = {font=\tiny},
              legend columns=2,
              height=5.5cm,
            ]
            \addplot+  [no marks, smooth,very thick, black] table[x=z, y=v, col sep=comma] {\datatable};
    
            \addplot+ [no marks, smooth,very thick, gray] table[x=z, y=f0, col sep=comma]{\datatable};
    
            \addplot+ [no marks, smooth,very thick,plotcolor2] table[x=z, y=fstarC, col sep=comma]{\datatable};
    
              \addplot+ [dashed, no marks, plotcolor1, domain=\pgfkeysvalueof{/pgfplots/xmin}:\pgfkeysvalueof{/pgfplots/xmax}] { \EqvC };

            \nextgroupplot[
              xlabel={z},
              ylabel={scaled density},
              xmin=0, xmax=6,
              ymin=0, ymax=1.0,
              xtick distance = 2,
              ytick = {0,.5,1},
              ticklabel style = {font=\tiny},
              legend columns=2,
              height=2.5cm,
            ]
    
              \addplot[name path=p, draw=none,forget plot] table[x=z, y=qA] {\datatable};
              \addplot[name path=zero, draw=none, forget plot] coordinates {(0,0) (6,0)};
              \addplot[fill=plotcolor1, draw=none] fill between[of=p and zero];
    
            \nextgroupplot[
              xlabel={z},
              xmin=0, xmax=6,
              ymin=0, ymax=1.0,
              xtick distance = 2,
              ytick = {0,.5,1},
              ticklabel style = {font=\tiny},
              legend columns=2,
              height=2.5cm,
            ]
    
              \addplot[name path=p, draw=none,forget plot] table[x=z, y=qB] {\datatable};
              \addplot[name path=zero, draw=none, forget plot] coordinates {(0,0) (6,0)};
              \addplot[fill=plotcolor1, draw=none] fill between[of=p and zero];
    
            \nextgroupplot[
              xlabel={z},
              xmin=0, xmax=6,
              ymin=0, ymax=1.0,
              xtick distance = 2,
              ytick = {0,.5,1},
              ticklabel style = {font=\tiny},
              legend columns=2,
              height=2.5cm,
            ]
    
              \addplot[name path=p, draw=none,forget plot] table[x=z, y=qC] {\datatable};
              \addplot[name path=zero, draw=none, forget plot] coordinates {(0,0) (6,0)};
              \addplot[fill=plotcolor1, draw=none] fill between[of=p and zero];
    
            \end{groupplot}
    
          \end{tikzpicture}
        \caption{Illustration of the effect of the default $f_0$}
        {\footnotesize Underlying specification and color-coding identical to \Cref{fig:three-panels-comparative-statics}; the default $f_0$ is the grey line.}
        \label{fig:default-effect-comparative-statics}
    \end{figure}

\section{Technical results and additional background}\label{online-app:theory}

    \subsection{Proof of technical lemmas}\label{online-app:technical-lemmas}

        \paragraph{Stability of spread under evolutionary operations}
        The following facts, which follow from standard facts on strongly log-concave distributions, are useful to characterize how the spread of population distributions evolves along the evolution process. 
        \begin{enumerate}[label=(\alph*)]
            \item \emph{Selection.} Reweighting a law with spread at most $s$ by $\varphi\bigl(\mathcal{L}(\cdot|\nu)\bigr)^t = e^{-t \tau \mathcal{L}(\cdot|\nu)}$, for $t \in [0,1]$, yields a law with spread at most $s / \bigl(1 - \tau\kappa s / (27\nu^2)\bigr)$, provided $\tau \kappa s < 27 \nu^2$. Indeed, minus the log-density increases by $t \tau \mathcal{L}(\cdot|\nu)$, whose second derivative is at least $-\tau\kappa/(27\nu^2)$.
            \item \emph{Combination and mutation.} Spreads add up under sums of independent random variables, and multiplying a random variable by a constant multiplies its spread by the square of that constant \citep[Theorem 3.7]{saumard2014log}. In particular, the average of two independent random variables with spread at most $s$ has spread at most $s/2$, and adding an independent Gaussian with variance $\rho \langle q, Cq \rangle$ adds $\rho \langle q, Cq \rangle$ to the spread.
            \item \emph{Variance bound.} If a random variable has spread at most $s$, the variance of any differentiable function of it is at most $s$ times the expected square of the derivative of that function \citep[Theorem 4.1]{brascamp1976extensions}.
        \end{enumerate}
        Applying (c) to the identity and to the squared deviation from the mean shows that a random variable with spread at most $s$ has variance at most $s$ and fourth central moment at most $5 s^2$.

        \paragraph{Proof of \Cref{lem:reduction}}    
        We argue by induction on $g$.
        At $g = 0$, $P_0 = \delta_{f_0}$ and the residual of $f_0$ is zero.
        Suppose (i) holds under $P_g$.
        Since $\Phi_g$ depends on $f$ only through $\Esp_q[f]$, reweighting $P_g$ by $\Phi_g$ changes the law of $\Esp_q[f]$ but leaves the residual independent of it, with an unchanged law; so (i) holds under $\tilde{P}_g$.
        Now let $f_1, f_2$ be independent draws from $\tilde{P}_g$ and $f'$ the new type of \Cref{assumption:mutation}, so that $f' - f_0 = (1-\rho\lambda)\bigl(\frac{f_1+f_2}{2} - f_0\bigr) + \sqrt{\rho} \, \varepsilon$.
        Integrating against $q$ gives the expression in (iii) for $\Esp_q[f']$, which depends only on $\Esp_q[f_1]$, $\Esp_q[f_2]$ and $\langle q, \varepsilon \rangle$.
        Subtracting, the residual of $f'$ equals $(1-\rho\lambda)$ times the average of the residuals of $f_1$ and $f_2$, plus $\sqrt{\rho}$ times the residual of the noise, $\varepsilon - \langle q, \varepsilon \rangle \frac{Cq}{\langle q, Cq \rangle}$.
        The latter is independent of $\langle q, \varepsilon \rangle$: the two are jointly Gaussian and uncorrelated, since $\mathrm{Cov}(\varepsilon(z), \langle q, \varepsilon \rangle) = Cq(z)$ and $\Var(\langle q, \varepsilon \rangle) = \langle q, Cq \rangle$.
        Hence the residual of $f'$ is a function of random variables independent of those which determine $\Esp_q[f']$, and its mean is $(1-\rho\lambda)$ times the mean residual under $\tilde{P}_g$, which is zero.
        This proves (i) at $g+1$, and (iii).
        For (ii), take expectations in the decomposition of $f$ into $\Esp_q[f]$ and its residual; for $\Sigma_g q = \Esp_{P_g}[(f-f_g)(\Esp_q[f] - m_g)]$, the residual contributes nothing by (i).
        \hfill $\square$

        \paragraph{Proof of \Cref{lem:one-step}}
        All losses below have second argument $\hat{\overline{v}}_g$, which we omit.
        For $t \in [0,1]$, consider the law of $\Esp_q[f]$ under $P_g$ reweighted by $e^{-t\tau\mathcal{L}}$, which interpolates between its law under $P_g$ ($t=0$) and under $\tilde{P}_g$ ($t=1$), and write $\Esp_t$, $\Var_t$, $\mathrm{Cov}_t$ for expectations, variances and covariances under it.
        Using fact (a) above, it has spread at most $2s$ for $s$ small enough.
        Differentiating in $t$ (which is justified since the loss grows at most linearly and laws with bounded spread have Gaussian tails):
        \begin{align*}
            \frac{d}{dt}\Esp_t\bigl[\Esp_q[f]\bigr] &= - \tau \, \mathrm{Cov}_t\bigl(\Esp_q[f], \mathcal{L}(\Esp_q[f])\bigr),
            \\
            \frac{d}{dt}\Var_t\bigl(\Esp_q[f]\bigr) &= - \tau \, \mathrm{Cov}_t\Bigl(\bigl(\Esp_q[f] - \Esp_t[\Esp_q[f]]\bigr)^2, \mathcal{L}(\Esp_q[f])\Bigr).
        \end{align*}
        By (c), the standard deviation of $\mathcal{L}(\Esp_q[f])$ under $\Esp_t$ is at most $\sqrt{2s} \sup|\mathcal{L}'|$, that of $\Esp_q[f]$ at most $\sqrt{2s}$, and that of the squared deviation at most $4s$ (by the moment bounds above).
        Using the Cauchy-Schwarz inequality then gives that the first covariance is $O(s)$ and the second is $O(s^{3/2})$.
        Integrating over $t \in [0,1]$ gives the first and third claims, and also shows that $\Var_t(\Esp_q[f]) = V_g + O(s^{3/2})$ and $\Esp_t[\Esp_q[f]] = m_g + O(s)$ for all $t$.
        For the second claim, write:
        \begin{align*}
            \mathrm{Cov}_t\bigl(\Esp_q[f], \mathcal{L}(\Esp_q[f])\bigr)
            = \mathcal{L}'(m_g) \Var_t\bigl(\Esp_q[f]\bigr) + \mathrm{Cov}_t\Bigl(\Esp_q[f], \mathcal{L}(\Esp_q[f]) - \mathcal{L}'(m_g) \Esp_q[f] \Bigr).
        \end{align*}
        The derivative of the function in the last covariance is $\mathcal{L}'(\cdot) - \mathcal{L}'(m_g)$, at most $\sup|\mathcal{L}''| \, |\cdot - m_g|$ in absolute value; so by (c) its variance is at most $2s \sup|\mathcal{L}''|^2 \Esp_t[(\Esp_q[f] - m_g)^2]$, which is $O(s^2)$ since $\Esp_t[(\Esp_q[f] - m_g)^2] = \Var_t(\Esp_q[f]) + (\Esp_t[\Esp_q[f]] - m_g)^2 = O(s)$.
        By the Cauchy-Schwarz inequality, the last covariance is $O(s^{3/2})$, so that $\mathrm{Cov}_t(\Esp_q[f], \mathcal{L}(\Esp_q[f])) = \mathcal{L}'(m_g) V_g + O(s^{3/2})$ for all $t$; integrating over $t$ gives the second claim.
        \hfill $\square$

        \paragraph{Proof of \Cref{lem:spread}}
        (i) We show by induction on $g \geq 1$ that the law of $\Esp_q[f]$ under $P_g$ has spread at most $3\rho\langle q, Cq \rangle$.
        In generation $1$, every new type is $f_0 + \sqrt{\rho}\,\varepsilon$, so the law of $\Esp_q[f]$ under $P_1$ is Gaussian with variance (hence spread) $\rho\langle q, Cq \rangle$.
        In later generations, this law is a convolution with the same Gaussian, so it has a positive, smooth density.
        Now assume the bound holds in generation $g$, and follow it through the three operations of one generation.
        Selection: for $\rho \leq 9\hat{\overline{v}}_g^2/(4\tau\kappa\langle q, Cq \rangle)$, which holds uniformly in $g$ for $\rho$ small enough since realized stakes are bounded below, (a) multiplies the bound by at most $\frac{4}{3}$, giving $4\rho\langle q, Cq \rangle$ under $\tilde{P}_g$.
        Combination: by \Cref{lem:reduction}(iii) and (b), averaging two independent draws halves the bound, and multiplying by $1-\rho\lambda$ and adding a constant does not increase it, giving $2\rho\langle q, Cq \rangle$.
        Mutation: adding the independent Gaussian $\sqrt{\rho}\,\langle q, \varepsilon \rangle$ adds its variance $\rho\langle q, Cq \rangle$, giving $3\rho\langle q, Cq \rangle$ in generation $g+1$.
        In words, the bound is carried from one generation to the next because combining two cultural models halves it, which more than offsets the widening due to selection.

        (ii) Integrating the exact recursion against $q$ gives $m_{g+1} - \Esp_q[f_0] = (1-\rho\lambda)(\tilde{m}_g - \Esp_q[f_0])$, and by \Cref{lem:one-step} and (i), $|\tilde{m}_g - m_g|$ is at most a constant times $\rho$.
        Hence $|m_{g+1} - \Esp_q[f_0]|$ is at most $(1-\rho\lambda)$ times $|m_g - \Esp_q[f_0]|$ plus that constant times $\rho$; by induction from $m_0 = \Esp_q[f_0]$, $|m_g - \Esp_q[f_0]|$ is at most that constant divided by $\lambda$.

        (iii) By \Cref{lem:reduction}(iii), the variance obeys exactly $V_{g+1} = \frac{1}{2}(1-\rho\lambda)^2 \tilde{V}_g + \rho \langle q, Cq \rangle$.
        Without selection ($\tilde{V}_g = V_g$), this is an affine recursion with contraction factor $\frac{1}{2}(1-\rho\lambda)^2 \leq \frac{1}{2}$ and fixed point $\rho \langle q, Cq \rangle / \bigl(1 - \frac{1}{2}(1-\rho\lambda)^2\bigr)$, which lies in $[\rho\langle q, Cq \rangle, 2\rho\langle q, Cq \rangle]$ and equals $2\rho\langle q, Cq \rangle + O(\rho^2)$.
        Selection adds $\frac{1}{2}(1-\rho\lambda)^2(\tilde{V}_g - V_g) = O(\rho^{3/2})$ in each generation, by \Cref{lem:one-step} and (i).
        So the distance of $V_g$ to the fixed point is at most halved in each generation, up to an error $O(\rho^{3/2})$.
        Starting from a distance at most $\rho\langle q, Cq \rangle$ in generation $1$ (since $V_1 = \rho\langle q, Cq \rangle$), it is at most $2^{1-g}\rho\langle q, Cq \rangle + O(\rho^{3/2})$ in generation $g$, by summing the geometric series of errors; adding the $O(\rho^2)$ distance from the fixed point to $2\rho\langle q, Cq \rangle$ gives (iii).
        \hfill $\square$

    \subsection{The Rational Inattention Problem}\label{subsec:rational-inattention-details}
    
        Most of the derivations for the rational inattention block are either fairly straightforward or standard in the rational inattention literature; the reader can refer to the survey by \cite{MackowiakMatejkaWiederholt2023} for a review of the literature and techniques used.
    
        Take a generic problem, let $\omega \sim \mathcal{N}(0,\sigma^2)$ be the unknown state and $u(a;\omega):=-v(\omega-a)^2$ the loss function (where we blackbox the stakes parameters $v>0$).
        The cost of information is given by:
            \(
            I(x,\omega) := H(\omega) - \Esp[H(\omega|x)]
            \)
        where $x$ denotes the signal about the state $\omega$ and $H$ denotes the entropy function; for any random variable $y \sim p$, $H(y)= - \int \log(p) dp$.
        It is usual to justify the rational inattention framework and its canonical entropy cost specification as a heuristic model for situations where attention can be flexibly allocated and there is enough time for attention to adapt and focus on relevant features of the environment. 
        In their recent survey of the literature, \citet{MackowiakMatejkaWiederholt2023} provide the following general intuition which captures well our context of analysis: \textit{"We consider RI to be an “as-if model” or a benchmark that applies well in repeated choice situations, or in choices over the long term. In these cases, the agent thinks about the optimal strategy once, and then applies it many times with little additional effort. Alternatively, it can be a strategy that the agent gradually learned through experience or stumbled upon it due to some evolutionary reasons."}
        
        The agent's rational inattention problem is a choice over signal structure $x$:\footnote{The factor $2$ in the cost is a normalization so that the cost for Gaussian signals reads $\kappa \log(\sigma^2/\xi)$ as in the main text.}
            \begin{align*}
                \max_{x : \Omega \rightarrow \Delta(X)} \Esp \biggl[ \max_a \Esp \bigl[ u(a;\omega) \bigm| x \bigr] \biggr] - 2\kappa I(x,\omega)
            \end{align*}   
        We can substantially simplify the problem above, and appeal to classical techniques in the rational inattention literature to obtain the closed form solution in \Cref{subsec:theory-model}.
        First, making $u$ explicit, exchanging integrals and observing that quadratic loss entails the optimal action for a given signal structure to be simply the conditional expectation, we get a representation in the form of a canonical Gaussian-Quadratic RI problem.
        Gaussian signals are optimal under a Gaussian prior and quadratic loss \citep{MackowiakWiederholt2009} which thanks to simplifications of the Shannon mutual information for Gaussian distributions allows us to recast both payoffs and costs first in terms of the induced posterior variance $\xi$, as stated in the main text.

\end{document}

%% file: preamble.tex
\usepackage[utf8]{inputenc}
\usepackage[T1]{fontenc}
\usepackage[english]{babel}

\usepackage{layout}
\usepackage{setspace}
\usepackage[top=2.5cm, bottom=2.5cm, left=2cm, right=2cm]{geometry}
\usepackage{ragged2e}
\usepackage{multirow}

\usepackage[bottom]{footmisc}     
\usepackage[all]{nowidow}         
\newcommand\blfootnote[1]{%
  \begingroup
  \renewcommand\thefootnote{}\footnote{#1}%
  \addtocounter{footnote}{-1}%
  \endgroup
}

\usepackage{eurosym}
\usepackage{lipsum}

\usepackage{titlesec}
\titleformat*{\paragraph}{\scshape\bfseries}

\usepackage{mathtools} 
\usepackage{amssymb} 
\usepackage{amsmath}
\usepackage{amsfonts} 
\usepackage{amsthm}  
\usepackage{thmtools} 
\usepackage{dsfont}
\usepackage{stmaryrd}
\usepackage{mathtools} 
\usepackage{amsbsy} 
\usepackage{bbm}  
\usepackage{units} 
\usepackage{bbm} 
\usepackage{empheq} 
\usepackage{cancel}

\theoremstyle{plain}
\newtheorem{theorem}{Theorem}
\newtheorem{proposition}{Proposition}
\newtheorem{lemma}{Lemma}

\theoremstyle{definition}

\newtheorem{assumption}{Assumption}

\theoremstyle{remark}

\DeclareMathOperator*{\argmin}{arg\,min}

\newcommand{\reals}{\mathbb{R}}
\newcommand{\Esp}{\mathbb{E}}
\newcommand{\Var}{\mathbb{V}}

\allowdisplaybreaks

\usepackage{threeparttablex}

\usepackage{enumitem}

\usepackage{graphicx} 
\usepackage{subcaption}
\usepackage{wrapfig} 
\usepackage[font=small, labelfont=sc]{caption}
\newcommand\fnote[1]{\captionsetup{font=footnotesize,width=0.8\textwidth}\caption*{#1}}
\usepackage{pdfpages}
\usepackage{tikz}
\usepackage{pgfplots}
\usepackage{pgfplotstable}
\usetikzlibrary{arrows.meta}
\usetikzlibrary{pgfplots.fillbetween}
\pgfplotsset{compat=1.18}
\usetikzlibrary{
  arrows.meta,   
  positioning,   
  calc           
}
\usepgfplotslibrary{groupplots}

\usepackage{xcolor}
\definecolor{Ccolor}{rgb}{.1,.3,1}

\definecolor{myblue}{rgb}{0.2 , 0.3 , 0.8}
\definecolor{mygreen}{RGB}{0, 150, 0}
\definecolor{myred}{RGB}{207,8,8}

    \definecolor{TolBlack}{HTML}{222222}
    \definecolor{TolDarkGrey}{HTML}{555555}
    \definecolor{TolGrey}{HTML}{BBBBBB}
    \definecolor{TolWhite}{HTML}{FFFFFF}

    \definecolor{TolMutedIndigo}{HTML}{332288}
    \definecolor{TolMutedCyan}{HTML}{88CCEE}
    \definecolor{TolMutedTeal}{HTML}{44AA99}
    \definecolor{TolMutedGreen}{HTML}{117733}
    \definecolor{TolMutedOlive}{HTML}{999933}
    \definecolor{TolMutedSand}{HTML}{DDCC77}
    \definecolor{TolMutedRose}{HTML}{CC6677}
    \definecolor{TolMutedWine}{HTML}{882255}
    \definecolor{TolMutedPurple}{HTML}{AA4499}

    \definecolor{TolBrightBlue}{HTML}{4477AA}
    \definecolor{TolBrightCyan}{HTML}{66CCEE}
    \definecolor{TolBrightGreen}{HTML}{228833}
    \definecolor{TolBrightYellow}{HTML}{CCBB44}
    \definecolor{TolBrightRed}{HTML}{EE6677}
    \definecolor{TolBrightPurple}{HTML}{AA3377}
    \definecolor{TolBrightGrey}{HTML}{BBBBBB}

    \definecolor{TolVibrantBlue}{HTML}{0077BB}
    \definecolor{TolVibrantCyan}{HTML}{33BBEE}
    \definecolor{TolVibrantTeal}{HTML}{009988}
    \definecolor{TolVibrantOrange}{HTML}{EE7733}
    \definecolor{TolVibrantRed}{HTML}{CC3311}
    \definecolor{TolVibrantMagenta}{HTML}{EE3377}
    \definecolor{TolVibrantGrey}{HTML}{BBBBBB}

    \colorlet{plotcolor1}{TolMutedCyan!80!black}
    \colorlet{plotcolor2}{TolMutedRose}
    \colorlet{plotcolor3}{TolMutedGreen}

\usepackage[hypertexnames=false]{hyperref} 
\usepackage[nameinlink,noabbrev]{cleveref}
\crefname{appendix}{appendix}{appendices}
\Crefname{appendix}{Appendix}{Appendices}
\crefname{assumption}{assumption}{assumptions}
\Crefname{assumption}{Assumption}{Assumptions}
\newcommand*{\hyref}[2]{\hyperref[#1]{#2 \ref{#1}}}
\hypersetup{
    pdffitwindow=false,                 
    pdfstartview={XYZ null null 1.00},          
    pdfnewwindow=true,                  
    colorlinks=true,                  
    linkcolor={red!50!black},             
    citecolor={blue!50!black},              
    urlcolor={red!50!black},                   
  linkbordercolor={white},              
}

%% file: run_062/plot_config.tex
\providecommand{\runlabels}{A,B,C}
\providecommand{\thetaA}{0.2}
\providecommand{\EqvA}{11.46804917599555}
\providecommand{\fstarColA}{fstarA}
\providecommand{\qColA}{qA}
\providecommand{\thetaB}{1.0}
\providecommand{\EqvB}{4.812950509057619}
\providecommand{\fstarColB}{fstarB}
\providecommand{\qColB}{qB}
\providecommand{\thetaC}{2.5}
\providecommand{\EqvC}{9.074252047231749}
\providecommand{\fstarColC}{fstarC}
\providecommand{\qColC}{qC}

%% file: sweep_001/plot_config.tex
\providecommand{\thetamin}{0.8288135593220339}
\providecommand{\Vmin}{5.7211322773498035}
\providecommand{\epsval}{2.6838668905646523e-06}
\providecommand{\deltaval}{0.0016022582851931819}
\providecommand{\thetamindelta}{0.8272113010368407}
\providecommand{\thetamlusdelta}{0.8304158176072272}

%% file: run_061/plot_config.tex
\providecommand{\runlabels}{A,B,C}
\providecommand{\thetaA}{0.2}
\providecommand{\EqvA}{11.46804917599555}
\providecommand{\fstarColA}{fstarA}
\providecommand{\qColA}{qA}
\providecommand{\thetaB}{1.0}
\providecommand{\EqvB}{4.812950509057619}
\providecommand{\fstarColB}{fstarB}
\providecommand{\qColB}{qB}
\providecommand{\thetaC}{2.5}
\providecommand{\EqvC}{9.074252047231749}
\providecommand{\fstarColC}{fstarC}
\providecommand{\qColC}{qC}